\documentclass[11pt,a4paper]{article} %{article} % Beginn der LaTeX-Datei
\pdfoutput=1
\usepackage{geometry}
\usepackage{amsmath,amssymb}  % erleichtert Mathe 
\usepackage{enumerate}% schicke Nummerierung
\usepackage{amsfonts}   % if you want the fonts
\usepackage{graphicx} % f??r Grafik-Einbindung
\usepackage{soul}

\mathchardef\ordinarycolon\mathcode`\:
\mathcode`\:=\string"8000
\begingroup \catcode`\:=\active
  \gdef:{\mathrel{\mathop\ordinarycolon}}
\endgroup

\usepackage{array,multirow}

\usepackage[english]{babel}
\usepackage[T1]{fontenc}
\usepackage{lmodern}
\usepackage[utf8]{inputenc}  % f??r Unix-Systeme
\usepackage{bbm}

\usepackage{lscape} % For matrix quer

\usepackage[numbers,sort&compress]{natbib}
\usepackage{slashed}

\usepackage{siunitx}

\usepackage{tikz}

\usepackage{float}

\usepackage{url}

\usepackage{hyperref}
\hypersetup{
  colorlinks   = true, %Colours links instead of ugly boxes
  urlcolor     = blue, %Colour for external hyperlinks
  linkcolor    = black, %Colour of internal links
  citecolor   = black %Colour of citations
}

\allowdisplaybreaks
\def    \beq            {\begin{equation}}
\def    \eeq            {\end{equation}}
\def    \bea           {\begin{eqnarray}}
\def    \eea           {\end{eqnarray}}

\newcommand\tb{\tan\beta}

\newcommand\ReDiag{\mathop{%
  \raise .5pt\hbox{[}%
  \widetilde{\mathrm{Re}}%
  \raise .5pt\hbox{]}}}
\newcommand\ReOffDiag{\mathop{%
  \raise .5pt\hbox{$\llbracket$}%
  \widetilde{\mathrm{Re}}%
  \raise .5pt\hbox{$\rrbracket$}}}

\newcommand\DRbar{\ensuremath{\smash{\overline{\mathrm{DR}}}}}

\newcommand\cL{{\cal L}}

\newcommand\SW{s_\mathrm{w}}
\newcommand\CW{c_\mathrm{w}}
\newcommand\MW{M_W}
\newcommand\MZ{M_Z}

\newcommand\MA{M_A}
\newcommand\MHp{M_{H^\pm}}

\newcommand\ino[1]{\chi_{#1}}

\newcommand\neu[1]{\ino{#1}^0}

\newcommand\refeq[1]{Eq.~(\ref{#1})}
\newcommand\refeqs[1]{Eqs.~(\ref{#1})}
\newcommand\refta[1]{Tab.~\ref{#1}}
\newcommand\refse[1]{Sect.~\ref{#1}}

\newcommand\citere[1]{Ref.~\cite{#1}}
\newcommand\citeres[1]{Refs.~\cite{#1}}

\newcommand{\mnSSM}{\ensuremath{\mu\nu\mathrm{SSM}}}

\newcommand{\CP}{{\cal CP}}
\newcommand{\cp}{{\CP}}

\newcommand{\eev}{\,\, \mathrm{eV}}
\newcommand{\tev}{\,\, \mathrm{TeV}}
\newcommand{\gev}{\,\, \mathrm{GeV}}

\newcommand\FA{\texttt{FeynArts}}
\newcommand\FC{\texttt{FormCalc}}
\newcommand\LT{\texttt{LoopTools}}
\newcommand\FH{\texttt{FeynHiggs}}
\newcommand\fh{\texttt{FeynHiggs}}

\def\order#1{\ensuremath{{\cal O}(#1)}}
\def\reffi#1{\mbox{Fig.~\ref{#1}}}

\def\als{\alpha_s}
\def\alt{\alpha_t}
\def\alb{\alpha_b}

\def\De{\Delta}
\def\de{\delta}
\def\la{\lambda}

\definecolor{Orange}{named}{orange}
\definecolor{Purple}{named}{purple}
\definecolor{Lightblue}{cmyk}{0.9,0.1,0.1,0.3}
\definecolor{dgelborange}{cmyk}{0.,0.3,0.5, 0.}
\definecolor{Lila}{rgb}{0.5,0.,1}

\usepackage{color}

\definecolor{white}{rgb}{1,1,1}
\definecolor{darkred}{rgb}{0.3,0,0}
\definecolor{darkgreen}{rgb}{0,0.3,0}
\definecolor{darkblue}{rgb}{0,0,0.3}
\definecolor{pink}{rgb}{0.78,0.09,0.51}
\definecolor{purple}{rgb}{0.28,0.24,0.55}
\definecolor{orange}{rgb}{1,0.6,0.0}
\definecolor{grey}{rgb}{0.4,0.4,0.4}
\definecolor{aquamarine}{rgb}{0.4,0.8,0.65}

\newcommand{\ii}{\text{i}}

\newcommand{\bal}{\begin{align}}
\newcommand{\eal}{\end{align}}

\allowdisplaybreaks
\usepackage{breqn}

\numberwithin{equation}{section}

\begin{document}
\thispagestyle{empty}

\def\thefootnote{\fnsymbol{footnote}}

\begin{flushright}
\mbox{}
IFT--UAM/CSIC--19-030 % \\
%arXiv:XXXX.XXXXX [hep-ph]
\end{flushright}

\vspace{0.5cm}

\begin{center}

{\large\sc 
{\bf Precise prediction for the Higgs-Boson Masses in the 
\boldmath{$\mu\nu$}SSM with three right-handed
neutrino superfields}}

\vspace{1cm}

{\sc
T.~Biek\"otter$^{1,2}$%
\footnote{email: thomas.biekotter@csic.es}
, S.~Heinemeyer$^{1,3,4}$%
\footnote{email: Sven.Heinemeyer@cern.ch}%
~and C.~Mu\~noz$^{1,2}$%
\footnote{email: c.munoz@uam.es}%
}

\vspace*{.7cm}

{\sl
$^1$Instituto de F\'isica Te\'orica UAM-CSIC, 
Cantoblanco, 28049, Madrid, Spain

\vspace*{0.1cm}

$^2$Departamento de F{\'i}sica Te{\'o}rica, Universidad Aut{\'o}noma 
de Madrid (UAM), \\
Campus de Cantoblanco, 28049 Madrid, Spain

\vspace*{0.1cm}

$^3$Campus of International Excellence UAM+CSIC, 
Cantoblanco, 28049, Madrid, Spain 

\vspace*{0.1cm}

$^4$Instituto de F\'isica de Cantabria (CSIC-UC), 
39005, Santander, Spain
}

\end{center}

\vspace*{0.1cm}

\begin{abstract}
\noindent
The \mnSSM\ is a simple supersymmetric extension of the Standard Model (SM)
capable of describing neutrino physics in agreement with experiments.
We perform the complete one-loop renormalization of the neutral scalar
sector of the \mnSSM\ with three generation of right-handed neutrinos in a
mixed on-shell/\DRbar\ scheme.
We calculate the full one-loop corrections to the neutral scalar masses
of the \mnSSM. The one-loop contributions are supplemented by available
MSSM higher-order corrections. We obtain numerical results
for a SM-like Higgs-boson mass consistent with experimental
bounds, while simultaneously agreeing with neutrino oscillation data.
We illustrate the distinct phenomenology of the \mnSSM\ in scenarios
in which one or more right-handed sneutrinos are
lighter than the SM-like Higgs boson, which might be substantially
mixed with them.
These scenarios are experimentally
accessible, on the one hand, through direct searches of the right-handed
sneutrinos decaying into SM particles, and on the other hand,
via the measurements of
the SM-like Higgs-boson mass and its couplings.
In this way the parameter space
of the \mnSSM\ can be probed without the need to
propose model dependent searches at colliders.
%dedicated searches.
Finally, we demonstrate how the \mnSSM\ can simultaneously accommodate
two excesses measured at LEP and LHC at $\sim 96\gev$ at the $1\sigma$ level,
while at the same time reproducing neutrino masses and mixings in
agreement with neutrino oscillation measurements.
\end{abstract}

\def\thefootnote{\arabic{footnote}}
\setcounter{page}{0}
\setcounter{footnote}{0}

\newpage

%%%%%%%%%%%%%%%%%%%%%%%%%%%%%%%%%%%%%%%%%%%%%%%%%%%%%%%%%%%%%%%%%%%%%%%%%%%%%%
%%%%%%%%%%%%%%%%%%%%%%%%%%%%%%%%%%%%%%%%%%%%%%%%%%%%%%%%%%%%%%%%%%%%%%%%%%%%%%

\section{Introduction}
\label{sec:intro}
 
The scalar particle at $\sim 125\gev$ discovered by the
ATLAS~\cite{Aad:2012tfa}
and CMS~\cite{Chatrchyan:2012xdj}
experiments has so far shown to be consistent with the
Standard Model (SM) Higgs-boson prediction. The Higgs
boson was the last missing piece in the description of electroweak symmetry
breaking (EWSB) and the generation of masses of fundamental
particles within the SM. The measurement of the mass
of this new state already reached a remarkable
precision~\cite{Aad:2015zhl}:\footnote{This value constitutes the last
ATLAS and CMS combination.
Newer measurements confirm the average within the given
uncertainties~\cite{Aaboud:2018wps,Sirunyan:2017exp}.}
\begin{equation}
m_H = 125.09 \pm 0.21 (\text{stat.}) \pm 0.11 (\text{syst.}) \; .
\end{equation}
However, other properties of the Higgs boson, while being in agreement with the
SM predictions, are still measured with relatively large
uncertainties~\cite{Sirunyan:2018koj,ATLAS-CONF-2018-031}.
Thus, even though any theory beyond the SM necessarily needs to
accommodate a state corresponding to a SM-like Higgs boson at $\sim 125\gev$,
there is still ample room for interpretations of the Higgs-boson signal with
sizable deviations w.r.t.\ the SM prediction.

Supersymmetry (SUSY) is one of the most studied beyond the Standard Model
(BSM) extensions. SUSY combines bosonic
and fermionic degrees of freedom of the fundamental fields and spacetime
itself. In particular, SUSY models predict two scalar particles
for each SM fermion and a fermion for each SM gauge boson.
% including a so-called Higgsino related to the Higgs field.
The simplest version of such models is the Minimal Supersymmetric
Standard Model (MSSM)~\cite{Nilles:1983ge,Haber:1984rc}.
Besides the doubling of the SM particle content due
to SUSY, the MSSM contains a second Higgs doublet which in the
$\cp$-conserving case leads to a physical spectrum of two $\cp$-even,
one $\cp$-odd and two charged Higgs bosons. Both the lighter or the heavier
$\cp$-even scalar can be interpreted as the SM-like Higgs boson at
$\sim 125\gev$~\cite{Heinemeyer:2011aa,Bechtle:2016kui,Bahl:2018zmf}.
Despite its simplicity, the MSSM is capable of fixing
a few shortcomings of the SM. If the breaking of SUSY takes place
not too far away from the electroweak scale, the
hierarchy problem~\cite{Weinberg:1975gm,tHooft:1979rat} is
solved by additional quantum corrections from the SUSY partners that
cancel large corrections to the Higgs mass from the heavy SM fermions.
Apart from that, the extended spectrum leads to the unification of
the three gauge couplings in a singular point at very high
energies~\cite{Georgi:1974sy}.
Due to the conservation of $R$-parity, the lightest supersymmetric
particle (LSP) is stable and can contribute to the dark matter relic
abundance~\cite{Goldberg:1983nd,Ellis:1983ew}.

However, the MSSM does not address all the open problems of the SM, and
also introduces new issues, motivating non-minimal SUSY
extensions of the SM. The most prominent example is the
Next-to-Minimal supersymmetric standard model
(NMSSM)~\cite{Ellwanger:2009dp,Maniatis:2009re}
which extends the particle content of the MSSM by a gauge-singlet superfield.
The $Z_3$-symmetric NMSSM provides a solution to the $\mu$-problem
by naturally associating an adequate scale to the $\mu$-parameter
in the MSSM superpotential~\cite{Ellis:1988er,Miller:2003ay}.
In the NMSSM, the fermionic component of the singlet superfield
(called singlino) extends the neutral fermion sector of the MSSM to
a total of five neutralinos. Assuming $\cp$-conservation, the complex
scalar component of the singlet superfield will extend the $\cp$-even
and the $\cp$-odd scalar sector by an additional particle state, respectively.

Neither the MSSM nor the NMSSM accommodate neutrino masses and lepton-flavor
violation in the neutrino sector. Therefore, a well motivated 
extension of the SM is the $\mu$-from-$\nu$ Supersymmetric
Standard Model (\mnSSM)~\cite{LopezFogliani:2005yw,Munoz:2009an}.
In this model, the particle content of the MSSM
is extended by right-handed neutrino superfields. Since they are gauge-singlets,
the $\mu$-problem can be solved in total analogy to the NMSSM.
Remarkably, in the \mnSSM\ it is possible to accommodate neutrino masses
and mixings in agreement with experiments via an electroweak seesaw
mechanism, dynamically generated during the
EWSB~\cite{LopezFogliani:2005yw,Escudero:2008jg,Ghosh:2008yh,Bartl:2009an,Fidalgo:2009dm,Ghosh:2010zi}.
In addition to the Higgs doublet fields also the right- and
left-handed scalar neutrinos
acquire a vacuum expectation value (vev). Thus, the \mnSSM\ solves the
$\mu$- and the $\nu$-problem (neutrino masses) simultaneously without
the need to introduce additional energy scales beyond the
SUSY-breaking scale. In contrast to the (N)MSSM, $R$-parity
and lepton number are not conserved,
leading to a completely different
phenomenology, characterized by distinct prompt or displaced
decays of the LSP~\cite{Ghosh:2017yeh,Lara:2018rwv,Lara:2018zvf,LaraPerez:2019thq}.
Although the LSP is not stable anymore, the \mnSSM\ can provide a
dark matter candidate with a gravitino with a lifetime longer
than the age of the
universe~\cite{Choi:2009ng,GomezVargas:2011ph,Albert:2014hwa,Gomez-Vargas:2016ocf}.
The breaking of $R$-parity is induced by a neutrino Yukawa
term, with the size of the couplings $Y^\nu_{ij} \leq 10^{-6}$
determined by the electroweak seesaw.
Because of the values of $Y^\nu_{ij}$, mixings between SM particles and their
supersymmetric partners are suppressed.
Nevertheless, the additional sources
of mixing effects induce a conceptually modified spectrum compared to the MSSM
and the NMSSM. The spectrum will be described in detail in
\refse{sec:mnSSM}.

\medskip

SUSY relates the quartic
couplings of the neutral scalar potential to the gauge couplings of
the underlying field theory.
Therefore, within SUSY the scalar masses can be predicted in terms
of other model parameters,
and the precise value of the SM-like Higgs-boson
mass is of particular significance. 
However, the SM-like Higgs-boson mass predictions strongly
depend on quantum corrections which can be calculated only
to certain order in perturbation theory. Missing higher-order
contributions lead to a sizable amount of uncertainty which
is usually of a few GeV (see below for details),
hence an order of magnitude
larger than the experimental uncertainty. This is why a lot of effort
is made to predict the Higgs mass to the highest
possible precision~\cite{kuts}.
We briefly summarize the status of Higgs-mass predictions in the MSSM,
the NMSSM, and the \mnSSM\ in the following.

In the MSSM the tree-level mass can be predicted by just two
SUSY parameters, i.e., the ratio of the vevs of the
Higgs doublets $\tan\beta$, and either the mass of the $\cp$-odd
Higgs boson $\MA$ or the mass of the charged Higgs boson $\MHp$,
leading to an upper bound given by the $Z$-boson mass.
Large loop corrections are needed to achieve a Higgs-boson mass
of $\sim 125\gev$. Beyond the one-loop level, the dominant two-loop
corrections of
$\order{\alt\als}$~\cite{Heinemeyer:1998jw,Heinemeyer:1998kz,Heinemeyer:1998np,Zhang:1998bm,Espinosa:1999zm,Degrassi:2001yf},
\order{\alt^2}~\cite{Espinosa:2000df,Brignole:2001jy},
$\order{\alb\als}$~\cite{Brignole:2002bz,Heinemeyer:2004xw} and
\order{\alt\alb}~\cite{Brignole:2002bz}
are known (here we use $\alpha_f = (Y^f)^2/(4\pi)$, with $Y^f$ denoting
the fermion Yukawa coupling).
These corrections, together with a
resummation of leading and subleading logarithms from the top/stop
sector~\cite{Hahn:2013ria} (see also~\cite{Draper:2013oza,Lee:2015uza} 
for more details on this type of approach),
a resummation of leading contributions from the bottom/sbottom
sector~\cite{Brignole:2002bz,Heinemeyer:2004xw,Hempfling:1993kv,Hall:1993gn,Carena:1994bv,Carena:1999py}
(see also~\cite{Noth:2008tw, Noth:2010jy}) and momentum-dependent two-loop
contributions~\cite{Borowka:2014wla, Borowka:2015ura} (see
also~\cite{Degrassi:2014pfa}) are included in the public
code~\fh~\cite{Heinemeyer:1998yj,Hahn:2009zz,Heinemeyer:1998np,Degrassi:2002fi,Frank:2006yh,Hahn:2013ria,Bahl:2016brp,Bahl:2017aev,Bahl:2018qog,feynhiggs-www}.
The most recent version of \fh\ contains an improved effective field
theory calculation relevant for large SUSY
scales~\cite{Bahl:2016brp,Bahl:2018qog,Bahl:2018ykj}.
The complete two-loop QCD contributions in the $\cp$-violating
MSSM were calculated in \citere{Borowka:2018anu}, but
not yet included in \fh.
A (nearly) full two-loop effective potential (EP) calculation,
including even the leading
three-loop corrections, has also been
published~\cite{Martin:2005eg,Martin:2007pg}, which is, however, not
publicly available as a computer code. Furthermore, another leading
three-loop calculation of \order{\alt\als^2}, depending on the various
SUSY mass hierarchies, has been
performed~\cite{Harlander:2008ju,Kant:2010tf}, resulting in the code 
{\tt H3m} and is now available as a stand-alone
  code~\cite{Harlander:2017kuc}.
It was proven that regularization by dimensional reduction
preserves supersymmetry at the required three-loop
order~\cite{Stockinger:2018oxe}.
A new calculation of the three-loop contributions of the $\order{\alt\als^2}$
extends the validity of these corrections to the whole parameter space
of the $\cp$-conserving MSSM~\cite{Reyes:2019ply}.
Most recently, the leading logarithmic terms of the $\order{\alt\als^3}$
have been obtained (see the updated version
of the public code \texttt{Himalaya})~\cite{Harlander:2018yhj}.
The theoretical uncertainty on the lightest
$\cp$-even Higgs-boson mass within the MSSM from unknown higher-order
contributions is still at the level of about $2-3~\gev$ for scalar top
masses at the TeV-scale, where the actual uncertainty
depends on the considered parameter region~\cite{Degrassi:2002fi,Heinemeyer:2004gx,Buchmueller:2013psa,Allanach:2018fif,Bahl:2018jom,Bahl:2019prep}.

In the NMSSM the full one-loop calculation including the momentum dependence
has been performed in the \DRbar\ renormalization scheme in
\citere{Degrassi:2009yq,Staub:2010ty}, or in a mixed on-shell (OS)-\DRbar\ 
scheme in \citere{Ender:2011qh,Graf:2012hh,Drechsel:2016jdg}.
Dominant two-loop contributions of \order{\alt\als,\alt^2}
have been calculated in the
leading logarithmic approximation~\cite{Yeghian:1999kr,Ellwanger:1999ji},
and of \order{\alt\als, \alb\als} in the \DRbar\ scheme in the EP
approach~\cite{Degrassi:2009yq}.
The two-loop corrections involving only
superpotential couplings were given in
\citere{Goodsell:2014pla}.
A two-loop calculation of the \order{\alt\als}
corrections with the top/stop sector renormalized in the OS scheme or in the
\DRbar\ scheme was provided in \citere{Muhlleitner:2014vsa},
while the two-loop corrections of \order{\alt^2} in the $\cp$-violating NMSSM
were calculated in a mixed OS-\DRbar\ scheme~\cite{Dao:2019qaz}.
These contributions are implemented in the public code
\texttt{NMSSMCalc}.
A consistent combination of a full one-loop calculation with all corrections
beyond one-loop in the MSSM approximation was given in
\citere{Drechsel:2016jdg}.
According to a comparison of the various two-loop contributions,
at present the theoretical uncertainties from
unknown higher-order corrections in the NMSSM are expected to be still
larger than for the MSSM~\cite{Staub:2015aea,Drechsel:2016htw,Dao:2019qaz}.

Beyond the MSSM and the NMSSM, only generic \DRbar-calculations of
Higgs-boson mass corrections exist publicly available.
An automated calculation of the
full one-loop corrections, supplemented by partial two-loop corrections
to neutral scalars~\cite{Goodsell:2014pla}
is implemented in the Mathematica
package \texttt{SARAH}~\cite{Staub:2013tta,Goodsell:2016udb},
which can be used to produce
a spectrum generator based on the public code
\texttt{SPheno}~\cite{Porod:2011nf}.
A hybrid Higgs-boson mass calculation combining effective
field theory and fixed-order calculations for
a generic class of SUSY models is
publicly available in the code
\texttt{FlexibleSUSY}~\cite{Athron:2017fvs},
also using the expression for the renormalization
group equations and fixed-order self-energies
as they are calculated by \texttt{SARAH}.

In a previous publication we presented the first calculation
of radiative corrections to the neutral scalars in
a mixed OS-\DRbar\ scheme for the \mnSSM\
with only one generation of right-handed neutrinos~\cite{Biekotter:2017xmf}.
We described in detail the renormalization of the scalar potential,
including the full one-loop quantum corrections.
We consistently combined the full one-loop corrections with the leading
MSSM-like two-loop contributions using \fh. We showed that the contributions
from the (s)top- and the (s)bottom-sector are also dominant in the \mnSSM,
therefore proving that the combination of the one-loop result together
with the two-loop contributions from \fh\ provides a calculation of the
SM-like Higgs-boson mass at a similar accuracy as the NMSSM prediction.
In this work, we go one step beyond and extend the calculation to the
\textit{full} \mnSSM\ with three generations of right-handed neutrinos.
A striking difference between the one- and the three-generation case is
that in the latter case the neutrino sector can be described in agreement
with experimental results without having to rely on the radiative generation
of neutrino masses. On account of this, we are able to present benchmark
scenarios of the \mnSSM\ accurately accommodating a SM-like Higgs boson
at $\sim 125\gev$, as well as correct neutrino mass differences and mixing
angles. In addition, we show that it is possible to simultaneously
explain two excesses measured at LEP and CMS at a mass of $\sim 96\gev$
at the $1\sigma$ level.
An earlier study in the \mnSSM, before the discovery of the Higgs boson,
discussing Higgs bounds and possible signals at the LHC, and suggesting
the re-analysis of the LEP data in light of the excess,
can be found in \citere{Fidalgo:2011ky}.

The paper is organized as follows. In \refse{sec:mnSSM} we describe the
model and explain the particle mixings in each sector.
In \refse{sec:renopot} we give details about the renormalization of the neutral
scalar potential at the one-loop level, including the full set of free
parameters of the \mnSSM. We present the renormalization conditions
applied to extract the parameter counterterms, either in the neutral
scalar or the neutral fermion sector.
In \refse{sec:getmasses} we explain the extraction of the one-loop
corrections to the $\cp$-even scalar masses, based on the renormalization
prescription introduced before. We also describe the incorporation of
higher-order contributions from the MSSM.
In \refse{sec:numanal} we discuss a set of benchmark scenarios with
several light Higgs bosons. We conclude in \refse{sec:concl}.

%%%%%%%%%%%%%%%%%%%%%%%%%%%%%%%%%%%%%%%%%%%%%%%%%%%%%%%%%%%%%%%%%%%%%%%%%%%%%%%
%%%%%%%%%%%%%%%%%%%%%%%%%%%%%%%%%%%%%%%%%%%%%%%%%%%%%%%%%%%%%%%%%%%%%%%%%%%%%%%

\section{\protect\boldmath The model: \mnSSM\ with three generations of right handed neutrinos}
\label{sec:mnSSM}

The superpotential of the \mnSSM\ with three generations of
right-handed neutrinos is written as
\begin{align}
W = \; & 
\epsilon_{ab} \left(
Y^e_{ij} \, \hat H_d^a\, \hat L^b_i \, \hat e_j^c +
Y^d_{ij} \, \hat H_d^a\, \hat Q^{b}_{i} \, \hat d_{j}^{c} 
+
Y^u_{ij} \, \hat H_u^b\, \hat Q^{a}_{i} \, \hat u_{j}^{c}
\right)
\nonumber\\
% &+&
% \epsilon_{ab} Y^{\nu}_{ij} \, \hat H_u^b\, \hat L^a_i \, \hat \nu^c_j -
+ \; &   
\epsilon_{ab} \left(
Y^{\nu}_{ij} \, \hat H_u^b\, \hat L^a_i \, \hat \nu^c_j 
-
%\epsilon_{ab}
\lambda_{i} \, \hat \nu^c_i\, \hat H_u^b \hat H_d^a
\right)
+
\frac{1}{3}
\kappa{_{ijk}} 
\hat \nu^c_i\hat \nu^c_j\hat \nu^c_k
\ ,
\label{superpotential2}
\end{align}
where $\hat H_d^T=(\hat H_d^0, \hat H_d^-)$ and 
$\hat H_u^T=(\hat H_u^+, \hat H_u^0)$ are the MSSM-like doublet Higgs
superfields, $\hat Q_i^T=(\hat u_i, \hat d_i)$ and 
$\hat L_i^T=(\hat \nu_i, \hat e_i)$ are the left-chiral
quark and lepton superfield doublets,
and $\hat u_{i}^{c}$, $\hat d_{i}^{c}$, $\hat e_i^c$ and $\hat{\nu}_i^c$ are
the right-chiral quark and lepton superfields.
$i$ and $j$ are family indices running from one to three and 
$a,b=1,2$ are indices of the fundamental representation of SU(2) 
with $\epsilon_{ab}$ the totally antisymmetric tensor and 
$\varepsilon_{12}=1$. The color indices are not written out. 
$Y^u$, $Y^d$ and $Y^e$ are the usual Yukawa couplings
also present in the MSSM. 
The trilinear singlet self couplings $\kappa_{ijk}$ and the trilinear 
coupling with the Higgs doublets $\lambda_i$ in the second row
are analogues to the couplings of the singlet in the 
superpotential of the $Z_3$-symmetric NMSSM.
The $\mu$-term is generated dynamically after the spontaneous EWSB,
when the right-handed sneutrinos obtain a vev.
The $\kappa$-term forbids a global U(1) symmetry
avoiding the existence of a Goldstone boson in the $\cp$-odd sector. 
The remarkable difference to the
NMSSM is the additional Yukawa coupling $Y^\nu$ which induces
explicit breaking of $R$-parity through the $\lambda$- and
$\kappa$-terms,
and which justifies the interpretation of the singlet
superfields as right-handed neutrino superfields.
It should be pointed out
that in this case lepton number is not conserved anymore, and also the
flavor symmetry in the leptonic sector is broken. 
A more complete motivation of this superpotential can be found in
\citeres{LopezFogliani:2005yw,Escudero:2008jg,Ghosh:2017yeh}.

Working in the framework of low-energy SUSY the
corresponding soft SUSY-breaking Lagrangian is given by
\begin{align}
-\mathcal{L}_{\text{soft}}  = \; & 
\epsilon_{ab} \left(
T^e_{ij} \, H_d^a  \, \widetilde L^b_{iL}  \, \widetilde e_{jR}^* +
T^d_{ij} \, H_d^a\,   \widetilde Q^b_{iL} \, \widetilde d_{jR}^{*} 
+
T^u_{ij} \,  H_u^b \widetilde Q^a_{iL} \widetilde u_{jR}^*
%^{^*} 
%\widetilde d_{jR}^{^{^*}} 
%^{^*}  
+ \text{h.c.}
\right)
\nonumber \\
+ \; &
\epsilon_{ab} \left(
T^{\nu}_{ij} \, H_u^b \, \widetilde L^a_{iL} \widetilde \nu_{jR}^*
%^{^*} 
- 
T^{\lambda}_{i} \, \widetilde \nu_{iR}^*
%^{^*}
\, H_d^a  H_u^b
+ \frac{1}{3} T^{\kappa}_{ijk} \, \widetilde \nu_{iR}^*
%^{^*} 
\widetilde \nu_{jR}^*
%^{^*} 
\widetilde \nu_{kR}^*
%^{^*}
\
+ \text{h.c.}\right)
\nonumber\\
+ \; & 
\left(m_{\widetilde{Q}}^2\right)_{ij} 
%\widetilde{Q}_{iL}^{a^{^*}}  
\widetilde{Q}_{iL}^{a*}
\widetilde{Q}^a_{jL}
+\left(m_{\widetilde{u}}^{2}\right)_{ij} \widetilde{u}_{iR}^*
%^{^*}  
\widetilde u_{jR}
+ \left(m_{\widetilde{d}}^2\right)_{ij}  \widetilde{d}_{iR}^*
%^{^{^*}}  
\widetilde d_{jR}
+
\left(m_{\widetilde{L}}^2\right)_{ij}  
%\widetilde{L}_{iL}^{a^{^*}}  
\widetilde{L}_{iL}^{a*}  
\widetilde{L}^a_{jL}
\nonumber\\
+ \; &
\left(m_{\widetilde{\nu}}^2\right)_{ij} \widetilde{\nu}_{iR}^*
%^{^*} 
\widetilde\nu_{jR} 
+
\left(m_{\widetilde{e}}^2\right)_{ij}  \widetilde{e}_{iR}^*
%^{^*}  
\widetilde e_{jR}
+ 
m_{H_d}^2 {H^a_d}^*
%{^*} 
H^a_d + m_{H_u}^2 {H^a_u}^*
%^{^*} 
H^a_u
+
\left( m_{H_d\widetilde{L}}^2 \right)_{i}H_d^{a*} \widetilde{L}_{iL}^a
\nonumber \\
+ \; &  \frac{1}{2}\, \left(M_3\, {\widetilde g}\, {\widetilde g}
+
M_2\, {\widetilde{W}}\, {\widetilde{W}}
+M_1\, {\widetilde B}^0 \, {\widetilde B}^0 + \text{h.c.} \right)\ .
\label{2:Vsoft}
\end{align}
In the first four lines
the fields denote the scalar components of the
corresponding superfields.
In the last line the fields denote the fermionic
superpartners of the gauge bosons.
The scalar trilinear parameters 
$T^{e,\nu,d,u,\lambda,\kappa}$ correspond to the trilinear couplings 
in the superpotential. The soft mass parameters 
$m_{\widetilde{Q},\widetilde{u},\widetilde{d},
\widetilde{L},\widetilde{\nu},\widetilde{e}}^2$ are hermitian
$3\times 3$ matrices in family space.
$m_{H_d,H_u}^2$ are the soft masses
of the doublet Higgs fields.

We will neglect flavor mixing in the squark
and the quark sector, so the soft masses will be diagonal and we write
$m_{\widetilde{Q}_{i} }^2$, $m_{\widetilde{u}_{i}}^2$ and
$m_{\widetilde{d}_{i}}^2$, as well as for the soft trilinears
$T^u_i=A^u_i
Y^u_i$, $T^d_i=A^d_i Y^d_i$, where the summation convention on repeated
indices is not implied, and the quark Yukawas $Y^u_{ii}=Y^u_i$ and
$Y^d_{ii}=Y^d_i$ are chosen to be diagonal.
For the sleptons we define 
$T^e_{ij}=A^e_{ij}Y^e_{ij}$ and $T^\nu_{ij}=A^\nu_{ij} Y^\nu_{ij}$,
again without summation over repeated indices.

$m_{H_d\widetilde{L}}^2$ is a 3-dimensional
vector in family space, which is always regarded to be absent in the
tree-level Lagrangian of the \mnSSM, because it spoils the electroweak
seesaw mechanism. We include it here, because the operator is generated
at the one-loop level and the parameters $m_{H_d\widetilde{L}}^2$ are needed
to renormalize the scalar potential.
The same reasoning applies for the non-diagonal elements of
the soft slepton masses $(m_{\widetilde{L}}^2)_{i \neq j}$
and $(m_{\widetilde{\nu}}^2)_{i \neq j}$.

Theoretically, the absence of soft mass
parameters mixing different fields
at tree level, $( m_{H_d\widetilde{L}}^2 )_{i}$,
$(m_{\widetilde{L}}^2)_{i \neq j}$,
$(m_{\widetilde{Q}}^2)_{i \neq j}$, etc., 
can be justified by the diagonal structure of the K\"ahler metric in
certain supergravity models, or when the dilaton field
is the source of SUSY breaking in string
constructions~\cite{Ghosh:2017yeh}.
Notice also that when the
down-type Higgs doublet superfield is interpreted as a fourth family
of leptons the parameters
$m_{H_d\widetilde{L}}^2$ can be seen as
off-diagonal elements
of $m_{\widetilde{L}}^2$~\cite{Lopez-Fogliani:2017qzj}.

After the EWSB the neutral scalar fields will acquire
a vev. This includes the left- and right-handed
sneutrinos, because they are not protected by lepton-number conservation as in
the MSSM and the NMSSM. We define the decomposition 
\bea
H_d^0 &=& \frac{1}{\sqrt 2} \left(H_{d}^{\mathcal{R}} + v_d +
\ii\, H_{d}^{\mathcal{I}}\right)\ ,
\\
\label{vevd}
\nonumber
\\  
H^0_u &=& \frac{1}{\sqrt 2} \left(H_{u}^{\mathcal{R}}  + v_u +
  \ii\, H_{u}^{\mathcal{I}}\right)\ ,  
\\
\label{vevu}
\nonumber 
\\
\widetilde{\nu}_{iR} &=&  
      \frac{1}{\sqrt 2} \left(\widetilde{\nu}^{\mathcal{R}}_{iR}+ v_{iR} +
        \ii\, \widetilde{\nu}^{\mathcal{I}}_{iR}\right),  
\\
\label{vevnuc}
\nonumber
\\
  \widetilde{\nu}_{iL} &=& \frac{1}{\sqrt 2} \left(\widetilde{\nu}_{iL}^{\mathcal{R}} 
  + v_{iL} +
    \ii\, \widetilde{\nu}_{iL}^{\mathcal{I}}\right)\ ,
\label{vevnu}
\eea
in such a way that after the EWSB they develop the real vevs 
\bea
\langle H_d^0 \rangle = \frac{v_d}{\sqrt 2}\ , \, \quad 
\langle H_u^0 \rangle = \frac{v_u}{\sqrt 2}\ , \,
\quad 
\langle \widetilde \nu_{iR}\rangle = \frac{v_{iR}}{\sqrt 2}\ , \,  \quad
\langle \widetilde \nu_{iL} \rangle = \frac{v_{iL}}{\sqrt 2}
\ ,
\label{vevs}
\eea
which is valid assuming $\cp$-conservation, as we will do
throughout this paper.

%%%%%%%%%%%%%%%%%%%%%%%%%%%%%%%%%%%%%%%%%%%%%%%%%%%%%%%%%%%%%%%%%%%%%%%%%%%%%%%

The neutral scalar potential of the \mnSSM\  is given at tree level
with all parameters chosen to be real by the soft terms and
the $F$- and $D$-term contributions of the superpotential. One finds
\begin{equation}
V^{(0)} = V_{\text{soft}} + V_F  +  V_D\ , 
\label{finalpotential}
\end{equation}
with
\begin{align}
V_{\text{soft}}= \; & 
\left(
T^{\nu}_{ij} \, H_u^0\,  \widetilde \nu_{iL} \, \widetilde \nu_{jR}^* 
- T^{\lambda}_{i} \, \widetilde \nu_{iR}^*\, H_d^0  H_u^0
+ \frac{1}{3} T^{\kappa}_{ijk} \, \widetilde \nu_{iR}^* \widetilde \nu_{jR}^* 
\widetilde \nu_{kR}^*\
+
\text{h.c.} \right)
\nonumber\\
+ \; &
%&&+
\left(m_{\widetilde{L}}^2\right)_{ij} \widetilde{\nu}_{iL}^* \widetilde\nu_{jL}
% \widetilde{L}_{Li}^{a^{^*}}  \widetilde{L}^a_{Lj} 
+
\left(m_{\widetilde{\nu}}^2\right)_{ij} \widetilde{\nu}_{iR}^* \widetilde\nu_{jR} +
m_{H_d}^2 {H^0_d}^* H^0_d + m_{H_u}^2 {H^0_u}^* H^0_u
\ ,
\label{akappa}
\\
\nonumber
\\
V_{F}  = \; &
 \lambda_{j}\lambda_{j} H^0_{d}H_d^0{^{^*}}H^0_{u}H_u^0{^{^*}}
 +
\lambda_{i}\lambda_{j} \tilde{\nu}^{*}_{iR}\tilde{\nu}_{jR}H^0_{d}H_d^0{^*}
 +
\lambda_{i}\lambda_{j}
\tilde{\nu}^{*}_{iR}\tilde{\nu}_{jR}  H^0_{u}H_u^0{^*}   
\nonumber\\                                              
+ \; &
\kappa_{ijk}\kappa_{ljm}\tilde{\nu}^*_{iR}\tilde{\nu}_{lR}
                                   \tilde{\nu}^*_{kR}\tilde{\nu}_{mR}
%  \nonumber\\
%   &-& 
- \left(\kappa_{ijk}\lambda_{j}\tilde{\nu}^{*}_{iR}\tilde{\nu}^{*}_{kR} H_d^{0*}H_u^{0*}                                      
 -Y^{\nu}_{ij}\kappa_{ljk}\tilde{\nu}_{iL}\tilde{\nu}_{lR}\tilde{\nu}_{kR}H^0_{u}
 \right.
 \nonumber\\
 + \; &
 \left.
 Y^{\nu}_{ij}\lambda_{j}\tilde{\nu}_{iL} H_d^{0*}H_{u}^{0*}H^0_{u}
% \nonumber \\
% &+&
+{Y^{\nu}_{ij}}\lambda_{k} \tilde{\nu}_{iL}^{*}\tilde{\nu}_{jR}\tilde{\nu}_{kR}^* H^0_{d}
 + \text{h.c.}\right) 
\nonumber \\
+ \; &
Y^{\nu}_{ij}{Y^{\nu}_{ik}} \tilde{\nu}^{*}_{jR}
\tilde{\nu}_{kR}H^0_{u}H_u^0{^*}                                                
 +
Y^{\nu}_{ij}{Y^{\nu}_{lk}}\tilde{\nu}_{iL}\tilde{\nu}_{lL}^{*}\tilde{\nu}_{jR}^{*}
                                  \tilde{\nu}_{kR}  
 +
Y^{\nu}_{ji}{Y^{\nu}_{ki}}\tilde{\nu}_{jL}\tilde{\nu}_{kL}^* H^0_{u}H_u^{0*}\, ,
\\
\nonumber
\\
V_D  = \; &
\frac{1}{8}\left(g_1^{2}+g_2^{2}\right)\left(\widetilde\nu_{iL}\widetilde{\nu}_{iL}^* 
+H^0_d {H^0_d}^* - H^0_u {H^0_u}^* \right)^{2}\, .
\label{dterms}
\end{align}

\subsection{The neutral scalar sector}
\label{sec:neutrscasec}

Using the decomposition from
\refeqs{vevd} - (\ref{vevnu}) 
the linear and bilinear terms
in the fields define the tadpoles $T_{\varphi}$ and the scalar $\cp$-even
and $\cp$-odd neutral mass matrices $m_{\varphi}^2$ and
$m_{\sigma}^2$ after electroweak symmetry breaking,
\begin{align}
V_H=\cdots - {T_{\varphi}}_i\varphi_i + \frac{1}{2} \varphi^T_i
\left( m_{\varphi}^2 \right)_{ij}
  \varphi_j + \frac{1}{2} \sigma^T_i
\left( m_{\sigma}^2 \right)_{ij}
  \sigma_j + \cdots \; .
\end{align}
where we collectively denote with 
$\varphi^T=(H_d^{\mathcal{R}},H_u^{\mathcal{R}},
\widetilde{\nu}_{1R}^{\mathcal{R}},
\widetilde{\nu}_{2R}^{\mathcal{R}},
\widetilde{\nu}_{3R}^{\mathcal{R}}
,\widetilde{\nu}_{1L}^{\mathcal{R}}
,\widetilde{\nu}_{2L}^{\mathcal{R}}
,\widetilde{\nu}_{3L}^{\mathcal{R}})$ 
and 
$\sigma^T=(H_d^{\mathcal{I}},H_u^{\mathcal{I}},
\widetilde{\nu}_{1R}^{\mathcal{I}},
\widetilde{\nu}_{2R}^{\mathcal{I}},
\widetilde{\nu}_{3R}^{\mathcal{I}}
,\widetilde{\nu}_{1L}^{\mathcal{I}}
,\widetilde{\nu}_{2L}^{\mathcal{I}}
,\widetilde{\nu}_{3L}^{\mathcal{I}})$ 
the $\cp$-even and $\cp$-odd scalar fields, respectively.
The linear terms are only allowed for $\cp$-even fields and given by
\begin{align}
T_{H_d^{\mathcal{R}}}=&-m_{H_d}^2v_d -
\left( m_{H_d\widetilde{L}}^2 \right)_{i} v_{iL} -
\frac{1}{8}\left(g_1^2+g_2^2\right)v_d \left( v_d^2 + v_{iL}v_{iL} - v_u^2\right)
\notag \\
&-\frac{1}{2} v_d v_u^2 \lambda_i \lambda_i +
\frac{1}{\sqrt{2}} v_u v_{iR} T^\lambda_i +
\frac{1}{2} v_u^2 Y^\nu_{ji} \lambda_i v_{jL} -
\frac{1}{2} v_d v_{iR} \lambda_i v_{jR} \lambda_j \notag \\
&+\frac{1}{2} v_u \kappa_{ikj} \lambda_i v_{jR} v_{kR} +
\frac{1}{2} v_{iR} \lambda_i v_{jL} Y^\nu_{jk} v_{kR} \; , \label{eq:tp1}\\[4pt]
T_{H_u^{\mathcal{R}}}=&-m_{H_u}^2v_u+
\frac{1}{8}\left( g_1^2+g_2^2\right)v_u\left( v_d^2+v_{iL}v_{iL}-v_u^2\right)
\notag \\
&-\frac{1}{2} v_d^2 v_u \lambda_i \lambda_i +
\frac{1}{\sqrt{2}} v_d v_{iR} T^\lambda_i +
v_d v_u Y^\nu_{ji} \lambda_i v_{jL} -
\frac{1}{\sqrt{2}} v_{iL} T^\nu_{ij} v_{jR} -
\frac{1}{2} v_u v_{iR} \lambda_i v_{jR} \lambda_j \notag \\
&-\frac{1}{2} v_u Y^\nu_{ji} Y^\nu_{ki} v_{jL} v_{kL} -
\frac{1}{2} v_u Y^\nu_{ij} Y^\nu_{ik} v_{jR} v_{kR} +
\frac{1}{2} v_d \kappa_{ijk} \lambda_i v_{jR} v_{kR} -
\frac{1}{2} Y^\nu_{li} \kappa_{ikj} v_{jR} v_{kR} v_{lL} \; , \label{eq:tp2}\\[4pt]
T_{\widetilde{\nu}_{iR}^{\mathcal{R}}}=
&-\left(m_{\widetilde{\nu}}^2\right)_{ij} v_{jR} -
\frac{1}{\sqrt{2}} v_u v_{jL} T^\nu_{ji} -
\frac{1}{2} v_u^2 Y^\nu_{ji} Y^\nu_{jk} v_{kR} +
v_d v_u \kappa_{ijk} \lambda_j v_{kR} -
\frac{1}{\sqrt{2}} T^\kappa_{ijk} v_{jR} v_{kR} \notag \\
&+\frac{1}{2} v_d v_{jL} Y^\nu_{ji} v_{kR} \lambda_k -
v_u Y^\nu_{lj} \kappa_{ijk} v_{kR} v_{lL} -
\frac{1}{2} v_{jL} Y^\nu_{ji} v_{kL} Y^\nu_{kl} v_{lR} -
\kappa_{ijm} \kappa_{jlk} v_{kR} v_{lR} v_{mR} \notag \\
&-\frac{1}{2} \left( v_d^2 + v_u^2 \right) \lambda_i \lambda_j v_{jR} +
\frac{1}{2} v_d v_{jL} Y^\nu_{jk} v_{kR} \lambda_i +
\frac{1}{\sqrt{2}} v_d v_u T^\lambda_i \; , \label{eq:tp3}\\[4pt]
T_{\widetilde{\nu}_{iL}^{\mathcal{R}}}=&-
\left(m_{\widetilde{L}}^2\right)_{ij}v_{jL} -
\left( m_{H_d\widetilde{L}}^2 \right)_{i}v_d -
\frac{1}{8}\left( g_1^2+g_2^2\right)v_{iL}\left( v_d^2+v_{jL}v_{jL}-v_u^2\right)
\notag \\
&+\frac{1}{2} v_d v_u^2 Y^\nu_{ij} \lambda_{j} -
\frac{1}{\sqrt{2}} v_u v_{jR} T^\nu_{ij} -
\frac{1}{2} v_u^2 Y^\nu_{ij} Y^\nu_{kj} v_{kL} +
\frac{1}{2} v_d v_{jR} Y^\nu_{ij} v_{kR} \lambda_k \notag \\
&-\frac{1}{2} v_u Y^\nu_{ij} \kappa_{jkl} v_{kR} v_{lR} -
\frac{1}{2} v_{jR} Y^\nu_{ij} v_{kL} Y^\nu_{kl} v_{lR} \label{eq:tp4} \; .
\end{align}
The tadpoles vanish in the true vacuum of the model.
We stress that the proportionality of $T^\nu$ and $Y^\nu$,
as used above,
assures that the condition $v_{iL}\ll v_d, v_u, v_{iR}$, is fulfilled
after solving the minimization conditions
in \refeq{eq:tp4}.
This is essential for the generation of the electroweak seesaw mechanism
and the origin for the smallness of the left-handed neutrino masses,
without introducing any further energy scale.
During the
renormalization procedure they will be treated as OS parameters,
i.e., finite corrections will be canceled by their corresponding
counterterms. This guarantees that the vacuum is
stable w.r.t.\ quantum corrections.

The bilinear terms 
\begin{align}
m_{\varphi}^2= 
  \left( \begin{array}{cccc} 
  m_{H_{d}^{\mathcal{R}}H_{d}^{\mathcal{R}}}^{2}& 
m_{H_{d}^{\mathcal{R}}H_{u}^{\mathcal{R}}}^{2} & 
m_{H_{d}^{\mathcal{R}}\widetilde{\nu}_{jR}^{\mathcal{R}}}^{2} & 
m_{H_{d}^{\mathcal{R}}\widetilde{\nu}_{jL}^{\mathcal{R}}}^{2}
\\
m_{H_{u}^{\mathcal{R}}H_{d}^{\mathcal{R}}}^{2} & 
m_{H_{u}^{\mathcal{R}}H_{u}^{\mathcal{R}}}^{2}& 
m_{H_{u}^{\mathcal{R}}\widetilde{\nu}_{jR}^{\mathcal{R}}}^{2}  & 
m_{H_{u}^{\mathcal{R}}\widetilde{\nu}_{jL}^{\mathcal{R}}}^{2}
\\
 m_{\widetilde{{\nu}}^{\mathcal{R}}_{iR} H_{d}^{\mathcal{R}}}& 
m_{\widetilde{{\nu}}^{\mathcal{R}}_{iR} H_{u}^{\mathcal{R}}}&   
m_{\widetilde{{\nu}}^{\mathcal{R}}_{iR} \widetilde{{\nu}}^{\mathcal{R}}_{jR}}^{2} &   
m_{\widetilde{{\nu}}^{\mathcal{R}}_{iR} \widetilde{{\nu}}^{\mathcal{R}}_{jL}}^{2} 
\\
m_{\widetilde{\nu}_{iL}^{\mathcal{R}} H_{d}^{\mathcal{R}}}^{2}  & 
 m_{\widetilde{\nu}_{iL}^{\mathcal{R}} H_{u}^{\mathcal{R}}}^{2} & 
m_{\widetilde{\nu}_{iL}^{\mathcal{R}} \widetilde{\nu}^{\mathcal{R}}_{jR}}^{2} &   
m_{\widetilde{\nu}_{iL}^{\mathcal{R}} \widetilde{\nu}_{jL}^{\mathcal{R}}}^{2} 
         \end{array} \right)\ ,  
\label{matrixscalar1}
\end{align}
and
\begin{align}
m^2_{\sigma}= 
  \left( \begin{array}{cccc} 
  m_{H_{d}^{\mathcal{I}}H_{d}^{\mathcal{I}}}^{2} & 
m_{H_{d}^{\mathcal{I}}H_{u}^{\mathcal{I}}}^{2} & 
m_{H_{d}^{\mathcal{I}}\widetilde{\nu}_{jR}^{\mathcal{I}}}^{2} & 
m_{H_{d}^{\mathcal{I}}\widetilde{\nu}_{jL}^{\mathcal{I}}}^{2}
\\
m_{H_{u}^{\mathcal{I}}H_{d}^{\mathcal{I}}}^{2} & 
m_{H_{u}^{\mathcal{I}}H_{u}^{\mathcal{I}}}^{2}& 
m_{H_{u}^{\mathcal{I}}\widetilde{\nu}_{jR}^{\mathcal{I}}}^{2}  & 
m_{H_{u}^{\mathcal{I}}\widetilde{\nu}_{jL}^{\mathcal{I}}}^{2}
\\
 m_{\widetilde{{\nu}}^{\mathcal{I}}_{iR} H_{d}^{\mathcal{I}}}^2& 
m_{\widetilde{{\nu}}^{\mathcal{I}}_{iR} H_{u}^{\mathcal{I}}}^2&   
m_{\widetilde{{\nu}}^{\mathcal{I}}_{iR} \widetilde{{\nu}}^{\mathcal{I}}_{jR}}^{2} &   
m_{\widetilde{{\nu}}^{\mathcal{I}}_{iR} \widetilde{{\nu}}^{\mathcal{I}}_{jL}}^{2} 
\\
m_{\widetilde{\nu}_{iL}^{\mathcal{I}} H_{d}^{\mathcal{I}}}^{2}  & 
 m_{\widetilde{\nu}_{iL}^{\mathcal{I}} H_{u}^{\mathcal{I}}}^{2} & 
m_{\widetilde{\nu}_{iL}^{\mathcal{I}} \widetilde{\nu}^{\mathcal{I}}_{jR}}^{2} &   
m_{\widetilde{\nu}_{iL}^{\mathcal{I}} \widetilde{\nu}_{jL}^{\mathcal{I}}}^{2}
         \end{array} \right)\ , 
\label{matrixscalar2} 
\end{align}
are $8\times 8$ matrices in family space whose rather lengthy entries
are given in the Apps.~\ref{app:cpeven} and~\ref{app:cpodd}.
We transform to the mass eigenstate basis
of the $\cp$-even scalars through a
unitary transformation defined by the matrix $U^H$
that diagonalizes the mass matrix $m_{\varphi}^2$,
\bea
U^H
m^2_{\varphi}\ {U^H}^{^T}=
m^2_{h}
\ ,
\label{eq:scalarhiggs}
\eea
with 
\begin{equation}
\varphi = {U^H}^{^T} h\, ,
\label{physscalarhiggses}
\end{equation}
where the $h_i$ are the $\cp$-even scalar fields in the mass eigenstate basis.
Without $\cp$-violation in the scalar sector the matrix $U^H$ is real. Similarly,
for the $\cp$-odd scalars we define the rotation matrix $U^A$ that diagonalizes
the mass matrix $m_{\sigma}^2$,
\bea
U^A
m^2_{\sigma}\ {U^A}^{^T}=
m^2_{A}
\ , \qquad \text{with } \sigma={U^A}^T A \; ,
\label{eq:scalaroddhiggs}
\eea
which includes the neutral Goldstone boson $A_1=G^0$.
Because of the smallness of the neutrino Yukawa couplings $Y^\nu_{ij}$, which also
implies that the left-handed sneutrino vevs $v_{iL}$ have to be small, so that the
tadpole coefficients vanish at tree level \cite{Escudero:2008jg}, the mixing of
the left-handed sneutrinos with the doublet fields and the singlets will be small. 

It is a well known fact that the quantum corrections to the Higgs
potential are highly significant in supersymmetric models, see
e.g.\ \citeres{Heinemeyer:2004ms,Heinemeyer:2004gx,Djouadi:2005gj} 
for reviews. As in the NMSSM~\cite{Ellwanger:2009dp}, the upper bound on 
the lowest Higgs mass squared at tree level is relaxed through additional
contributions from the 
right-handed sneutrinos~\cite{Escudero:2008jg};
\begin{equation}
\label{eq:approhiggs}
m_{h_1}^{(0)} \leq
M_Z^2 \left( \cos^2 2\beta +
\frac{2 \pmb{\lambda}^2}{g_1^2+g_2^2}
	\sin^2 2\beta \right) \; , \quad \text{with} \quad
\pmb{\lambda}^2 := \lambda_1^2+\lambda_2^2+\lambda_3^2 \; .
\end{equation}
Nevertheless, quantum corrections were 
still shown to contribute significantly especially to the prediction of 
the SM-like Higgs-boson
mass~\cite{Draper:2016pys,Ellwanger:1993hn,Elliott:1993uc,Elliott:1993bs,Ender:2011qh,Drechsel:2016htw,Drechsel:2016ukp,Drechsel:2016jdg,Biekotter:2017xmf}.
In \citere{Biekotter:2017xmf} we already investigated how
important the unique loop corrections of the \mnSSM\ beyond the NMSSM 
are in realistic scenarios, considering only one generation of right-handed
neutrinos, finding only negligible differences compared to the NMSSM-like
corrections. This is related to the small size of the neutrino Yukawas $Y^\nu$
compared to the other couplings in the superpotential.
In this paper we go beyond \citere{Biekotter:2017xmf} and
investigate the  complete \mnSSM\
with three right-handed neutrinos. Thus, genuine effects from the \mnSSM\ are
guaranteed to play a role in the prediction of the SM-like Higgs-boson mass
just by the presence of additional singlets, whose couplings to the Higgs
doublet fields are not suppressed by the size of $Y^\nu$.
Furthermore, the model can accommodate neutrino data at tree level, so we
will be able to describe the phenomenology related to both the scalar
and the fermionic sector (and their interplay) more precisely.

If the mixing of the $\cp$-even and -odd sneutrinos with
the doublet fields is small,
which is always the case for the left-handed sneutrinos,
one can obtain approximate analytical
expressions for the tree-level masses of the sneutrinos.
For the left-handed sneutrinos the dominant terms are proportional
to the inverse of their vevs. In particular,
assuming that only diagonal elements of $Y^\nu_{ij}$ and
$\kappa_{ijk}$ are non-zero,
one finds
for the diagonal
entries of the mass matrix $m_\varphi^2$ corresponding to the $\cp$-even left-handed
sneutrinos,
\begin{equation}
\label{eq:nuLapprx}
 m_{\widetilde{\nu}_{iL}^{\mathcal{R}} \widetilde{\nu}_{iL}^{\mathcal{R}}}^{2} \sim
 \frac{Y^\nu_{ii}}{2 v_{iL}} \left[
   v_d v_u^2 \lambda_i + \sqrt{2} v_d v_{iR} \mu
   - v_u v_{iR} \left(
                  \sqrt{2} A^\nu_{ii} + \kappa_{iii} v_{iR}
                \right)
 \right] \; ,
\end{equation}
where we defined the effective $\mu$-term as
\begin{equation}
\mu = \frac{1}{\sqrt{2}}
\left( v_{1R} \lambda_1 + v_{2R} \lambda_2
  + v_{3R} \lambda_3 \right) \; .
\end{equation}
Note that the first term in \refeq{eq:nuLapprx} can usually be neglected
as long as $v_{iR}\gg v_d,v_u$.
Each $\cp$-odd left-handed sneutrino is nearly
degenerate with the corresponding $\cp$-even one, though they
are slightly lighter due to different contributions proportional
to the gauge couplings,
\begin{equation}
\label{eq:nuLapprxdiff}
 m_{\widetilde{\nu}_{iL}^{\mathcal{I}} \widetilde{\nu}_{iL}^{\mathcal{I}}}^{2} - 
 m_{\widetilde{\nu}_{iL}^{\mathcal{R}} \widetilde{\nu}_{iL}^{\mathcal{R}}}^{2} =
 - \frac{1}{4} \left( g_1^2 + g_2^2 \right) v_{iL}^2 \; .
\end{equation}
For the $\cp$-even right-handed sneutrinos we find, under the assumptions that
non-diagonal elements of $\kappa_{ijk}$ and $(m_{\widetilde{\nu}_R}^2)_{ij}$
vanish, for the $3\times 3$ submatrix
\begin{equation}
\label{eq:nuRapprx}
 m_{\widetilde{\nu}_{iR}^{\mathcal{R}} \widetilde{\nu}_{jR}^{\mathcal{R}}}^{2} \sim
 \frac{1}{2} v^2 \lambda_i \lambda_j
   + \delta_{ij} \left[
     \frac{1}{\sqrt{2}} v_{iR} \kappa_{iii} A^\kappa_{iii}
     + 2 \kappa_{iii}^2 v_{iR}^2
   - \frac{\lambda_i}{\sqrt{2} v_{iR}}
     \left(
       \mu v^2 - v_d v_u A^\lambda_i
     \right)
   \right] \; .
\end{equation}
Furthermore, in case of universal values $\kappa_{iii}=\kappa$,
$A^\kappa_{iii}=A^\kappa$, $v_{iR}=v_R$, $\lambda_i=\lambda$
and $A^\lambda_i=A^\lambda$,
this matrix has the form
\begin{equation}
m_{\widetilde{{\nu}}^{\mathcal{R}}_{iR} \widetilde{{\nu}}^{\mathcal{R}}_{jR}}^{2} =
\begin{pmatrix}
a & b & b \\
b & a & b \\
b & b & a
\end{pmatrix} \; . %, \quad \text{with} \; a \gg b \; .
\end{equation}
The eigenvalues of such a matrix are $a-b$ , $a-b$ and $a+2b$, and only the
mass eigenstate corresponding to the latter eigenvalue mixes with the
SM-like Higgs boson~\cite{Ghosh:2014ida}. Later we will make use of this
fact to simplify the accommodation of SM-like Higgs boson properties, even
when all three right-handed neutrinos have masses close to or even
below $125\gev$, because two right-handed sneutrinos will
be conveniently decoupled from the
remaining scalars, and interact very weakly with SM particles.
In extensions of the NMSSM with several gauge singlets, this decoupling
can lead to practically stable particles. In the \mnSSM\ this is
not possible, because the decoupling cannot be exact, even when $\kappa_{ijk}$
and $(m_{\widetilde{\nu}_R}^2)_{ij}$ are diagonal.
We stress that the universality
of the $\kappa_{iii}$ is stable with respect to the RGE running, if
also the $\lambda_i$ parameters are universal.
In this case, one can deduce from the
explicit form of the one-loop counterterms $\delta \kappa_{ijk}$
shown in the App.~\ref{app:explicitCT}
that differences in the running are exclusively
generated by terms proportional to $(Y^\nu)^2$, which are negligible
in realistic scenarios.
For $v_{iR} \gg v_d,v_u$ \refeq{eq:nuRapprx} further simplifies to
\begin{equation}
 m_{\widetilde{\nu}_{iR}^{\mathcal{R}} \widetilde{\nu}_{iR}^{\mathcal{R}}}^{2} \sim
   \frac{1}{\sqrt{2}} v_{iR} \kappa_{iii} A^\kappa_{iii}
   + 2 \kappa_{iii}^2 v_{iR}^2 \; ,
\end{equation}
while for the $\cp$-odd right-handed sneutrinos one finds
\begin{equation}
  m_{\widetilde{\nu}_{iR}^{\mathcal{I}} \widetilde{\nu}_{iR}^{\mathcal{I}}}^{2} \sim
   - \frac{3}{\sqrt{2}} A^\kappa_{iii} \kappa_{iii} v_{iR} \; .
\end{equation}
Thus, to avoid tachyons both in the $\cp$-even and -odd scalar spectrum,
we will follow the sign convention $A^\kappa_{iii} < 0$,
$\kappa_{iii} > 0$ and $v_{iR} > 0$.

Before we come to the one-loop renormalization of the neutral scalar potential
we briefly describe the other relevant sectors of the \mnSSM.

%%%%%%%%%%%%%%%%%%%%%%%%%%%%%%%%%%%%%%%%%%%%%%%%%%%%%%%%%%%%%%%%%%%%%%%%%%%%%%%
%%%%%%%%%%%%%%%%%%%%%%%%%%%%%%%%%%%%%%%%%%%%%%%%%%%%%%%%%%%%%%%%%%%%%%%%%%%%%%%

\subsection{Squark sector}

The numerically most important one-loop corrections to the scalar
potential are expected from the stop/top-sector, analogous to the
(N)MSSM~\cite{ELLIS199183,Ellwanger:1993hn,Elliott:1993ex,Elliott:1993uc,Elliott:1993bs,Pandita:1993tg}, due to the huge Yukawa
coupling of the (scalar) top. The tree-level mass matrices of the squarks differ
slightly from the ones in the MSSM. Neglecting flavor mixing in the
squark sector, one finds for the up-type squark mass matrix
$M^{\widetilde{u}}_i$ of generation $i$, 
\begin{align}
    \left(M^{\widetilde{u}}_i\right)_{11}&=
    	\left(m_{\widetilde{Q}}^2\right)_i + \frac{1}{24}(3g_2^{2}-g_1^2)(v_d^2+v_{jL}v_{jL}-v_u^2)+\frac{1}{2}v_u^2{Y^{u}_i}^2 \; ,
    \label{eq:usquarks11} \\
    \left(M^{\widetilde{u}}_i\right)_{12}&=\frac{1}{2}\left(\sqrt{2} v_u A^u_i
    	+ v_{jL} Y^\nu_{jk} v_{kR}
    	- v_d v_{jR} \lambda_j\right)Y^u_i \; , \label{eq:usquarks12} \\
    \left(M^{\widetilde{u}}_i\right)_{22}&=
    	\left(m_{\widetilde{u}}^2\right)_i+\frac{1}{6}g_1^2(v_d^2+v_{jL}v_{jL}-v_u^2)+\frac{1}{2}v_u^2{Y^u_i}^2 \; .
    \label{eq:usquarks22}
\end{align}
It should be noted that in the non-diagonal element explicitly
appear the neutrino 
Yukawa couplings. This term arises in the F-term contributions of the
squark potential through the quartic coupling of up-type quarks and one 
left-handed and the right-handed sneutrinos after EWSB.
The mass eigenstates $\widetilde{u}_{i1}$ and $\widetilde{u}_{i2}$ are
obtained by the unitary transformation
\begin{equation}
\begin{pmatrix}
 \widetilde{u}_{i1} \\
 \widetilde{u}_{i2}
\end{pmatrix}
=U^{\widetilde{u}}_i
\begin{pmatrix}
 \widetilde{u}_{iL} \\
 \widetilde{u}_{iR}
\end{pmatrix}\; ,
\qquad U^{\widetilde{u}}_i{U^{\widetilde{u}}_i}^\dagger=\mathbbm{1} \; .
\end{equation}
Similarly, for the down-type squarks it is
\begin{align}
    \left(M^{\widetilde{d}}_i\right)_{11}&=
    	\left(m_{\widetilde{Q}}^2\right)_i
    	- \frac{1}{24}(3g_2^{2}+g_1^2)(v_d^2+v_{jL}v_{jL}-v_u^2)
    	+ \frac{1}{2}v_d^2{Y^{d}_i}^2 \; ,
    \label{eq:dsquarks11} \\
    \left(M^{\widetilde{d}}_i\right)_{12}&=
    	\frac{1}{2}\left(\sqrt{2} v_d A^d_i
    	- v_u \lambda_j  v_{jR}\right) Y^d_i \; , \\
    \left(M^{\widetilde{d}}_i\right)_{22}&=
    	\left(m_{\widetilde{d}}^2\right)_i
    	- \frac{1}{12}g_1^2(v_d^2+v_{jL}v_{jL}-v_u^2)
    	+ \frac{1}{2}v_d^2{Y^d_i}^2 \; .
    \label{eq:dsquarks22}
\end{align}
The mass eigenstates $\widetilde{d}_{i1}$ and $\widetilde{d}_{i2}$ are obtained by the unitary transformation
\begin{equation}
\begin{pmatrix}
 \widetilde{d}_{i1} \\
 \widetilde{d}_{i2}
\end{pmatrix}
=U^{\widetilde{d}}_i
\begin{pmatrix}
 \widetilde{d}_{iL} \\
 \widetilde{d}_{iR}
\end{pmatrix}\; ,
\qquad U^{\widetilde{d}}_i{U^{\widetilde{d}}_i}^\dagger=\mathbbm{1} \; .
\end{equation}

%%%%%%%%%%%%%%%%%%%%%%%%%%%%%%%%%%%%%%%%%%%%%%%%%%%%%%%%%%%%%%%%%%%%%%%%%%%%%%%

\subsection{Charged scalar sector}

Since $R$-parity, lepton-number conservation and lepton-flavor universality are broken,
the six charged left- and right-handed sleptons mix with each other and with
the two charged scalars from the Higgs doublets. In the basis $C^T=
({H^-_d}^*,{H^+_u},\widetilde{e}_{iL}^*,\widetilde{e}_{jR}^*)$ we find
the following mass terms in the Lagrangian:
\begin{equation}
\cL_{C} \; = \; 
-{C^*}^T {m}^2_{H^+} C\, .
\label{matrix122}
\end{equation}
Assuming $\CP$ conservation ${m}^2_{H^+}$ is a symmetric
matrix of dimension 8,
\begin{align}
m_{H^+}^2= 
  \left( \begin{array}{cccc} 
  m_{H_{d}^-{H^{-}_d}^{*}}^{2} & m_{H_{d}^- H_{u}^+}^{2} & 
m_{{H_d^-} \widetilde{e}^*_{jL}}^2 & m_{{H_d^-} \widetilde{e}^*_{jR}}^2  \\
 m_{{H_{u}^+}^* {H_d^-}^*}^{2}  & m_{{H^{+}_u}^{*} H_{u}^{+}}^{2} & 
m_{{H_u^+}^*\widetilde{e}^*_{jL}}^{2} & m_{{H_u^+}^*\widetilde{e}^*_{jR}}^{2} \\
 m_{\widetilde{e}_{iL} {{H^{-}_d}}^{*}}^2 & m_{\widetilde{e}_{iL} H_u^+}^2 &   
m_{\widetilde{e}_{iL} \widetilde{e}_{jL}^{*}}^{2} &   
m_{\widetilde{e}_{iL} \widetilde{e}_{jR}^{*}}^{2}\\
 m_{\widetilde{e}_{iR} {{H^{-}_d}}^{*}}^2 & m_{\widetilde{e}_{iR} H_u^+}^2 & 
  m_{\widetilde{e}_{iR} \widetilde{e}_{jL}^{*}}^{2} &   
m_{\widetilde{e}_{iR} \widetilde{e}_{jR}^{*}}^{2}
         \end{array} \right)\ .  
         \label{matrixcharged2}
\end{align}
The entries are given
in App.~\ref{app:charged}.
The mass matrix is diagonalized by an orthogonal matrix $U^+$:
\bea
U^+
m^2_{H^+}\ {U^+}^{^T}
=
\left(m^2_{H^+}\right)^{\text{diag}}
\ ,
\label{scalarhiggs22}
\eea
where the diagonal elements of $\left(m^2_{H^+}\right)^{\text{diag}}$ are the squared masses of the mass eigenstates
\begin{equation}
H^+=U^+\; C \; ,
\end{equation}
which include the charged Goldstone boson $H^+_1=G^\pm$.

%%%%%%%%%%%%%%%%%%%%%%%%%%%%%%%%%%%%%%%%%%%%%%%%%%%%%%%%%%%%%%%%%%%%%%%%%%%%%%%

\subsection{Charged fermion sector}

The charged leptons mix with the charged gauginos and the charged higgsinos.
Following the notation of \citere{Ghosh:2017yeh}, we write the relevant part
of the Lagrangian in terms of two-component spinors 
$(\chi^-)^T = \left({({e_{iL})}^{c}}^{^*}, \widetilde{W}^-,
\widetilde{H}^-_d\right)$
and 
$(\chi^+)^T = \left(({e_{jR}})^c, \widetilde{W}^+, \widetilde{H}^+_u\right)$:
\begin{equation}
\cL_{\chi^\pm} \; = \; -({\chi^{-}})^T 
{m}_{e}
\chi^+ + \mathrm{h.c.}\, .
\label{matrixcharginos0}
\end{equation}
The $5\times5$ mass matrix $m_e$ is defined by
\begin{equation}
m_e=
\renewcommand*{\arraystretch}{1.5}
\begin{pmatrix}
    \frac{v_d Y^e_{11}}{\sqrt{2}} & \frac{v_d Y^e_{12}}{\sqrt{2}} & \frac{v_d Y^e_{13}}{\sqrt{2}} & \frac{g_2 v_{1L}}{\sqrt{2}} &
    	-\frac{v_{iR} Y^\nu_{i1}}{\sqrt{2}} \\
    \frac{v_d Y^e_{21}}{\sqrt{2}} & \frac{v_d Y^e_{22}}{\sqrt{2}} & \frac{v_d Y^e_{23}}{\sqrt{2}} & \frac{g_2 v_{2L}}{\sqrt{2}} & 
    	-\frac{v_{iR} Y^\nu_{i2}}{\sqrt{2}} \\
    \frac{v_d Y^e_{31}}{\sqrt{2}} & \frac{v_d Y^e_{32}}{\sqrt{2}} & \frac{v_d Y^e_{33}}{\sqrt{2}} & \frac{g_2 v_{3L}}{\sqrt{2}} &
    	-\frac{v_{iR} Y^\nu_{i3}}{\sqrt{2}} \\
    0 & 0 & 0 & M_2 & \frac{g_2 v_{u}}{\sqrt{2}} \\
    -\frac{v_{iL}Y^e_{1i}}{\sqrt{2}} & -\frac{v_{iL}Y^e_{2i}}{\sqrt{2}} & -\frac{v_{iL}Y^e_{3i}}{\sqrt{2}} & \frac{g_2 v_{d}}{\sqrt{2}} & \mu
\end{pmatrix} \; .
\label{eq:chargferm}
\end{equation}
The mass matrix is diagonalized by two unitary matrices $U^e_L$ and $U^e_R$,
\begin{equation}
{U_R^e}^{^*} m_{e} {U_L^e}^{^\dagger} = m_{e}^{\text{diag}}
\, ,
\label{diagmatrixneutralinosn}
\end{equation}
where $m_{e}^{\text{diag}}$ contains the masses of the charged fermions in the mass eigenstate base
\begin{eqnarray}
\chi^{+} = {U_L^e}^{^\dagger} \lambda^+\, ,
\\
\chi^{-} = {U_R^e}^{^\dagger} \lambda^-\, .
\label{physcharginos}
\end{eqnarray}
Note that terms mixing the SM leptons and the MSSM-like charginos
in \refeq{eq:chargferm} are suppressed by the size of
the left-handed vevs $v_{iL}$ or the neutrino Yukawa couplings $Y^\nu_{ij}$.
The smallness of $v_{iL}$ in comparison to the other vevs and $M_2$
assures the decoupling of the three leptons from the Higgsino and the wino,
prohibiting substantial lepton-flavor-universality and lepton-number
violation in the charged fermion sector.

%%%%%%%%%%%%%%%%%%%%%%%%%%%%%%%%%%%%%%%%%%%%%%%%%%%%%%%%%%%%%%%%%%%%%%%%%%%%%%%

\subsection{Neutral fermion sector}
\label{sec:neuferm}
The three left-handed neutrinos and the right-handed neutrinos mix with
the neutral Higgsinos and gauginos. Again, following
\citere{Ghosh:2017yeh}, we write the relevant part of the Lagrangian in
terms of two-component spinors
$({\chi^{0}})^T=\left({(\nu_{iL})^{c}}^{^*},\widetilde B^0, 
\widetilde W^{0},\widetilde H_{d}^0,\widetilde H_{u}^0,\nu_{jR}^*\right)$
as
\begin{equation}
\cL_{\chi^0} \; = \;-\frac{1}{2} ({\chi^{0}})^T 
{m}_{\nu} 
\chi^0 + \mathrm{h.c.}\, ,
\label{matrixneutralinos}
\end{equation}
where ${m}_{\nu}$ is the $10\times 10$ symmetric mass matrix. The neutral fermion mass matrix is determined by
\begin{align}
{m}_{\nu}=
\left( \begin{array}{ccccccc}
0 & 0 & 0 & -\frac{g_1 v_{1L}}{2} & \frac{g_2 v_{1L}}{2} &
 0 & \frac{v_{iR} Y^\nu_{1i}}{\sqrt{2}} \\
0 & 0 & 0 & -\frac{g_1 v_{2L}}{2} & \frac{g_2 v_{2L}}{2} &
 0 & \frac{v_{iR} Y^\nu_{2i}}{\sqrt{2}} \\
0 & 0 & 0 & -\frac{g_1 v_{3L}}{2} & \frac{g_2 v_{3L}}{2} &
 0 & \frac{v_{iR} Y^\nu_{3i}}{\sqrt{2}} \\
-\frac{g_1 v_{1L}}{2} & -\frac{g_1 v_{2L}}{2} & -\frac{g_1 v_{3L}}{2} &
 M_1 & 0 & -\frac{g_1 v_{d}}{2} & \frac{g_1 v_{u}}{2} \\
\frac{g_2 v_{1L}}{2} & \frac{g_2 v_{2L}}{2} & \frac{g_2 v_{3L}}{2} &
 0 & M_2 & \frac{g_2 v_{d}}{2} & -\frac{g_2 v_{u}}{2} \\
0 & 0 & 0 & -\frac{g_1 v_{d}}{2} & \frac{g_2 v_{d}}{2} & 0 &
 -\mu \\
\frac{v_{iR} Y^\nu_{1i}}{\sqrt{2}} & \frac{v_{iR} Y^\nu_{2i}}{\sqrt{2}} &
 \frac{v_{iR} Y^\nu_{3i}}{\sqrt{2}} & \frac{g_1 v_u}{2} & -\frac{g_2 v_u}{2} &
 -\mu & 0 \\
\frac{v_u Y^\nu_{11}}{\sqrt{2}} & \frac{v_u Y^\nu_{21}}{\sqrt{2}} &
 \frac{v_u Y^\nu_{31}}{\sqrt{2}} & 0 & 0 & -\frac{v_u \lambda_1}{\sqrt{2}} &
 \frac{- v_d \lambda_1 + v_{iL} Y^\nu_{i1}}{\sqrt{2}} \\
\frac{v_u Y^\nu_{12}}{\sqrt{2}} & \frac{v_u Y^\nu_{22}}{\sqrt{2}} &
 \frac{v_u Y^\nu_{32}}{\sqrt{2}} & 0 & 0 & -\frac{v_u \lambda_2}{\sqrt{2}} &
 \frac{- v_d \lambda_2 + v_{iL} Y^\nu_{i2}}{\sqrt{2}} \\
\frac{v_u Y^\nu_{13}}{\sqrt{2}} & \frac{v_u Y^\nu_{23}}{\sqrt{2}} &
 \frac{v_u Y^\nu_{33}}{\sqrt{2}} & 0 & 0 & -\frac{v_u \lambda_3}{\sqrt{2}} &
 \frac{- v_d \lambda_3 + v_{iL} Y^\nu_{i3}}{\sqrt{2}}
\end{array} \right. \cdots \notag \\
\cdots \left. \begin{array}{ccc}
\frac{v_{u}Y^\nu_{11}}{\sqrt{2}} &
 \frac{v_{u}Y^\nu_{12}}{\sqrt{2}} & \frac{v_{u}Y^\nu_{13}}{\sqrt{2}} \\
\frac{v_{u}Y^\nu_{21}}{\sqrt{2}} &
 \frac{v_{u}Y^\nu_{22}}{\sqrt{2}} & \frac{v_{u}Y^\nu_{23}}{\sqrt{2}} \\
\frac{v_{u}Y^\nu_{31}}{\sqrt{2}} &
 \frac{v_{u}Y^\nu_{32}}{\sqrt{2}} & \frac{v_{u}Y^\nu_{33}}{\sqrt{2}} \\
0 & 0 & 0 \\
0 & 0 & 0 \\
-\frac{v_{u} \lambda_1}{\sqrt{2}} &
  -\frac{v_{u} \lambda_2}{\sqrt{2}} & -\frac{v_{u} \lambda_3}{\sqrt{2}} \\
\frac{-v_d \lambda_1 + v_{iL}Y^\nu_{i1}}{\sqrt{2}} &
 \frac{-v_d \lambda_2 + v_{iL}Y^\nu_{i2}}{\sqrt{2}} &
 \frac{-v_d \lambda_3 + v_{iL}Y^\nu_{i3}}{\sqrt{2}} \\
\sqrt{2} v_{iR} \kappa_{11i} & \sqrt{2} v_{iR} \kappa_{12i} &
 \sqrt{2} v_{iR} \kappa_{13i} \\
\sqrt{2} v_{iR} \kappa_{12i} & \sqrt{2} v_{iR} \kappa_{22i} &
 \sqrt{2} v_{iR} \kappa_{23i} \\
\sqrt{2} v_{iR} \kappa_{13i} & \sqrt{2} v_{iR} \kappa_{23i} &
 \sqrt{2} v_{iR} \kappa_{33i}
\end{array} \right)
\end{align}
Because of the Majorana nature of the neutral fermions we can 
diagonalize ${m}_{\nu}$ with the help of just a single - but complex - 
unitary matrix $U^V$,
\begin{equation}
{U^V}^{^*} m_{\nu}\ {U^V}^{^\dagger} = m_{\nu}^{\text{diag}}\, ,
\label{diagmatrixneutralinos}
\end{equation}
with
\begin{equation}
\chi^{0} = {U^V}^{^\dagger} \lambda^0\, ,
\label{physneutralinos}
\end{equation}
where $\lambda^0$ are the two-component spinors in the mass basis. The eigenvalues of the diagonalized mass matrix $m_{\nu}^{\text{diag}}$ are the masses of the neutral fermions in the mass eigenstate basis.

The mass matrix has a seesaw
structure, assuring that the three lightest eigenvalues will be
very small, so that the mass eigenstates $\lambda_{1,2,3}^0$ can
practically be identified with the SM left-handed neutrinos.
Components from the MSSM-like neutralinos and the right-handed
neutrinos are negligible for the three lightest states.
Thus, the left-handed neutrino mixing can in very good
approximation (using diagonal $Y^e$) be expressed in the usual PMNS
formalism~\cite{Pontecorvo:1967fh,Maki:1962mu} by
the three mixing angles $\theta_{12}$, $\theta_{13}$ and
$\theta_{23}$,
\begin{align}
\left(\begin{array}{c}
\lambda^0_{1} \\
\lambda^0_{2} \\
\lambda^0_{3}
\end{array}\right)
=
\left(\begin{array}{ccc}
c_{12}c_{13} & -s_{12}c_{23}-c_{12}s_{23}s_{13} & s_{12}s_{23}-c_{12}c_{23}s_{13} \\
s_{12}c_{13} & c_{12}c_{23}-s_{12}s_{23}s_{13} & -c_{12}s_{23}-s_{12}c_{23}s_{13} \\
s_{13} & s_{23}c_{13} & c_{23}c_{13}
\end{array}\right)
\left(\begin{array}{c}
\chi^0_{1} \\
\chi^0_{2} \\
\chi^0_{3}
\end{array}\right)
= U^{V}_{ij}
\left(\begin{array}{c}
\chi^0_{1} \\
\chi^0_{2} \\
\chi^0_{3}
\end{array}\right) \; ,
\end{align}
with $i,j=\{1,2,3\}$, and we used the short-hand
notation $c_x=\cos\theta_x$ and $s_x=\sin\theta_x$.
In our numerical studies we fitted the experimentally well measured
quantities
\begin{align}
s_{13}^2 = |U^V_{31}|^2 \; , \qquad s_{12}^2 = \frac{|U^V_{21}|^2}{1-s_{13}^2} \; ,
\qquad s_{23}^2 = \frac{|U^V_{32}|^2}{1-s_{13}^2} \; , \\
\delta m_{12}^2 = m_{\lambda_2^0}^2 - m_{\lambda_1^0}^2 \; , \qquad
%\Delta m_{13}^2 = m_{\lambda_3^0}^2 -
%  \frac{m_{\lambda_1^0}^2 + m_{\lambda_2^0}^2}{2} \; .
\Delta m_{13}^2 \sim \Delta m_{23}^2 = m_{\lambda_3^0}^2 -
  m_{\lambda_{1,2}^0}^2
\end{align}
We restricted ourselves in the neutrino sector to a tree-level analysis,
because the one-loop corrections turn out to be moderate in size
(in the normal hierarchy pattern)~\cite{Ghosh:2010zi} and
can always be compensated by a small shift in the neutrino
Yukawa couplings $Y^\nu_{ij}$ without affecting the conclusions drawn
in the scalar sector, in particular for the observables related to the SM-like
Higgs boson.

To reduce the parameter space in our analysis,
we usually assume the couplings $Y^\nu$
to be diagonal, as we do for the lepton Yukawa couplings $Y^e$. We
emphasize that non-diagonal $Y^\nu_{ij}$ are not required to reproduce
the correct neutrino mixing, because sizable flavor mixing is always
present after the diagonalization of $m_\nu$, generated by the mixing
terms of the left-handed neutrino states with the gauginos, Higgsinos
and right-handed neutrinos. Quantitatively, this can be illustrated
assuming universal parameters $\lambda := \lambda_i$, $v_{R} := v_{iR}$,
$\kappa := \kappa_{iii}$ and $Y^\nu_{i} : = Y^\nu_{ii}$ ($\kappa_{ijk}=0$
and $Y^\nu_{ij}=0$ otherwise), by the formula~\cite{Fidalgo:2009dm}
\begin{eqnarray}
\label{eq:neumixeff}
(m^{\text{eff}}_{\nu})_{ij} 
\simeq \frac{Y^{\nu}_iY^{\nu}_jv_u^2}
{6\sqrt 2 \kappa v_{R}}
                   (1-3 \delta_{ij})-\frac{v_{iL} v_{jL}}{4M^{\text{eff}}}
%&  
-\frac{1}{4M^{\text{eff}}}\left[\frac{v_d\left(Y^{\nu}_iv_{jL}
   +Y^{\nu}_jv_{iL}\right)}{3\lambda}
   +\frac{Y^{\nu}_iY^{\nu}_jv_d^2}{9\lambda^2 }\right]\ ,
   \nonumber\\
  \end{eqnarray}     
with
\begin{eqnarray}
\label{effectivegauginomass}
 M^{\text{eff}}\equiv \frac{M_1 M_2}{g_1^2 M_2 + g_2^2 M_1}
   -\frac{v^2}{2\sqrt 2 \left(\kappa v_R^2+\lambda v_u v_d\right)
        \ 3 \lambda v_R}\left(2 \kappa v_R^{2} \frac{v_u v_d}{v^2}
        +\frac{\lambda v^2}{2}\right)\ .
%\nonumber\\
\end{eqnarray} 
\refeq{eq:neumixeff} demonstrates that substantial flavor mixing
is practically unavoidable
in the \mnSSM. The first two terms are of particular importance.
The first term can be attributed to the mixing with
the right-handed neutrinos and higgsinos, and the other terms
also include the gaugino mixing.
Note that for moderate values of $\tan\beta$ and
not too small values of $\lambda$
the first two terms are the dominant contributions.
They contain diagonal and non-diagonal contributions that
can easily be adjusted by an appropriate choice of the parameters $Y^\nu_i$,
$v_{iL}$ and the soft gaugino mass parameters $M_1$ and $M_2$.
These parameters play only a minor role in the predictions for the
SM-like Higgs boson mass and its mixing with the right-handed sneutrinos.
Thus, the above mentioned parameters will be used to reproduce neutrino
physics in agreement with experimental limits, without having to worry
about spoiling the properties of the SM-like Higgs boson.

%%%%%%%%%%%%%%%%%%%%%%%%%%%%%%%%%%%%%%%%%%%%%%%%%%%%%%%%%%%%%%%%%%%%%%%%%%%%%%%

\section{Renormalization of the Higgs potential at One-Loop}
\label{sec:renopot}
At tree level the part of the Higgs potential relevant for the
masses of the scalars is given by the tadpole coefficients in
\refeqs{eq:tp1}-(\ref{eq:tp4}) and the $\CP$-even and $\CP$-odd scalar mass
matrix elements in \refeqs{matrixscalar1} and (\ref{matrixscalar2}).
We want to employ a renormalization procedure as close as possible to the ones
used in the (N)MSSM. Therefore, we define in the following subsection
certain replacements to obtain a new set of free parameters.
The new set of free parameter will permit us to make use of
a mixed On-Shell (OS)/\DRbar\ renormalization scheme. The precise
definition of the counterterms of the free parameters will be
given in \refse{sec:counters}.
Finally, we describe the
renormalization conditions applied on each parameter and the extraction
of the counterterms in \refse{sec:condis}.

\subsection{Parameter replacements}
\label{sec:replacements}
The vevs of the doublet Higgs fields $v_u$ and $v_d$ are substituted by the
MSSM-like parameters $\tb$ and $v$ according to
\begin{equation}\label{eq:tanbetadef}
\tan\beta=\frac{v_u}{v_d} \qquad \text{and} \qquad
v^2=v_d^2+v_u^2+v_{iL}v_{iL} \; .
\end{equation}
Note that the definition of $v^2$ differs from the one in the MSSM by
the term $v_{iL}v_{iL}$. This allows to maintain the relations
between $v^2$ and the gauge-boson masses as they are in the
MSSM. Numerically, the difference in the definition of $v^2$ is
negligible.
Maintaining the
functional form of $\tan\beta$ as it is in the (N)MSSM is convenient
to facilitate the comparison of the quantum corrections in the
\mnSSM\ and the (N)MSSM, as the one-loop
counterterm of $\tan\beta$ is expressed without having to
include the counterterms for the left-handed sneutrino
vevs~\cite{Biekotter:2017xmf}.
The gauge couplings $g_1$ and $g_2$ will be replaced by the
gauge-boson masses $\MW$ and $\MZ$,
\begin{equation}\label{eq:gaguebosonmasses}
\MW^2=\frac{1}{4}g_2^2v^2 \qquad \text{and} \qquad
\MZ^2=\frac{1}{4}\left( g_1^2+g_2^2\right) v^2 \; .
\end{equation}
This is reasonable, because the gauge-boson masses are well measured
physical observables, so we can define them as OS parameters.
The explicit dependence of the quantum corrections on the
mass counterterm for $\MW^2$ drops out at the one-loop level,
but it will contribute implicitly in the definition of the counterterm
for $v^2$.
The scalar soft masses $m_{H_d}^2$ and $m_{H_u}^2$ and the diagonal elements
of the soft slepton mass matrices $m_{\widetilde{L}}^2$ and
$m_{\widetilde{\nu}}^2$ are replaced by the tadpole coefficient in which
they appear. Using the tadpole coefficients as input parameters facilitates
the absorption of quantum corrections that would spoil the true vacuum of the
potential.
Alternatively, one could also trade the vevs for the tadpole
coefficients, and keep the soft masses as input parameters.
However, it is computationally much more convenient to use the vevs as input
and solve the tadpole equations for the squared soft masses, because
they appear linearly, while solving the tadpole equations for the vevs,
using the soft masses as input, is a complex non-linear problem with
multiple solutions.
The complete set of independent parameters is summarized
in \refta{tab:setpara}.

\begin{table}
\centering
\begin{tabular}{c c c c c}
Soft masses & vevs & Gauge cpl. & Superpot. & Soft trilinears \\
\hline
$m_{H_d}^2$, $m_{H_u}^2$, ${m_{\widetilde{\nu}_{R}}^2}_{ij}$,
${m_{\widetilde{L}}^2}_{ij}$,
			${m_{H_d\widetilde{L}}^2}_i$ 
 & $v_d$, $v_u$, $v_{iR}$, $v_{iL}$ &
$g_1$, $g_2$ &
$\lambda_i$, $\kappa_{ijk}$, $Y^\nu_{ij}$ &
$T^\lambda_i$, $T^\kappa_{ijk}$, $T^\nu_{ij}$ \\
% & & & & \\
$\downarrow$ & $\downarrow$ & $\downarrow$ &  &  \\
 $T_{H_d^{\mathcal{R}}}$, $T_{H_u^{\mathcal{R}}}$,
 $T_{\widetilde{\nu}_{iR}^{\mathcal{R}}}$,
 $T_{\widetilde{\nu}_{iL}^{\mathcal{R}}}$,&
 $\tan\beta$, $v$, $v_{iR}$, $v_{iL}$ &
 $\MW$, $\MZ$ &
  &
  \\
 ${m_{\widetilde{\nu}}^2}_{i\neq j}$,
 ${m_{\widetilde{L}}^2}_{i\neq j}$,
 ${m_{H_d\widetilde{L}}^2}_i$ & & & & 
\end{tabular}
\caption{Set of independent parameters initially entering the tree-level Higgs potential of the \mnSSM\ in the first row, and final choice of free parameters after the substitutions defined in the text.}
\label{tab:setpara}
\end{table}

\subsection{Counterterms}
\label{sec:counters}

The entries of the neutral scalar mass matrices are functions of the independent
parameters,
\begin{align}
m_{\varphi}^2&=m_{\varphi}^2\left( \MZ^2, v^2, \tan\beta,
	\lambda_i, \dots \right) \; , \\
m_{\sigma}^2&=m_{\sigma}^2\left( \MZ^2, v^2, \tan\beta, 
	\lambda_i, \dots \right) \; ,
\end{align}
and we define their renormalization as
\begin{align}
m_{\varphi}^2&\to m_{\varphi}^2+\delta m_{\varphi}^2 \; , \\
m_{\sigma}^2&\to m_{\sigma}^2+\delta m_{\sigma}^2 \; .
\end{align}
The mass counterterms $\delta m_{\varphi}^2$ and $\delta m_{\sigma}^2$
enter the renormalized one-loop scalar self-energies. They are given
as a linear combination of the counterterms of the independent
parameters, which we define as
\begin{equation}
  \begin{split}
T_{H_d^{\mathcal{R}}} &\to
T_{H_d^{\mathcal{R}}}+\delta T_{H_d^{\mathcal{R}}} \; , \\
T_{H_u^{\mathcal{R}}} &\to
T_{H_u^{\mathcal{R}}}+\delta T_{H_u^{\mathcal{R}}} \; , \\
T_{\widetilde{\nu}_{iR}^{\mathcal{R}}} &\to
T_{\widetilde{\nu}_{iR}^{\mathcal{R}}}+\delta T_{\widetilde{\nu}_{iR}^{\mathcal{R}}}
\; , \\
T_{\widetilde{\nu}_{iL}^{\mathcal{R}}} &\to
T_{\widetilde{\nu}_{iL}^{\mathcal{R}}}+ \delta T_{\widetilde{\nu}_{iL}^{\mathcal{R}}}
\; , \\
{m_{\widetilde{\nu}}^2}_{i\neq j} &\to
{m_{\widetilde{\nu}}^2}_{i\neq j}+ \delta {m_{\widetilde{\nu}}^2}_{i\neq j}
\; , \\
{m_{\widetilde{L}}^2}_{i\neq j} &\to
{m_{\widetilde{L}}^2}_{i\neq j}+ \delta {m_{\widetilde{L}}^2}_{i\neq j}
\; , \\
{m_{H_d\widetilde{L}}^2}_i &\to
{m_{H_d\widetilde{L}}^2}_i+\delta {m_{H_d\widetilde{L}}^2}_i \; ,
  \end{split}
\qquad
  \begin{split}
\tan\beta &\to \tan\beta+\delta \tan\beta \; , \\
v^2 &\to v^2+\delta v^2 \; , \\
v_{iR}^2 &\to v_{iR}^2 + \delta v_{iR}^2 \; , \\
v_{iL}^2 &\to v_{iL}^2 + \delta v_{iL}^2 \; , \\
\MW^2 &\to \MW^2 +\delta \MW^2 \; , \\
\MZ^2 &\to \MZ^2 +\delta \MZ^2 \; , \\
  \end{split}
 \qquad
  \begin{split}
\lambda_i &\to \lambda_i + \delta \lambda_i \; , \\
\kappa_{ijk} &\to \kappa_{ijk} + \delta \kappa_{ijk} \; , \\
Y^\nu_{ij} &\to Y^\nu_{ij} + \delta Y^\nu_{ij} \; , \\
T^\lambda_i &\to T^\lambda_i + \delta T^\lambda_i \; , \\
T^\kappa_{ijk} &\to T^\kappa_{ijk} + \delta T^\kappa_{ijk} \; , \\
T^\nu_{ij} &\to T^\nu_{ij} + \delta T^\nu_{ij} \; .
  \end{split}
\end{equation}
The divergent parts of the
counterterms are fixed to cancel the UV divergences.
The finite pieces, and thus the meaning of the parameters, have to be
fixed by renormalization conditions. 
We will adopt a mixed renormalization scheme, where tadpoles
and gauge boson masses are fixed OS, and the other parameters are
fixed in the \DRbar\ scheme. The exact renormalization
conditions will be given in \refse{sec:condis}. The dependence of the
mass counterterms $\delta m_{\varphi}^2$ and $\delta
m_{\sigma}^2$ on the counterterms of the free parameters is given
at the one-loop level by the first order expansion w.r.t. the free parameters,
\begin{equation}
\delta m_{\varphi}^2=\sum_{X\in\; \text{free param.}} \left(
\frac{\partial}{\partial X}  m_{\varphi}^2 \right)\delta X \; , 
\qquad
\delta m_{\sigma}^2=\sum_{X\in\; \text{free param.}} \left(
\frac{\partial}{\partial X}  m_{\sigma}^2 \right)\delta X \; . 
\label{eq:masscounterderive}
\end{equation}
We define the mixing matrices $U^H$ and $U^A$ to diagonalize the
renormalized mass matrices, so they do not have to be renormalized.
%, because they are defined exclusively by renormalized quantities.
The expressions for the counterterms of the scalar mass matrices in the mass eigenstate basis are then given by
\begin{equation}\label{eq:masscounterrot}
\delta m_{h}^2=U^H \delta m_{\varphi}^2 {U^H}^T \; , 
\qquad
\delta m_{A}^2=U^A \delta m_{\sigma}^2{U^A}^T \; ,
\end{equation}
where we emphasize that $\delta m_{h}^2$ and $\delta m_{A}^2$ are not
diagonal, as they would be in a purely OS renormalization procedure
which is often used in theories with flavor mixing \cite{Grimus:2016hmw}.

The field renormalization required to obtain finite
scalar self-energies at arbitrary momentum, is defined by
\begin{equation}\label{eq:fieldrenodef}
\begin{pmatrix}
H_d \\ H_u \\ \widetilde{\nu}_{iR} \\ \widetilde{\nu}_{iL}
\end{pmatrix}
\to \sqrt{Z} 
\begin{pmatrix}
H_d \\ H_u \\ \widetilde{\nu}_{iR} \\ \widetilde{\nu}_{iL}
\end{pmatrix}=
\left( \mathbbm{1} + \frac{1}{2} \delta Z \right)
\begin{pmatrix}
H_d \\ H_u \\ \widetilde{\nu}_{iR} \\ \widetilde{\nu}_{iL}
\end{pmatrix} \; ,
\end{equation}
where $\sqrt{Z}$ and $\delta Z$ are $8 \times 8$ dimensional matrices
and the equal sign is valid at the one-loop level.
In contrast to the (N)MSSM, the field renormalization is not diagonal in the
interaction basis.
The reason is that the \mnSSM\ explicitly breaks lepton-number conservation
and lepton-flavor universality,
resulting in kinetic mixing terms at one-loop order.\footnote{As was
argued in \citere{Bijnens:2018rqw},
non-diagonal field renormalization constants are not
necessary if one only demands physical quantities to be UV finite, permitting
UV divergences in non-diagonal 2-point Green's functions to remain.
These would then be canceled by the additional mixing effects on the outer legs
of S-matrix elements following the LSZ theorem.}
For the $\cp$-even and $\cp$-odd neutral scalar fields the 
definition in \refeq{eq:fieldrenodef} implies the following field
renormalization in the mass eigenstate basis: 
\begin{equation}
h \to \left( \mathbbm{1} + \frac{1}{2} \delta Z^H \right) h \; , 
\qquad
A \to \left( \mathbbm{1} + \frac{1}{2} \delta Z^A \right) A \; ,
\end{equation}
with
\begin{equation}\label{eq:fieldrenorotate}
\delta Z^H = U^H \left(\delta Z \right) {U^H}^T \; \qquad \text{and} \qquad
\delta Z^A = U^A \left(\delta Z \right) {U^A}^T \; .
\end{equation}

\subsection{Renormalization conditions}
\label{sec:condis}
In this section we briefly describe our choice for the renormalization
conditions. 
We start with the OS conditions for the gauge boson mass parameters and the
tadpole coefficients followed by our definitions for the
\DRbar\ renormalized parameters, including the field renormalization.
All counterterms are extracted diagrammatically by calculating one-loop
corrections to linear, bilinear or trilinear terms of the Lagrangian, and
identifying the part of the corrections that had to be absorbed individually
by the counterterms of the parameters appearing in the tree-level expression of
the term. We generated the Feynman diagrams using our \FA~\cite{Hahn:2000kx}
model file, which
was initially created with \texttt{SARAH}
version 4.12.0~\cite{Staub:2009bi,Staub:2013tta}. We modified
the model file by hand to be able to use \FC~\cite{Hahn:1998yk}
for further evaluations 
and to improve the analytical and
numerical evaluation of the rather large expressions.
Since the divergent parts of one-loop counterterms can in principle also be
derived from the one-loop beta functions, for which generic analytical
formulas exist~\cite{Martin:1993zk,Yamada:1994id,Machacek:1983fi,
Luo:2002ti,Machacek:1983tz,Fonseca:2013bua},
the diagrammatic calculation of the counterterms
was an excellent test for the correctness of our \FA\ model file.

\begin{figure}
  \centering
  \includegraphics[width=0.98\textwidth]{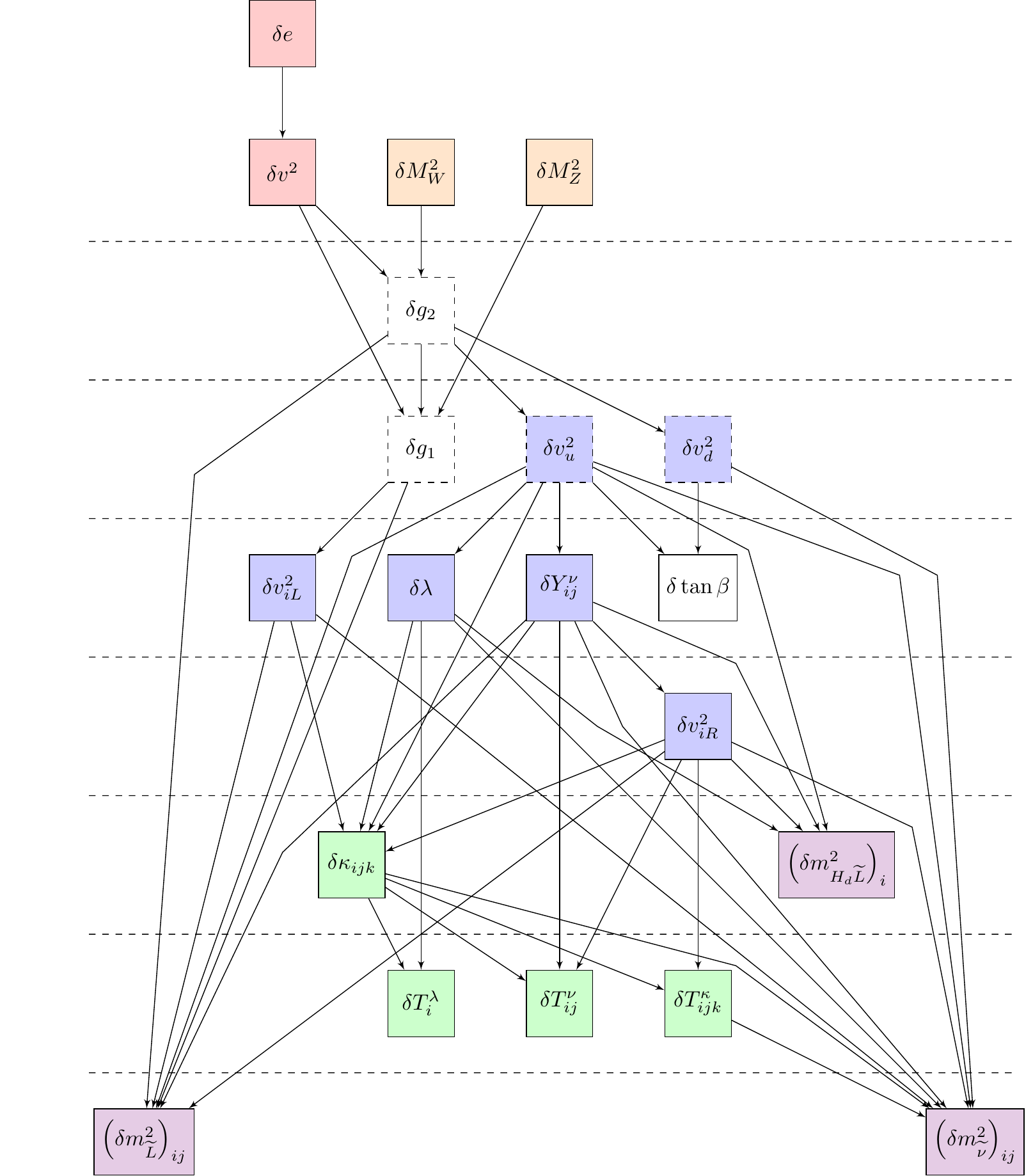}
    \caption{Strategy for extracting the counterterms needed for renormalizing
  the neutral scalar potential. The arrows indicate the order in which
  the counterterms were obtained, while the colors stand for
  the sector that was used to extract the counterterms.
  \textit{Red}: Renormalization of electromagnetic
  coupling. \textit{Violet}: Renormalization of $\cp$-odd self-energies.
  \textit{Yellow}: Renormalization of  gauge boson self-energies.
  \textit{Blue}: Renormalization of neutral fermion self-energies.
  \textit{Green}: Renormalization of $\CP$-even scalar trilinear couplings.
  \textit{White}: Completely fixed by the dependence on other counterterms.
  The counterterms in the dashed boxes do not belong to the set of independent
  parameters, but their counterterms were calculated as an intermediate step.
  The counterterms below one of the horizontal dashed lines could be
  extracted only after the counterterms above the same horizontal line
  were determined.}
  \label{ctoverview}
\end{figure}

The determination of the counterterms for the set of independent parameters
was done in a specific order, because in some cases the definition of 
the renormalization condition of one
counterterm depends on other counterterms, that necessarily had to be
determined before. In \reffi{ctoverview} we give an overview of the strategy for the extraction
of the counterterms. We also highlight in color the sectors of
the \mnSSM\ in which the corresponding counterterm was extracted (see caption).
The exact definition of the counterterms and their final analytic expressions
in terms of UV divergences for the \DRbar\
counterterms are listed in App.~\ref{app:counters}.

Therein, divergent parts are expressed
proporional to $\De$, 
\begin{equation}
\Delta = \frac{1}{\varepsilon}-\gamma_E+\ln{4\pi} \; ,
\end{equation}
where loop integral are solved in $4-2\varepsilon$ dimensions and
$\gamma_E=0.5772\dots$ is the Euler-Mascharoni constant.
Since the field renormalization constants contribute only via divergent
parts, they do not contribute to the finite result after canceling 
divergences in the self-energies.
As regularization
scheme we choose dimensional reduction~\cite{SIEGEL1979193,CAPPER1980479}
which was shown to be SUSY
conserving at the one-loop level~\cite{Stockinger:2005gx}. In contrast to the
OS renormalization scheme our field renormalization matrices are
hermitian.
%, since we do not account for $\cp$-violation.
This holds also
true for the field renormalization in the mass eigenstate basis, because
as already mentioned the rotations in \refeq{eq:scalarhiggs} and
\refeq{eq:scalaroddhiggs} diagonalize the renormalized tree-level scalar
mass matrices, so \refeqs{eq:fieldrenorotate} do not introduce
non-hermitian parts into the field renormalization that would have to
be canceled by a renormalization of the mixing matrices $U^H$ and $U^A$
themselves.

\subsubsection{OS conditions}
The SM gauge boson masses are renormalized OS requiring
\begin{equation}
\text{Re}\left[{\hat{\Sigma}_{ZZ}}^T\left( \MZ^2 \right) \right]=0 \qquad
\text{and} \qquad
\text{Re}\left[{\hat{\Sigma}_{WW}}^T\left( \MW^2 \right) \right]=0 \; ,
\end{equation}
where $\hat{\Sigma}^T$ stands for the transverse part of the renormalized gauge boson self-energy. For their mass counterterms these conditions yield
\begin{equation}
\delta \MZ^2 = \text{Re}\left[ {\Sigma}_{ZZ}^T\left( \MZ^2 \right) \right] \qquad
\text{and} \qquad
\delta \MW^2 = \text{Re}\left[ {\Sigma}_{WW}^T\left( \MW^2 \right) \right] \; .
\end{equation}
Here the ${\Sigma}^T$ (without the hat) denote the transverse part of the unrenormalized gauge boson self-energies.

\begin{figure}
  \centering
  \includegraphics[width=0.6\textwidth]{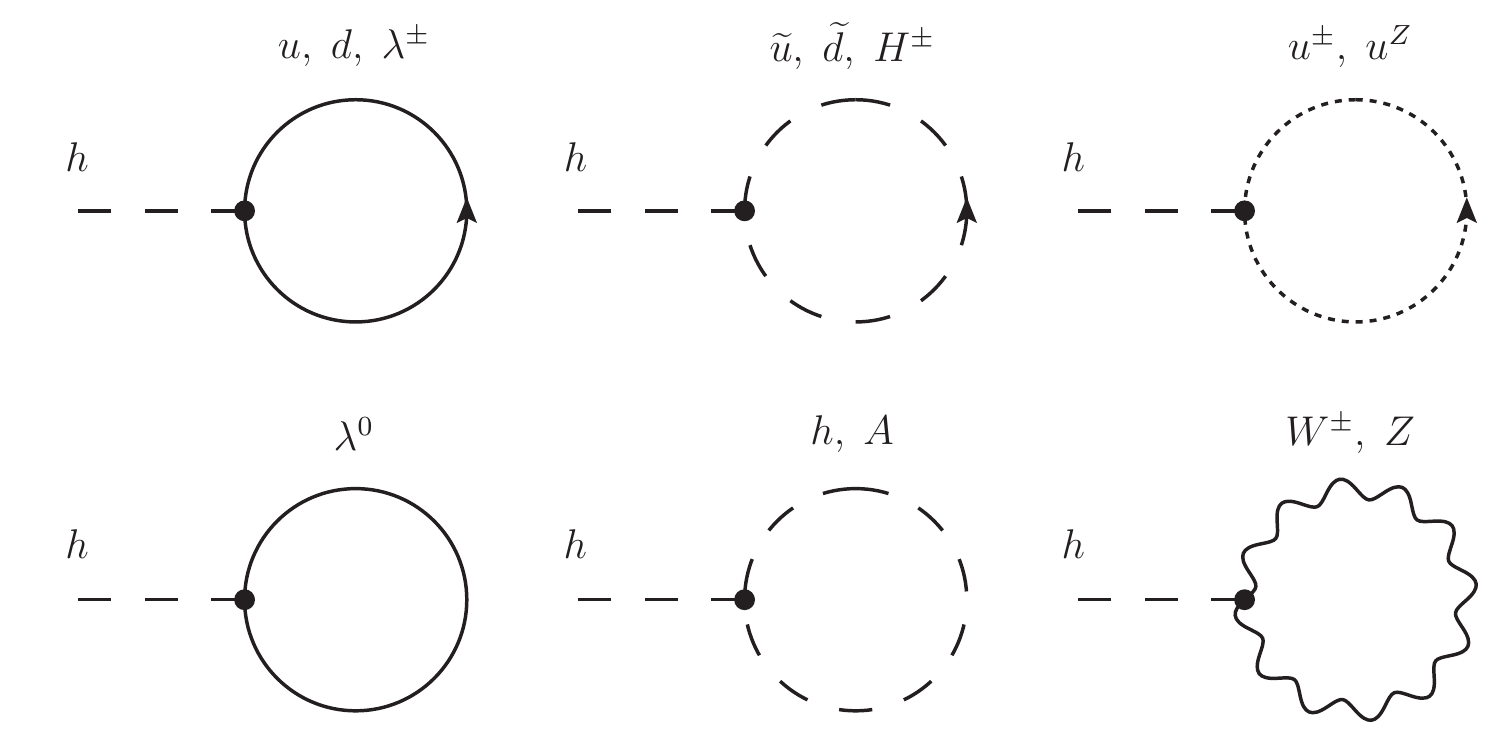}
  \caption{Generic Feynman diagrams for the tadpoles $T_{h_i}$.}
  \label{fig:tadpoles}
\end{figure}

For the tadpole coefficients $T_{\varphi_i}$ the OS conditions read
\begin{equation}
%T_{\varphi_i}+T_{\varphi_i}^{(1)}+\delta T_{\varphi_i}=0 \; ,
T_{\varphi_i}^{(1)}+\delta T_{\varphi_i}=0 \; ,
\end{equation}
where $T_{\varphi_i}^{(1)}$ are the one-loop contributions to the linear terms of the scalar potential, stemming from tadpole diagrams shown in \reffi{fig:tadpoles}. The tadpole diagrams are calculated in the mass eigenstate basis $h$. The one-loop tadpole contributions in the interaction basis $\varphi$ are then obtained by the rotation
\begin{equation}
T_\varphi^{(1)} = {U^H}^T T_h^{(1)} \; .
\end{equation}
Accordingly we find for the one-loop tadpole counterterms
\begin{equation}
\delta T_{\varphi_i}=-T_{\varphi_i}^{(1)} \; .
\end{equation}
%%%%%%%%%%%%%%%%%%%%%%%%%%%% F I G U R E %%%%%%%%%%%%%%%%%%%%%%%%%%%%%%%%%%%%%%
%%%%%%%%%%%%%%%%%%%%%%%%%%%% F I G U R E %%%%%%%%%%%%%%%%%%%%%%%%%%%%%%%%%%%%%%

%%%%%%%%%%%%%%%%%%%%%%%%%%%%%%%%%%%%%%%%%%%%%%%%%%%%%%%%%%%%%%%%%%%%%%%%%%%%%%%
%%%%%%%%%%%%%%%%%%%%%%%%%%%%%%%%%%%%%%%%%%%%%%%%%%%%%%%%%%%%%%%%%%%%%%%%%%%%%%%

\subsubsection{\DRbar\ conditions}
For practical purposes we
decided to renormalize all remaining parameters in the $\DRbar$
scheme, reflecting the fact that there are no physical observables yet that
could be directly related to them. 
The counterterms of each parameter were obtained by
calculating the divergent parts of one-loop corrections to different
scalar and fermionic two- and three-point functions. 
We sketch the determination of the counterterms in the (possible) order 
in which they can be successively derived (see
\reffi{ctoverview}).

The general strategy for extracting the counterterms of the free parameters
is the following. At first one finds a relatively simple tree-level expression,
containing the parameter whose counterterm one wants to extract,
and, apart from that, exclusively parameters whose countertems 
are already known. In our case we used bilinear and trilinear couplings
in the neutral scalar and fermionic
sector for this. Then one calculates the one-loop corrections to the
term by evaluating the corresponding Feynman diagrams. As we only
need the divergent parts of the loop corrections for \DRbar\ conditions,
we are able to calculate the diagrams in the gauge eigenstate basis, where
the number of diagrams is drastically reduced.
Once the divergent contributions are known, the counterterm to be
identified directly follows from the expression of the renormalized
Green's functions.

In this chapter we only state the general formulas for the renormalized
two- and three-point Green's functions. The exact conditions used to
extract each countertem is listed in App. \ref{app:paracounters}.
Therein, we also show the resulting analytic expressions for the counterterms
renormalized in the \DRbar\ scheme.\footnote{An exceptions is the renormalization
of the SM vev $v$, which we extract from the counterterm of the electromagnetic
gauge coupling in the Thomson limit (see \citere{Biekotter:2017xmf}
for details).} A more detailed discussion in the case of the
\mnSSM\ with one generation of right-handed neutrinos can be found
in \citere{Biekotter:2017xmf}.

\paragraph{\protect\boldmath Neutral fermion sector\\}
We derived most of the counterterms in the neutral fermion sector.
The mass matrix elements of the neutral fermions
$\left(m_\nu\right)_{ij}$ get one-loop corrections via the neutral
fermion self-energies ${\sum}_{\neu{i} \neu{j}}$, that for Majorana
fermions can be decomposed as
\begin{equation}
\Sigma_{\neu{i} \neu{j}}\left( p^2 \right) = \slashed{p}\Sigma_{\neu{i} \neu{j}}^F\left( p^2 \right)+\Sigma_{\neu{i} \neu{j}}^S\left( p^2 \right) \; .
\end{equation}
Defining for the renormalized mass matrix
\begin{equation}
\left(m_\nu\right)_{ij} \rightarrow \left(m_\nu\right)_{ij}
+ \delta \left(m_\nu\right)_{ij} \; ,
\end{equation}
the renormalized scalar part of the self-energies at
zero momentum is given by
\begin{equation}\label{eq:neufermreno}
\hat{\Sigma}_{\neu{i} \neu{j}}^S\left(0\right)=\Sigma_{\neu{i} \neu{j}}^S\left(0\right)-\frac{1}{2}
  \left( \delta Z_{ki}^\chi \left( m_\nu \right)_{kj} +
      \left( m_\nu \right)_{ik} \delta Z_{kj}^\chi \right)
    -\delta\left(m_\nu\right)_{ij} \; .
\end{equation}
The field renormalization constants can be obtained
in the \DRbar\ scheme by calculating the divergent
part of the fermionic piece,
\begin{equation}
\delta Z_{ij}^\chi=-\left.\Sigma_{\neu{i}\neu{j}}^F\right|^{\rm div} \; .
\end{equation}
The divergent parts of the self-energies of the
neutral fermions were calculated diagrammatically
in the interaction basis,
where diagrams with mass insertions have to be included.
If $\left(m_\nu\right)_{ij}$ containes
just a single parameter whose counterterm is unknow,
\refeq{eq:neufermreno} provides
a definition for the missing counterterm once the mass counterterm
$\delta\left(m_\nu\right)_{ij}$ is expressed in terms of the counterterms
of the fundamental parameters.
Unfortunately, the right-handed vevs $v_{iR}$
always appear in sums over the family index
and never isolated. Therefore we calculated loop corrections to the
three elements $\left(m_\nu\right)_{i,7}= - Y^\nu_{ij} v_{jR}/\sqrt{2}$,
which provides us with a linear system of three independent equations;
\begin{align}
Y^\nu_{ij} \delta v_{jR} =
\sqrt{2} \left. \Sigma_{\nu_{iL} \widetilde{H}^0_u}^S \right|^{\rm div} -
\frac{1}{2} \left(
            \delta Z_{ij}^\chi v_{kR} Y^\nu_{jk} -
            \delta Z_{i6}^\chi v_{jR} \lambda_j +
            \delta Z_{77}^\chi v_{jR} Y^\nu_{ij}
            \right) -
            v_{jR} \delta Y^\nu_{ij} \; ,
\end{align}
that can be solved analytically for the three counterterms
$\delta v_{iR}^2 = 2 v_{iR} \; \delta v_{iR}$.

\paragraph{\protect\boldmath Neutral scalar trilinear couplings\\}
General one-loop scalar three-point functions can be renormalized by wavefunction
counterterms and the specific vertex counterterm as
\begin{align}\label{eq:triplerenor}
\hat{\Gamma}_{\varphi_i \varphi_j \varphi_k}^{(1)} =
\Gamma_{\varphi_i \varphi_j \varphi_k}^{(0)} + 
\Gamma_{\varphi_i \varphi_j \varphi_k}^{(1)} -
\frac{1}{2} \biggl(
      \Gamma_{\varphi_l \varphi_j \varphi_k}^{(0)} \delta Z_{li} +
      \Gamma_{\varphi_i \varphi_l \varphi_k}^{(0)} \delta Z_{lj} +
      \Gamma_{\varphi_i \varphi_j \varphi_l}^{(0)} \delta Z_{lk}
      \biggr) -
\delta \Gamma_{\varphi_i \varphi_j \varphi_k}^{(1)} \; ,
\end{align}
where $\delta Z_{ij}$ are the scalar field renormailzation constants defined
in \refeq{eq:fieldrenodef}, $\Gamma_{\varphi_i \varphi_j \varphi_k}^{(0)}$
are the tree-level couplings, $\Gamma_{\varphi_i \varphi_j \varphi_k}^{(1)}$
are the one-loop corrections obtained by evaluating the non-irreducible
one-loop three-point diagrams,
and $\delta \Gamma_{\varphi_i \varphi_j \varphi_k}^{(1)}$ is the coupling
counterterm given as a function of the counterterms of the
independent parameters.
The field renormalization constants are defined as \DRbar-parameters.
We calculate the UV-divergent part of the derivative of the scalar $\cp$-even self-energies in the interaction basis and define
\begin{equation}\label{eq:drfieldcounterdef}
\delta Z_{ij}=-\left.\frac{d}{dp^2}\Sigma_{\varphi_i \varphi_j}
  \right|^{\rm div} \; .
\end{equation}
As before, if at tree level $\Gamma_{\varphi_i \varphi_j \varphi_k}^{(0)}$
just contains a single parameter whose counterterm is still unknown, the
counterterm can be extracted from the divergent part of the one-loop
corrections $\Gamma_{\varphi_i \varphi_j \varphi_k}^{(1)}$ demanding
that the renormalized quantity is finite.
Similarly to the vevs $v_{iR}$,
for the parameters $\kappa_{iij}$ it is not possible to find a
tree-level expression where each element appears isolated.
However, using the renormalized expression in \refeq{eq:triplerenor} for the vertex
\begin{equation}
\Gamma_{\widetilde{\nu}_{iR}^{\mathcal{R}} \widetilde{\nu}_{kR}^{\mathcal{R}}
\widetilde{\nu}_{jL}^{\mathcal{R}}}^{(0)} =
\frac{1}{2} v_d \lambda_k Y^\nu_{ji} +
\frac{1}{2} v_d \lambda_i Y^\nu_{jk} -
\frac{1}{2} v_{lL} Y^\nu_{li} Y^\nu_{jk} -
\frac{1}{2} v_{lL} Y^\nu_{lk} Y^\nu_{ji} -
v_u \kappa_{ikl} Y^\nu_{jl} \; ,
\end{equation}
we can extract the counterterms for
the three subsets $(\delta \kappa_{11j},\delta \kappa_{22j},\delta \kappa_{33j})$
by renormalizing the subset of vertices
($\Gamma_{\widetilde{\nu}_{1R}^{\mathcal{R}} \widetilde{\nu}_{1R}^{\mathcal{R}}
\widetilde{\nu}_{jL}^{\mathcal{R}}}$,
$\Gamma_{\widetilde{\nu}_{2R}^{\mathcal{R}} \widetilde{\nu}_{2R}^{\mathcal{R}}
\widetilde{\nu}_{jL}^{\mathcal{R}}}$,
$\Gamma_{\widetilde{\nu}_{3R}^{\mathcal{R}} \widetilde{\nu}_{3R}^{\mathcal{R}}
\widetilde{\nu}_{jL}^{\mathcal{R}}}$).
Thus, for each subset ($\kappa_{iij}\, ; i=1,2,3$) we
get a linear system of three ($j=1,2,3$) equations
to extract the counterterms $\delta \kappa_{iij}$ from the condition that the
renormalized one-loop three-point function is finite.

%%%%%%%%%%%%%%%%%%%%%%%%%%%% F I G U R E %%%%%%%%%%%%%%%%%%%%%%%%%%%%%%%%%%%%%%
\begin{figure}
  \centering
  \includegraphics[width=0.76\textwidth]{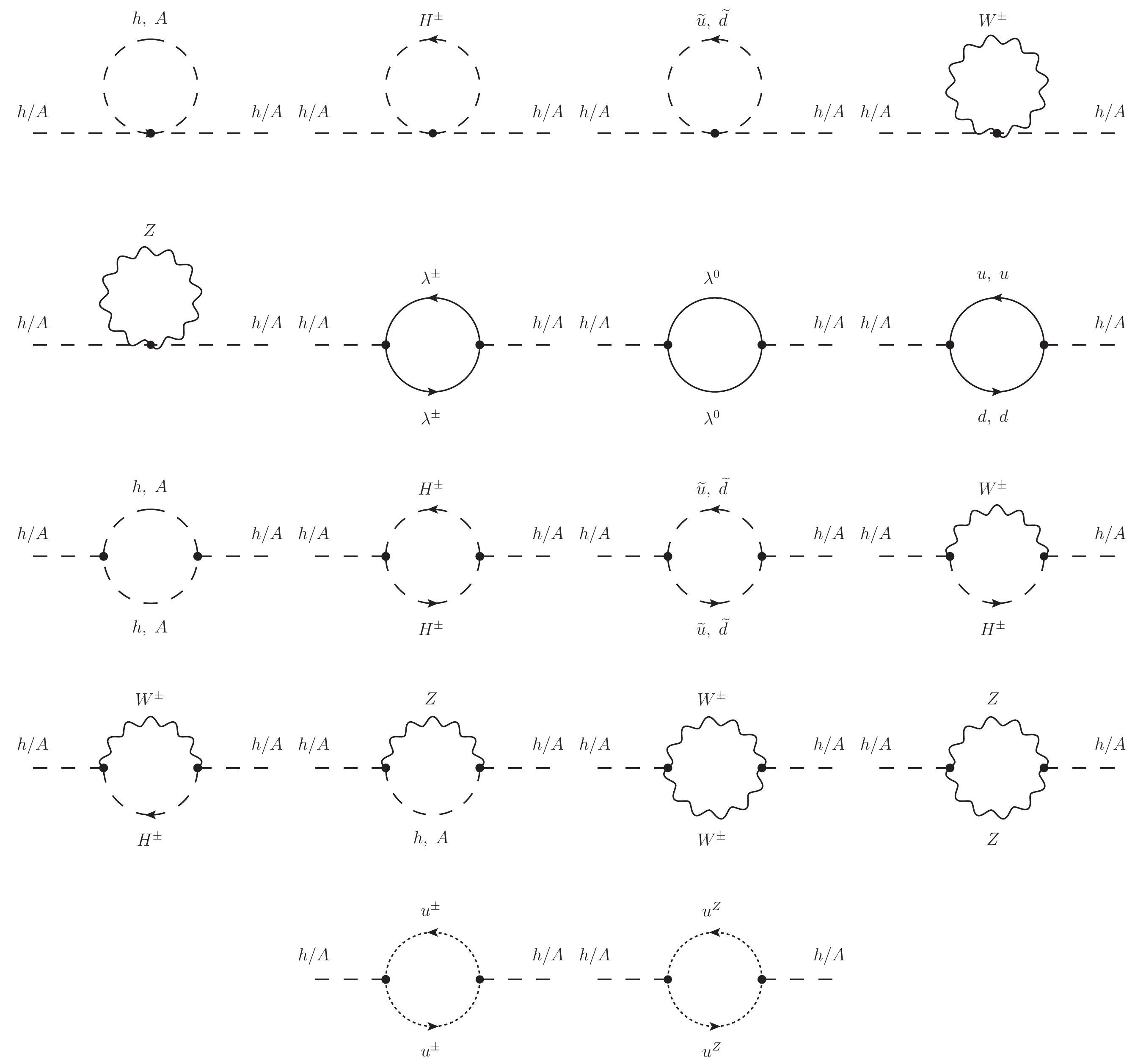}
  \caption{Generic diagrams for the $\cp$-even ($h$) and 
  $\cp$-odd ($A$) scalar self-energies in the mass
    eigenstate basis.}
\label{fig:scalarselfs}
\end{figure}
%%%%%%%%%%%%%%%%%%%%%%%%%%%% F I G U R E %%%%%%%%%%%%%%%%%%%%%%%%%%%%%%%%%%%%%%
\paragraph{\protect\boldmath Neutral scalar masses\\}
The soft scalar masses appear in the bilinear terms of the Higgs
potential. They can be renormalized by calculating radiative corrections
to scalar self-energies. Since our final aim is to obtain loop corrections for
the $\CP$-even scalars, we used the $\CP$-odd scalar sector to extract the
counterterms of the soft masses to have an independent crosscheck of
both neutral scalar sectors.

The general form of the renormalized scalar self-energies at the one-loop level is
\begin{align}
\hat{\Sigma}_{X_i X_j}\left( p^2 \right) =&
\Sigma_{X_i X_j}\left( p^2 \right)+
\frac{1}{2} p^2 \left( \delta Z_{ji} + \delta Z_{ij} \right) \notag \\
-&\frac{1}{2} \left( \delta Z_{ki} \left( m_{X}^2 \right)_{kj} +
\left( m_{X}^2 \right)_{ik} \delta Z_{kj} \right) -
\delta \left( m_{X}^2 \right)_{ij} \; ,
\end{align}
where $X=(\varphi,\sigma)$ represents either the $\cp$-even or the $\cp$-odd scalar fields and we made use of the fact that the field renormalization constants $\delta Z$ and the mass matrix $m_{X}^2$
are real.
Demanding that the renormalized self-energies $\hat{\Sigma}_{A_i A_j}$ are finite in the mass eigenstate basis we can define the divergent parts of the mass counterterms via
\begin{equation}
\left.\delta \left( m_{A}^2 \right)_{ij}\right|^{\rm div} =
\left.\Sigma_{A_i A_j}\left( 0 \right) \right|^{\rm div} -
\frac{1}{2}\left( \left(\delta Z^A\right)_{ji} m_{A_j}^2 +
m_{A_i}^2 \left(\delta Z^A\right)_{ij} \right)\; 
\label{eq:oddcounters} ,
\end{equation}
where the field counterterms in the mass eigenstate basis were defined
in \refeq{eq:fieldrenorotate}, and renormalized as \DRbar\ parameters
like the ones for the
$\cp$-even scalars (see \refeq{eq:drfieldcounterdef}),
and the masses squared $m_{A_i}^2$ are the
eigenvalues of the diagonal
$\cp$-odd scalar mass matrix $m_{A}^2$. In \reffi{fig:scalarselfs} we show
the diagrams that have to be calculated to
obtain the quantum corrections to scalar
self-energies at the one-loop level in the mass eigenstate basis.

We calculated all diagrams in the 't Hooft-Feynman gauge in which the Goldstone bosons $A_1$ and $H^\pm_1$ and the ghost fields $u^\pm$ and $u^Z$ have the same masses as the corresponding gauge bosons.
Calculating the $\cp$-odd self-energies $\Sigma_{A_i A_i}$
diagrammatically, we obtain the mass counterterms in mass
eigenstate basis through the \refeq{eq:oddcounters}.
Now inverting the rotation in 
\refeq{eq:masscounterrot} yields the mass
counterterms for the $\cp$-odd self-energies in
the interaction basis, 
\begin{equation}\label{eq:oddcountersrot}
\left.\delta m_{\sigma}^2\right|^{\rm div}=
{U^A}^T \left.\delta m_{A}^2\right|^{\rm div} U^A \; .
\end{equation}
Analytically, following the expansion
in \refeq{eq:masscounterderive},
some of the mass counterterms $\delta m_\sigma^2$ depend on
the counterterms of the soft mass parameters.
We use this dependences to extract
the counterterms of $(m_{H_d \widetilde{L}}^2)_i$ and the
counterterms of the non-diagonal elements of
$(m_{\widetilde{L}}^2)_{ij}$
and $(m_{\widetilde{\nu}}^2)_{ij}$.

\section{Loop corrected scalar masses}
\label{sec:getmasses}

In the previous section we have defined an
OS/\DRbar\ renormalization scheme for the \mnSSM\ neutral scalar sector. This
can be applied (via the \FA\ model file, in which the counterterms
are implemented) to any higher-order
correction in the \mnSSM. As a first application, we evaluate the full
one-loop corrections to the $\cp$-even scalar sector in the \mnSSM.
In the following we emphasize the differences w.r.t the analysis with just one
right-handed neutrino from \citere{Biekotter:2017xmf}.

\subsection{Evaluation at the one-loop level} 
The one-loop renormalized self-energies in the mass eigenstate basis are
given by
\begin{equation}\label{eq:renomself}
\hat{\Sigma}_{h_i h_j}^{(1)}\left( p^2 \right) = \Sigma_{h_i h_j}^{(1)}\left( p^2 \right) + \delta Z^H_{ij}
\left( p^2 - \frac{1}{2}\left( m_{h_i}^2 + m_{h_j}^2 \right) \right)
- \left(\delta m_{h}^2\right)_{ij} \; ,
\end{equation}
with the field renormalization constants $\de Z^H$ 
and the mass counter terms $\delta m_{h}^2$
in the mass eigenstate basis
defined by the rotations in \refeq{eq:fieldrenorotate}
and \refeq{eq:masscounterrot}, respectively.
$\Sigma_{h_i h_j}$ is the unrenormalized self-energy obtained by
calculating the diagrams shown in \reffi{fig:scalarselfs} 
with the $\cp$-even states $h$ on the external legs.
The self-energies were calculated in the 't Hooft-Feynman
gauge, so that gauge-fixing terms do not yield counterterm contributions
in the Higgs sector at the one-loop level. The loop integrals were regularized
using dimensional reduction \cite{SIEGEL1979193,CAPPER1980479} and
numerically evaluated for arbitrary real momentum using
\LT~\cite{Hahn:1998yk}. The contributions from complex values of $p^2$
were approximated using a Taylor expansion with respect to the imaginary part
of $p^2$ up to first order. 

In \refeq{eq:renomself} we already made use of the fact that $\delta
Z^H$ is real and symmetric in our renormalization scheme. The mass
counterterms are defined as functions of the counterterms of the free
parameters following \refeq{eq:masscounterderive} and
\refeq{eq:masscounterrot}. They contain finite contributions from the
tadpole counterterms and from the counterterm for the gauge boson mass
$\MZ^2$. The matrix $\delta m_{h}^2$ is real and symmetric.

The renormalized self-energies enter the inverse propagator matrix
\begin{equation}\label{eq:renoselfdef}
\hat{\Gamma}_{h}=\ii \left[ p^2\; \mathbbm{1} -
\left( m_{h}^2 - \hat{\Sigma}_{h}\left(p^2\right) \right)
\right] \; , \qquad  \text{with }
\left(\hat{\Sigma}_{h}\right)_{ij} = 
\hat{\Sigma}_{h_i h_j} \; .
\end{equation}
The loop-corrected scalar masses squared are the zeroes of the
determinant of the inverse propagator matrix. The determination of
corrected masses has to be done numerically when one wants to account for
the momentum-dependence of the renormalized self-energies. This is done
by an iterative method that has to be carried out for each of the six
squared loop-corrected masses.\footnote{Details
about the numerical algorithm used can be found
in \citere{Fuchs:2015jwa}.}

%%%%%%%%%%%%%%%%%%%%%%%%%%%%%%%%%%%%%%%%%%%%%%%%%%%%%%%%%%%%%%%%%%%%%%%%%%%%%%%

\subsection{Inclusion of higher orders}\label{sec:higherorders}

In \refeq{eq:renoselfdef} we did not include the superscript $^{(1)}$ in
the self-energies. Restricting the numerical evaluation to a pure
one-loop calculation would lead to very large theoretical uncertainties.
These can be avoided by the inclusion of corrections beyond the one-loop
level. Here we follow the approach of
\citeres{Drechsel:2016jdg,Biekotter:2017xmf}
and supplement the \mnSSM\ one-loop results by higher-order corrections in
the MSSM limit as provided by \fh\ (version 2.13.0)~\cite{Heinemeyer:1998yj,Hahn:2009zz,Heinemeyer:1998np,Degrassi:2002fi,Frank:2006yh,Hahn:2013ria,Bahl:2016brp,feynhiggs-www}.\footnote{Using the latest version 2.14.3 would have
a minor impact on our numerical analysis.}
In this way the leading and subleading two-loop corrections are
included, as well as a resummation of large logarithmic terms, see the
discussion in \refse{sec:intro}, 
\begin{equation}
\hat{\Sigma}_{h}\left( p^2 \right) = 
\hat{\Sigma}_{h}^{(1)}\left( p^2 \right) +
\hat{\Sigma}_{h}^{(2')} +
\hat{\Sigma}_{h}^{\rm resum} \; .
\end{equation}
In the partial two-loop contributions $\hat{\Sigma}_{h}^{(2')}$ we take
over the corrections of
\order{\alpha_s\alpha_t,\alpha_s\alpha_b,\alpha_t^2,\alpha_t\alpha_b}, 
assuming that the MSSM-like corrections
approximate numerically well the corresponding
\mnSSM\ corrections. This assumption 
is reasonable since the only difference between the squark sector of the
\mnSSM\ in comparison to the MSSM are numerically suppressed terms proportional to 
$v_{jL} Y^\nu_{jk} v_{kR}$ in the non-diagonal element of the up-type squark mass 
matrices (see \refeq{eq:usquarks12})
and proportional to $v_{iL}v_{iL}$ in the diagonal
elements of the up- and down-type squark mass matrices
(see \refeqs{eq:usquarks11}, (\ref{eq:usquarks22}),
(\ref{eq:dsquarks11}) and (\ref{eq:dsquarks22})).
Furthermore, in \citeres{Drechsel:2016ukp,Drechsel:2016jdg}
the quality of the MSSM
approximation was tested in the NMSSM, showing that the genuine NMSSM
contributions are in most cases sub-leading. Our results in
\citere{Biekotter:2017xmf} confirm that the same holds true in the
\mnSSM\ for the SM-like Higgs-boson mass.
The same is expected for
the contributions stemming from the resummation of large logarithmic
terms given by $\hat{\Sigma}_{h}^{\rm resum}$.

%%%%%%%%%%%%%%%%%%%%%%%%%%%%%%%%%%%%%%%%%%%%%%%%%%%%%%%%%%%%%%%%%%%%%%%%%%%%%%%
%%%%%%%%%%%%%%%%%%%%%%%%%%%%%%%%%%%%%%%%%%%%%%%%%%%%%%%%%%%%%%%%%%%%%%%%%%%%%%%

\section{Numerical analysis}
\label{sec:numanal}
In the following we will present several benchmark points (BPs) that
illustrate the phenomenology of the scalar sector of the \mnSSM.
We concentrate on scenarios
in which a right-handed sneutrino is mixed
with the SM-like Higgs boson.
By setting $\kappa_{ijk}=\kappa\;\delta_{ij}\delta_{jk}$,
as explained in \refse{sec:neutrscasec}, we achieve that only
a single right-handed sneutrino
substantially mixes with the SM-like Higgs boson.
Naturally, the mass scale of the right-handed
sneutrinos will then be of the order of the SM-like Higgs boson.
However, scenarios in which the decay of the SM-like Higgs boson to two
right-handed sneutrinos is kinematically allowed, are experimentally
very constrained~\cite{Ghosh:2014ida}.

In contrast to most of the previous studies of the \mnSSM\
with three generations of right-handed
neutrinos~\cite{Escudero:2008jg,Ghosh:2008yh,Fidalgo:2009dm,Ghosh:2010zi,Ghosh:2014ida},
we will not always make the simplifying
assumption that genuine low-energy \mnSSM-parameters have
universal values independent of the family index.
In \refse{sec:lamsqcons} we elaborate on the
effect of non-universal $\lambda_i$ on the SM-like Higgs-boson mass, while
keeping $\pmb{\lambda}^2=\lambda_i\lambda_i$ constant.
Since we know from \refeq{eq:approhiggs} that the tree-level mass
of the SM-like Higgs boson strongly depends on $\pmb{\lambda}^2$, it
will be discussed whether the loop corrections increase the dependence on the
individual values $\lambda_i$.

We consider the following experimental constraints on the scenarios presented:
\begin{itemize}
  \item[-] We use the public code
           \texttt{HiggsBounds v.5.2.0}~\cite{Bechtle:2008jh,Bechtle:2011sb,Bechtle:2013gu,Bechtle:2013wla,Bechtle:2015pma}
           to determine
           whether a BP has been excluded by cross section limits
           from Higgs searches at LEP, LHC or Tevatron. These searches are
           mostly sensitive to the heavy Higgs and the right-handed
           sneutrinos, if these are substantially mixed with the SM-like
           Higgs boson. The production of the left-handed sneutrinos is
           much smaller at the LHC, and signals from their decay usually
           demand dedicated searches~\cite{Lara:2018rwv},
           especially if the left-handed
           sneutrino is the LSP~\cite{Lara:2018zvf,Ghosh:2017yeh}.
  \item[-] The properties of the SM-like Higgs boson, i.e., its mass
           and signal rates at LHC and Tevatron, are checked using
           the public code
           \texttt{HiggsSignals v.2.2.1}~\cite{Bechtle:2013xfa,Stal:2013hwa,Bechtle:2014ewa}.
           Here we assume
           a theoretical mass uncertainty of $3\gev$.
           \texttt{HiggsSignals} provides
           us with a $\chi^2$-analysis of $n_{\text{obs}}=106$
           observables in the 7+8 TeV
           data package and $n_{\text{obs}}=101$ observables
           in the 13 TeV data package.
           In our plots we show the reduced
           $\chi^2_{\mathrm{red}}=\chi^2/n_{\mathrm{obs}}$,
           where a value of
           $\chi_{\mathrm{red}}^2=1$ means that on average the signal rates of the
           SM-like Higgs boson are at the level of the
           $\pm 1\sigma$ range of the measurements.
  \item[-] The properties of the neutrino sector are in agreement with
           the measurements of the mass-squared differences and the
           mixing angles obtained from neutrino oscillation
           experiments. We check
           that our predictions are within the $\pm 3\sigma$ bands published
           by the NuFit collaboration~\cite{Esteban:2016qun,nufit-www},
           \begin{align}
              6.80\eev^2  \leq \delta m_{12}^2/10^{-5} \leq  8.02\eev^2 \, , \quad
              2.399\eev^2 \leq \Delta m_{13}^2/10^{-3} \leq 2.593\eev^2 \, , \\
              0.0198 \leq s_{13}^2 \leq 0.0244 \, , \quad
              0.272  \leq s_{12}^2 \leq 0.346  \, , \quad
              0.418  \leq s_{23}^2 \leq 0.613  \, ,
           \end{align}
           where we considered the normal mass ordering
           which is now favored by experiments~\cite{Esteban:2018azc}.
           A genetic algorithm was used to find parameter
           points that minimize the sum of squared deviations between
           theoretical prediction and experimental values
           specified above~\cite{DEAP_JMLR2012}.
           Even though the \mnSSM\ allows for
           flavor-violating decays of leptons, the existing experimental
           bounds (for instance on $\mu\rightarrow e\gamma$) are
           automatically fulfilled when the constrains on neutrino
           masses are taken into account~\cite{Bartl:2009an}.
\end{itemize}
For the necessary input of \texttt{HiggsBounds} and \texttt{HiggsSignals}
we compute the decays of the scalars at leading order, but with the
loop-corrected mixing matrix elements inserted in the expressions of
the scalar couplings. In the limit of vanishing external momentum, which
we used in the determination of the mixing matrix elements for the couplings,
this method corresponds to include the finite wave-function
renormalization factors ($Z$-factors) for each external
scalar~\cite{Goodsell:2017pdq,Frank:2006yh}. For loop-induced decays and
off-shell decays to vector bosons we implemented analytic results 
from the MSSM well
known in the
literature~\cite{Djouadi:1995gv,Djouadi:2005gi,Spira:2016ztx,Djouadi:1997rp,PhysRevD.22.722}, and scaled the expressions with effective couplings
defined by the mixing matrix elements and $\tan\beta$
to obtain the result for the scalars in the \mnSSM.
For the coupling to $b$ quarks we included the running bottom mass
and for the decay to gluons the running of $\alpha_s$
from $M_Z$ to the mass of the decaying scalar,
and finally add leading higher-order
QCD corrections~\cite{memospira,Spira:2016ztx}.

As described in \refse{sec:neuferm} we use the left-handed vevs $v_{iL}$,
the soft gaugino masses $M_1$ and $M_2$, and
the neutrino Yukawa couplings $Y^\nu_{ij}$ to fit the neutrino masses and
mixings accurately, making use of the fact that
they can be modified without spoiling the
properties of the SM-like Higgs boson.
Besides for the scenario presented in \refse{sec:lep},
 it will be sufficient
to just consider diagonal non-zero elements of $Y^\nu_{ij}$.
Because we concentrate here on
the scalar sector of the \mnSSM, and since the fitting has to be done
numerically, we do the fitting in our scans 
just in one particular point for each analysis.
By varying a parameter, the prediction for
the neutrino properties can be outside
the experimentally allowed range in some points.
We indicate in our plots when this is the case.
Since the neutral fermion mass matrix is of dimension 10, with large
hierarchies between the neutrino sector and the remaining part,
including one-loop corrections is time-consuming and numerically very
challenging. Therefore we stick to a tree-level analysis for the
neutrinos. However, we checked for several points that the one-loop
corrections are sub-leading and can in principle be compensated by
a slight change of the parameters.\footnote{See also \citere{Ghosh:2010zi}
for a detailed discussion of radiative corrections to the neutrino masses.}

\subsection{Scan over $\lambda$}
\label{sec:scanlam}
\begin{table}
\centering
\renewcommand{\arraystretch}{1.3}
\begin{tabular}{c c c c c c c c}
 $\tan\beta$ & $\lambda$ & $\kappa$ & $v_{1,3R}$ & $v_{2R}$ & 
   $A^\lambda$ & $A^\kappa$ & $A^\nu$ \\
 \hline
 $5$ & $[0.13, 0.18]$ & $0.5$ & $1000$ & $765$ &
   $1000$ & $-1000$ & $-1000$ \\
 \hline
 \hline
 $A^u_3$ & $A^u_{1,2}$ & $A^{d,e}$ & 
   $m_{\widetilde{Q},\widetilde{u},\widetilde{d}}$ & $m_{\widetilde{e}}$ &
   $M_3$ & & \\
 \hline
 $-2000$ & $-1500$ & $-1500$ & $1500$ & $200$ & $2700$ & & \\
 \hline
 \hline
 $v_{1L}/10^{-4}$ & $v_{2L}/10^{-4}$ & $v_{3L}/10^{-4}$ & $Y^\nu_{11}/10^{-7}$ &
   $Y^\nu_{22}/10^{-7}$ & $Y^\nu_{33}/10^{-7}$ &
   $M_1$ & $M_2$ \\
 \hline
 $1.390$ & $6.215$ & $4.912$ & $4.181$ & $1.756$ & $6.306$ &
   $1228$ & $2814$
\end{tabular}
\caption{Low-energy values for the parameters, as defined in the text,
of the scan over $\lambda$.
Dimensionful parameters are given in GeV.
The parameters in the last row are fitted to neutrino oscillation data.}
\label{tab:scanlam}
\renewcommand{\arraystretch}{1.0}
\end{table}
\begin{figure}
  \centering
  \includegraphics[width=0.9\textwidth]{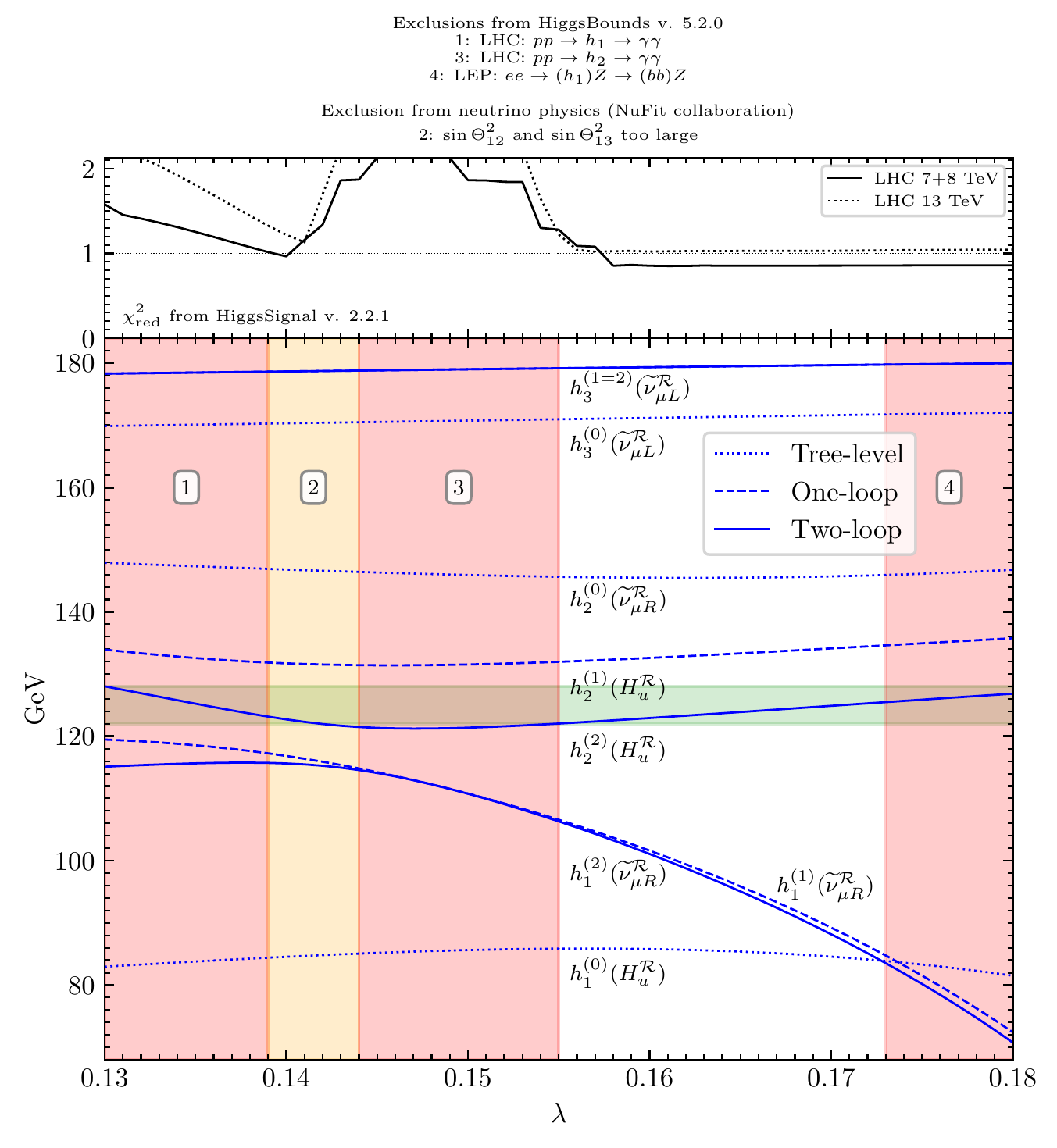}
  \caption{$\cp$-even scalar spectrum in the scan over $\lambda$ at
  tree level, one-loop level and partial two-loop level.
  We show in the brackets
  the dominant composition of the tree-level, one-loop
  and two-loop mass eigenstates $h^{(0)}$, $h^{(1)}$ and $h^{(2)}$, in
  the experimentally allowed region of the plot.
  The desired SM-like Higgs-boson mass is indicated with the horizontal green
  band, assuming a theory uncertainty of $3\gev$.
  The red regions are excluded by
  direct searches for additional scalars. In the yellow region the prediction
  for the mixing angles of the neutrinos lies outside of the 3$\sigma$ band
  of the experimental measurement. On top we show
  $\chi_{\mathrm{red}}^2$ for various Higgs-boson
  signal strength measurements at LHC.}
  \label{fig:scanlam}
\end{figure}
The first scenario we are presenting is one with a light right-handed
$\mu$-sneutrino that mixes substantially with the SM-like Higgs boson.
We show the chosen parameters in \refta{tab:scanlam}.
To simplify the notation we define $\lambda=\lambda_i$,
$A^\lambda = A^\lambda_i$,
$A^\nu = A^\nu_{ii}$,
$\kappa=\kappa_{iii}$ and
$A^\kappa = A^\kappa_{iii}$ and vanishing otherwise.
The soft parameters are given in terms of
$A^d=A^d_i$, $A^e=A^e_{ii}$,
$m_{\widetilde{Q}}=m_{\widetilde{Q}_i}$,
$m_{\widetilde{u}}=m_{\widetilde{u}_i}$,
$m_{\widetilde{d}}=m_{\widetilde{d}_i}$,
and $m_{\widetilde{e}}=m_{\widetilde{e}_{ii}}$
and vanishing otherwise.
We vary over the universal
parameter $\lambda$, while keeping the remaining parameters fixed.
For the right-handed $e$- and $\tau$-sneutrino vevs we chose
$v_{1,3R}=1\tev$, but set a smaller value of $v_{2R}=765\gev$ for the
$\mu$-sneutrino vev to decrease the mass of the $\cp$-even
$\mu$-sneutrino to the range around
the SM-like Higgs-boson mass.
The choice to pick $\widetilde{\nu}_{\mu R}$ as the light
right-handed sneutrino is of no relevance.
The large value of $\kappa=0.5$
assures that the other two right-handed sneutrinos will have masses between
$300$ and $400\gev$, well above $125\gev$.
Because the SM-like Higgs boson mass will get additional
contributions from the mixing with
$\widetilde{\nu}_{\mu R}^{\mathcal{R}}$, $\tan\beta$ can
be chosen rather low.

As mentioned in the previous subsection,
we fit the properties of the neutrinos
in just one particular point of the parameter scan. In this scenario,
this was done for $\lambda = 0.168$, leading to the values of $v_{iL}$,
$Y^\nu_{ii}$, $M_1$ and $M_2$ shown in \refta{tab:scanlam}.
We emphasize that this effectively leaves just the trilinear parameters
$A^\nu_{ii}$ to adjust the masses of the left-handed sneutrinos.
For the prediction of the masses of the right-handed sneutrinos and
the SM-like Higgs boson, the fitted parameters only play a minor role.

In \reffi{fig:scanlam} we show the resulting spectrum of
the light $\cp$-even scalars. The remaining $\cp$-even scalars not shown
in the plot have masses above $300\gev$ and do not play a role
in the following discussion.
The dotted lines represent the tree-level masses, the dashed lines the
masses including the full one-loop corrections, and the solid lines the
one-loop + partial two-loop + resummed
(referred to as two-loop in the following)
corrected masses, as explained in \refse{sec:higherorders}.
We mark four regions in the plot which are
excluded either by \texttt{HiggsBounds} (red),
or by not being in agreement with the neutrino oscillation data (yellow).
We stress that region 2 is just excluded for the precise choice
of parameters shown in \refta{tab:scanlam}. A new fit of the neutrino
properties for each value of $\lambda$ could easily accommodate predictions
for the properties of the neutrinos in agreement with experiments.
However, since this would exclusively affect the phenomenology of the
heavier left-handed sneutrinos in the scalar sector,
we do not apply the fit for each value of $\lambda$.

\begin{figure}
  \centering
  \includegraphics[width=0.8\textwidth]{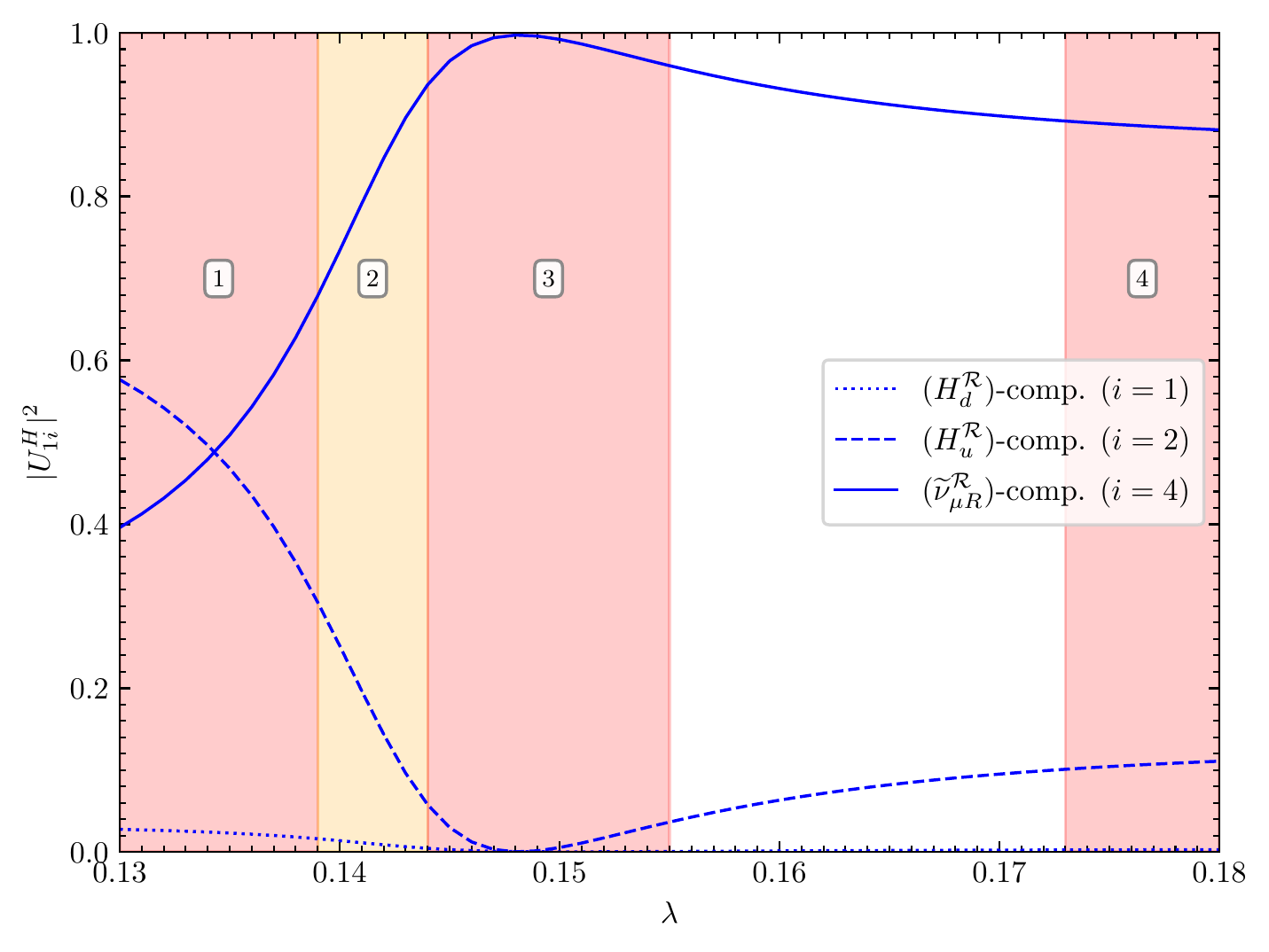}
  \caption{Doublet components ($H_d^{\mathcal{R}}$, $H_u^{\mathcal{R}}$) and
  the right-handed $\mu$-sneutrino component
  ($\widetilde{\nu}_{\mu R}^{\mathcal{R}}$)
  of the lightest $\cp$-even mass eigenstate $h_1$, which are
  defined by $|U^H_{1i}|^2$ with $i=1,2$ and $i=4$ respectively.}
  \label{fig:scanlammix}
\end{figure}

This spectrum is characterized by the interplay between the light
$\widetilde{\nu}_{\mu R}^{\mathcal{R}}$ and the SM-like Higgs boson.
For small $\lambda$
the two lightest loop-corrected mass
eigenstates $h_1$ and $h_2$ have roughly an equal amount of
$H_u^{\mathcal{R}}$- and
$\widetilde{\nu}_{\nu R}^{\mathcal{R}}$-admixture
(see also \reffi{fig:scanlammix}).
Consequently, region 1 is excluded by direct searches at the LHC,
because the diphoton resonance search for a SM-like higgs boson
excludes $h_1$ via its decay to photons~\cite{CMS:aya}.
%As $\lambda$ increases the
%tree-level mass corresponding to $\widetilde{\nu}_{\mu R}^{\mathcal{R}}$
%becomes smaller.
At $\lambda \sim 0.14$ the point is reached where
the mass of $h_1$ drops well below $125\gev$.
Thus, beyond that point $h_1$ can be identified with
$\widetilde{\nu}_{\mu R}^{\mathcal{R}}$, as the
doublet-component of $h_1$ shrinks to values
of roughly $\sim 10\%$.
$h_2$, on the other hand, sheds its sneutrino admixture,
so that it can be identified as the SM-like
Higgs boson, and the large quantum corrections from the
top/stop sector dominantly contribute to the
mass of $h_2$.
This yields an increase of the SM-like Higgs boson mass
of several GeV, so that beyond region 3 it
agrees with the experimental value,
assuming a theoretical uncertainty of $3\gev$.

An interesting observation is that in the
allowed region of $\lambda$
the large one-loop corrections change the order of
$\widetilde{\nu}_{\mu R}^{\mathcal{R}}$
and the SM-like Higgs boson.
While the large shift of the SM-like Higgs-boson
mass from $\sim 83\gev$ at tree level to
$\sim 125\gev$ at two-loop level
are familiar from the MSSM,
the large one-loop corrections to
$\widetilde{\nu}_{\mu R}^{\mathcal{R}}$,
with a tree-level mass of $\sim 147 \gev$
and a two-loop mass below $100\gev$,
emphasize the importance of accurately
taking into account the full
parameter space of the \mnSSM.

\begin{figure}
  \centering
  \includegraphics[width=0.8\textwidth]{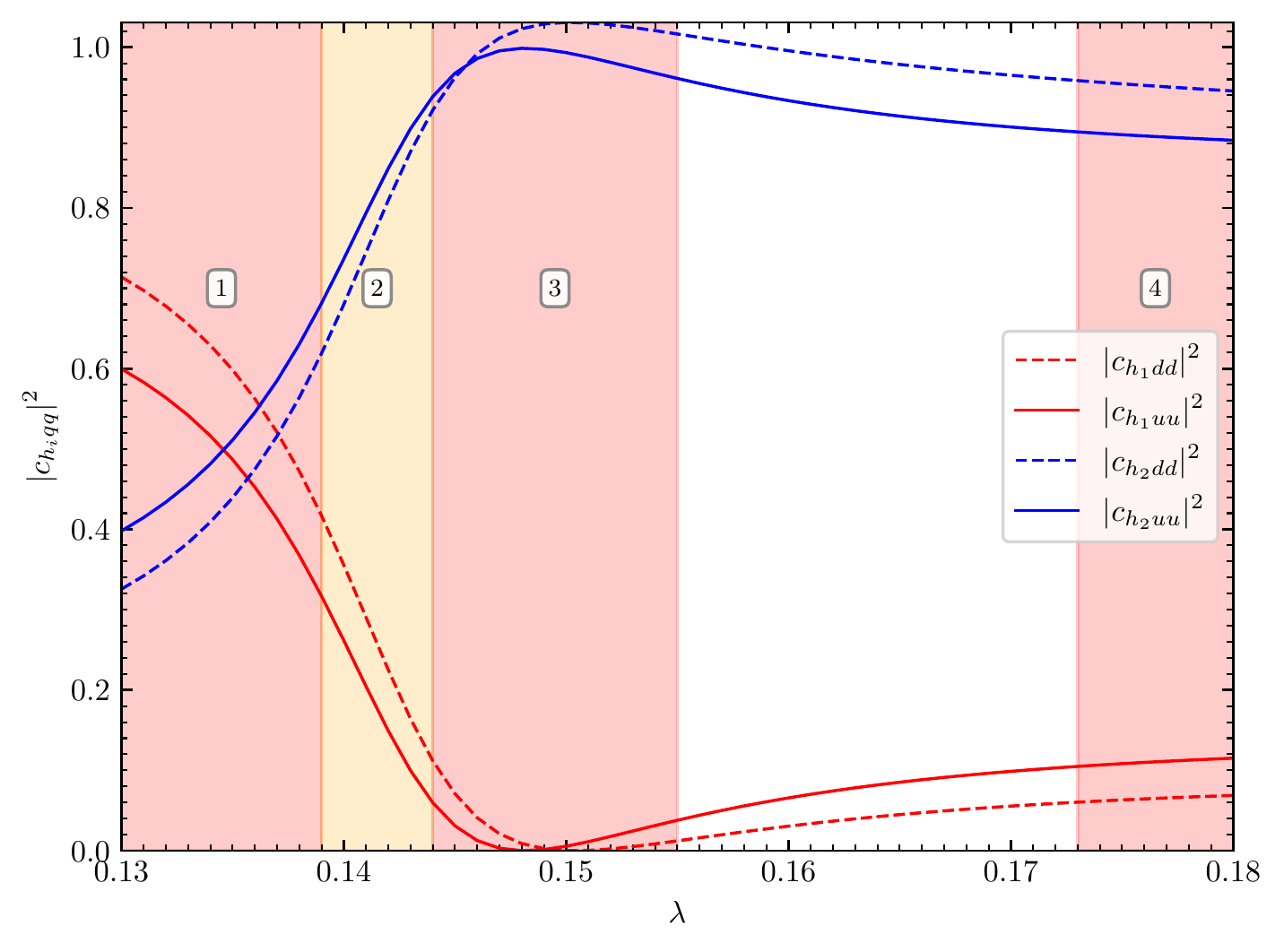}
  \caption{Effective couplings of the two light $\cp$-even
  scalar mass eigenstates $h_1^{(2)}$
  (\textit{red}) and $h_2^{(2)}$ (\textit{blue})
  to up-type quarks (\textit{solid}) and down-type
  quarks (\textit{dashed}), normalized
  to the SM prediction.}
  \label{fig:scanlamcpls}
\end{figure}

In the allowed region the doublet component of
$\widetilde{\nu}_{\mu R}^{\mathcal{R}}$
reaches values of approximately $10\%$, which can be seen
in \reffi{fig:scanlammix}, where we plot the down- and up-type
doublet component $H_d^{\mathcal{R}}$ and
$H_u^{\mathcal{R}}$,
and the
$\widetilde{\nu}_{\mu R}^{\mathcal{R}}$-component
of the lightest $\cp$-even scalar mass
eigenstates $h_1^{(2)}$.
Naturally, this mixing will also affect the SM-like Higgs-boson properties.
In this way, scenarios like the one shown here will be tested by experiments
in two different and complementary ways, both caused by the
mixing of $\widetilde{\nu}_{\mu R}^{\mathcal{R}}$ and the SM-like Higgs boson:
Firstly, direct searches for
additional Higgs bosons can be applied to
$\widetilde{\nu}_{\mu R}^{\mathcal{R}}$, because it
is directly coupled to SM particles.
Secondly, precise measurements of the SM-like Higgs-boson couplings can detect (or
exclude) possible variations from SM predictions.
To illustrate the possible modifications,
we show in \reffi{fig:scanlamcpls} the effective
coupling of the two light $\cp$-even
scalar mass eigenstates to up- and down-type quarks
normalized to the SM-prediction which
in good approximation can be expressed via the loop-corrected mixing matrix
elements $U^{H(2)}_{ij}$ and~$\beta$;
\begin{equation}
\label{eq:effcpl}
  c_{h_i d d} = \frac{U_{i1}^{H(2)}}{\cos\beta} \; , \qquad 
  c_{h_i u u} = \frac{U_{i2}^{H(2)}}{\sin\beta} \; .
\end{equation}
In the experimentally allowed region the effective coupling of
the SM-like Higgs boson to up-type quarks shows deviations of
roughly $10\%$. This is of the order of precision expected by
measurements of the SM Higgs boson couplings at the
High-Luminosity LHC~\cite{Cepeda:2019klc},
and (depending on the center-of-mass energy deployed)
an order of magnitude larger than
the uncertainty expected for these kind of measurements
at a possible future $e^+e^-$ collider like the ILC
~\cite{Dawson:2013bba,Fujii:2017vwa,Bambade:2019fyw}.
Comparing to \reffi{fig:scanlam} we can see that
the region where the effective couplings are closest to one, meaning
equal to the SM prediction, does not coincide with the region where
the $\chi_{\mathrm{red}}^2$ from \texttt{HiggsSignals} is minimized.
This is because the mass of the SM-like Higgs boson is slightly too small
in this range of $\lambda$, so even including a theoretical uncertainty
of $3\gev$ some signal strength measurements implemented in
\texttt{HiggsSignals} are not accounted for by $h_2$ and $\chi_{\mathrm{red}}^2$
becomes worse.

\subsection{Scan over $v_R$}
\label{sec:scanvr}
In \refse{sec:scanlam} we showed that light right-handed sneutrinos with
masses in the vicinity of the SM-like Higgs boson are theoretically
possible and can induce measurable modifications of the SM-like
Higgs-boson properties. Using data of direct searches and measurements
of the couplings of the SM-like Higgs boson, the parameter space of 
these scenarios can be constrained effectively.
In this section we present a scenario
that is not excluded by
current searches in which all three of the $\cp$-even right-handed
sneutrinos will have masses below $125\gev$.
We chose the parameters appearing in the mass
terms of the $\widetilde{\nu}_{i R}^{\mathcal{R}}$ to be universal, i.e.,
$\lambda := \lambda_i$, $v_R := v_{iR}$, $\kappa := \kappa_{iii}$,
$A^\lambda := A^\lambda_i$, $A^\nu := A^\nu_{ii}$ and
$A^\kappa := A^\kappa_{iii}$. As explained
at the beginning of \refse{sec:numanal},
the universality of $\kappa$
assures that only one of the $\widetilde{\nu}_{i R}^{\mathcal{R}}$
mixes substantially with the SM-like Higgs boson, while the other two
are practically decoupled. This makes it easier to control the total admixture
of the doublet components of the $\widetilde{\nu}_{i R}^{\mathcal{R}}$.

\begin{table}
\centering
\renewcommand{\arraystretch}{1.3}
\begin{tabular}{c c c c c c c c}
 $\tan\beta$ & $\lambda$ & $\kappa$ & $v_{R}$ & 
   $A^\lambda$ & $A^\kappa$ & $A^\nu$ & \\
 \hline
 $9.0$ & $0.08$ & $0.3$ & $[1210,1270]$ &
   $1000$ & $-1000$ & $-1000$  & \\
 \hline
 \hline
 $A^{u,d,e}$ &
   $m_{\widetilde{Q},\widetilde{u},\widetilde{d}}$ & $m_{\widetilde{e}}$ &
   $M_3$ & & & & \\
 \hline
 $-1000$ & $1000$ & $200$ & $2700$ & & & & \\
 \hline
 \hline
 $v_{1L}/10^{-5}$ & $v_{2L}/10^{-5}$ & $v_{3L}/10^{-4}$ & $Y^\nu_{11}/10^{-7}$ &
   $Y^\nu_{22}/10^{-7}$ & $Y^\nu_{33}/10^{-8}$ &
   $M_1$ & $M_2$ \\
 \hline
 $1.466$ & $8.520$ & $1.855$ & $2.963$ & $5.337$ & $5.902$ &
   $175.6$ & $188.0$
\end{tabular}
\caption{The same as in \refta{tab:scanlam} for the scan over $v_R=v_{iR}$.}
\label{tab:scanvr}
\renewcommand{\arraystretch}{1.0}
\end{table}

The complete set of free parameters is shown in \refta{tab:scanvr}.
In this scenario we scan over $v_R$, because they appear linearly in the
Majorana-like mass terms of the $\widetilde{\nu}_{i R}^{\mathcal{R}}$, so it is
a convenient parameter to control their masses. Compared to the scan over
$\lambda$ in \refse{sec:scanlam} the overall behavior of the SM-like Higgs
boson is aligned more to the SM predictions
by decreasing $\lambda$. Consequently,
because at tree level the additional contribution proportional to
$\pmb{\lambda}^2$ is smaller, $\tan\beta$ is larger
to increase the quantum corrections to the
SM-like Higgs-boson mass.
We also decrease $\kappa$ to make the masses of the
$\widetilde{\nu}_{i R}^{\mathcal{R}}$ smaller.
As before, the parameters in the
last row of \refta{tab:scanvr} were fitted to accurately predict the
left-handed neutrino masses and mixings. The fit was done in the point
$v_R=1226\gev$, but in this case the neutrino data is described accurately
over the whole range of $v_R$ at tree level.

\begin{figure}
  \centering
  \includegraphics[width=0.9\textwidth]{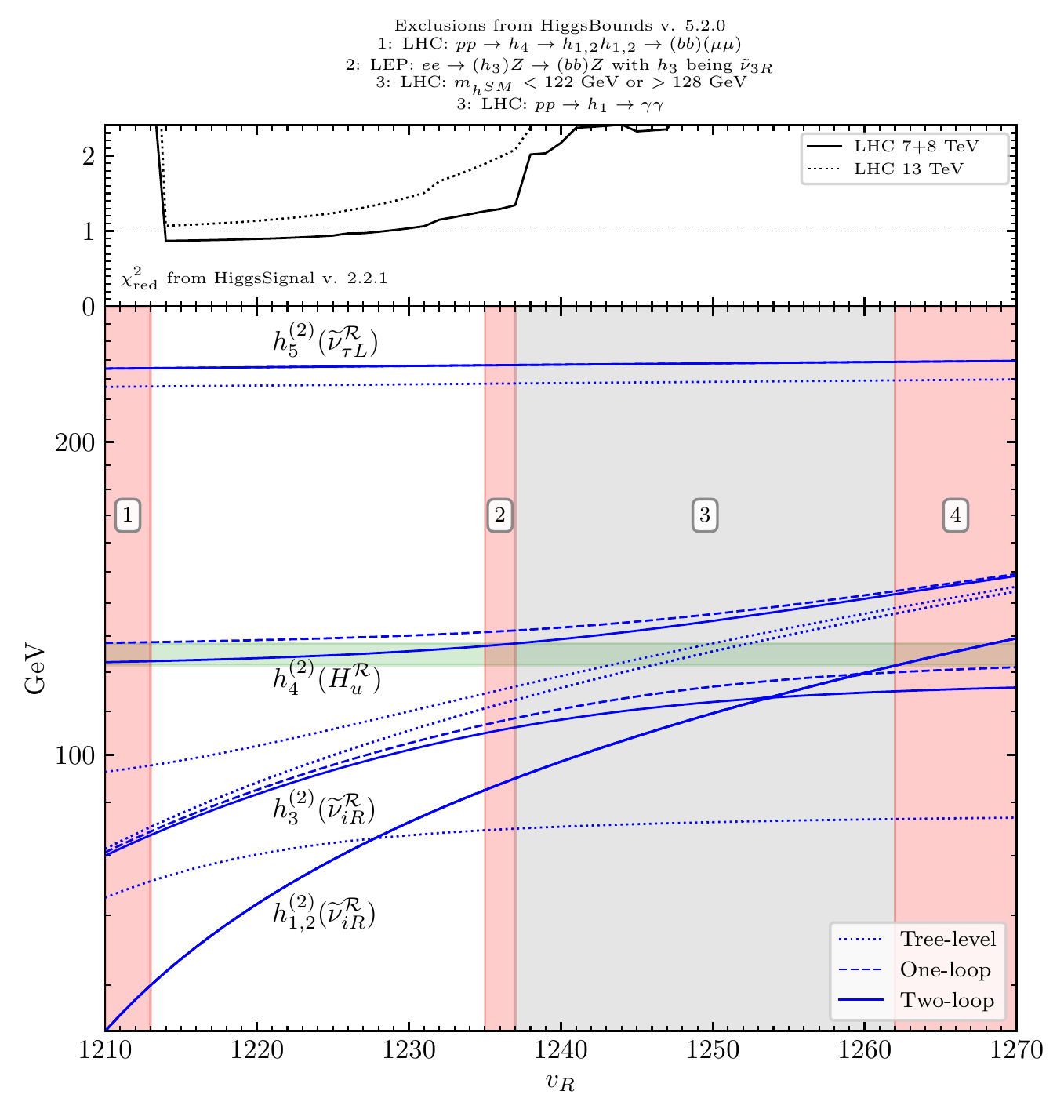}
  \caption{Light $\cp$-even scalar spectrum in the scan over $v_R$.
  Shown are the masses at tree level (\textit{dotted}),
  at the one-loop level (\textit{dashed}) and at the
  partial two-loop level (\textit{solid}).
  We show in the brackets the dominant composition
  of the loop-corrected mass eigenstates $h^{(2)}$
  in the experimentally allowed region of $v_R$.
  The desired SM-like Higgs-boson mass is indicated
  with the horizontal green
  band, assuming a theory uncertainty of $3\gev$.
  The red regions are excluded by
  direct searches for additional scalars.
  In the gray region the SM-like Higgs-boson mass is not predicted
  accurately. On top we show
  $\chi_{\rm red}^2$ for various Higgs-boson signal
  strength measurements at LHC.}
  \label{fig:scanvr}
\end{figure}

In \reffi{fig:scanvr} we show the resulting light $\cp$-even scalar spectrum.
In the experimentally allowed region
($1213\gev \leq v_R \leq 1235\gev$) the lightest
mass eigenstates are two almost degenerate right-handed sneutrinos.
The third right-handed sneutrino is roughly $20\gev$ heavier, and it acquires
substantial mixing with the SM-like Higgs boson.
Naturally, the $\widetilde{\nu}_{i R}^{\mathcal{R}}$
increase their masses when $v_R$
becomes larger, but also the SM-like Higgs boson mass increases, because
the mixing with the $\widetilde{\nu}_{3 R}^{\mathcal{R}}$
gives additional contributions.

The scenario is excluded experimentally for very small values of $v_R$,
because the two lightest mass eigenstates $h_{1,2}$ become lighter than
half the mass of the SM-like Higgs boson $h_4$, so the decays of $h_4$ into
$h_{1,2}$ opens up. Experimental searches for the decay of the SM-like
Higgs boson into two lighter scalars that subsequently decay into two $b$-jets
and a pair of $\mu$-leptons~\cite{Khachatryan:2017mnf}
exclude region 1 in \reffi{fig:scanvr}. These additional decay channels
of the SM-like Higgs boson are also the reason why the $\chi^2_{\mathrm{red}}$
rapidly increases in region 1, because it suppresses ordinary
SM-like decays of $h_4$.

When $v_R$ increases above $1235\gev$ further constrains from direct
searches for additional Higgs bosons
and measurement of the properties of the SM-like Higgs boson
become relevant. $\chi_{\mathrm{red}}^2$ quickly increases above 2 at
$v_R \sim 1237\gev$. Already at $v_R \sim 1235\gev$ the scenario is excluded by
LEP searches~\cite{Barate:2003sz}.
Note, that in the red region 2 the mixing of
$\widetilde{\nu}_{3 R}^{\mathcal{R}}$ with the SM-like Higgs boson enlarges, while
$\widetilde{\nu}_{3 R}^{\mathcal{R}}$ is kinematically still in reach
of being produced at LEP via the
Higgsstrahlung-process. Consequently, the channel
$ee\rightarrow (h_3)Z \rightarrow (b\bar{b})Z$,
where $h_3$ is identified with
$\widetilde{\nu}_{3R}^{\mathcal{R}}$,
excludes this interval.
Interestingly, in the experimentally allowed region, where
the mass of $\widetilde{\nu}_{3 R}^{\mathcal{R}}$
is even smaller, LEP data cannot rule out this scenario.
The reasons for this is not only the smaller mixing of
$\widetilde{\nu}_{3 R}^{\mathcal{R}}$
with the SM-like Higgs boson, but also that in the mass range below $100\gev$
LEP saw a slight excess over the SM background
(see also \refse{sec:lep})~\cite{Barate:2003sz}.

Beyond region 2 the current scenario is
experimentally excluded by the measurement of the SM-like Higgs-boson
mass in the gray region 3 and by the LHC cross section measurement
of the process $pp \rightarrow h_1 \rightarrow \gamma\gamma$ in
region 4~\cite{CMS:ril}.
This is because in region 3 the cross-over point is reached, in which the
masses of the $\widetilde{\nu}_{i R}^{\mathcal{R}}$
become larger than the SM-like Higgs-boson
mass. Through the interference effects the SM-like Higgs-boson mass is pushed
to lower values beyond that point. In region 4 the mass eigenstate corresponding
to the SM-like Higgs boson is
the lightest one at just about $118\gev$.
Even though there are two scalars in the mass range of the experimentally
measured Higgs-boson mass, there is no contribution to any signal-strength
measurement at the LHC, reflected by the fact that the $\chi_{\rm red}^2$ is
huge in region 4. The reason is that these states correspond to the
practically singlet like right-handed neutrino states. The third right-handed sneutrino carrying the doublet admixture taken from the SM-like Higgs boson
has a mass of over $140\gev$. Hence, it also does not contribute to signal-strength
measurements of the SM-like Higgs boson.

On a side note we briefly discuss the remaining light
scalar $h_5$ in \reffi{fig:scanvr},
which is the left-handed $\tau$-sneutrino at roughly $235$--$240\gev$.
%Similar to the scenario before (see \reffi{fig:scanlam}),
The fit to the
neutrino oscillation data generated a hierarchy between the
vevs of the left-handed sneutrinos, with $v_{3L}$ being the largest.
As a result, since dominant tree-level contributions to the
$\widetilde{\nu}_{iL}^{\mathcal{R}}$-masses scale with inverse
of $v_{iL}$ (see \refeq{eq:nuLapprx}),
the $\widetilde{\nu}_{\tau L}^{\mathcal{R}}$
is the lightest $\cp$-even left-handed sneutrino.
%In this case it is the LSP with a phenomenology very genuine to the \mnSSM.
It is rather invisible to usual searches for additional Higgs bosons at
colliders, but a dedicated analysis of LHC data was proposed to search for
light left-handed
sneutrinos in the framework of the \mnSSM~\cite{Lara:2018rwv}.
However, the analysis in \citere{Lara:2018rwv} concentrated on
$\tau$-sneutrinos as the LSP, whereas here there are even lighter
SUSY particles in the spectrum. For a detailed discussion of distinct
signatures at the LHC related to left-handed sneutrinos within the \mnSSM\
we refer to the literature~\cite{Ghosh:2017yeh,Lara:2018rwv,Lara:2018zvf}.
%and thereby gaining sensitivity by making use
%of the larger $\tau$-Yukawa coupling compared to the $\mu$-Yukawa coupling.

\subsection{Scan over $\lambda_i$ while $\pmb{\lambda}^2
=\lambda_i\lambda_i=\text{const.}$}
\label{sec:lamsqcons}
As already explained at the beginning of this section,
there is no theoretical
reason to choose the \mnSSM-like parameters universal w.r.t.\ the
family index. While the tree-level upper bound on the lightest $\cp$-even
scalar mass approximately depends on the term $\pmb{\lambda}^2
=\lambda_i\lambda_i$, and not on the individual $\lambda_i$, this
is not the case for the SM-like Higgs-boson mass,
as soon as mixing-effects induced by
the right-handed sneutrinos are considered. This effects can enter at
tree level, or via radiative corrections proportional to $\lambda_i$.
These radiative corrections depend on the masses and the mixing
of each of the right-handed sneutrino. Since it is not the
case that all three $\widetilde{\nu}_{iR}$ are degenerate, the radiative
corrections are expected to depend strongly on the individual values of
$\lambda_i$. Also, when $\pmb{\lambda}^2$ is fixed, the $\mu$-term which
is dynamically generated after EWSB and linearly dependent on $\lambda_i$
cannot be constant when the
$\lambda_i$ are varied. This can be another source of corrections
to the SM-like Higgs-boson mass that explicitly depend on the individual
values of the $\lambda_i$.

However, the loop corrections proportional to $\lambda_i$
are an order of magnitude smaller than the ones stemming from the
(s)top-sector,
partially because quantum contributions to the SM-like Higgs-boson mass
at the one-loop level proportional to $\lambda_i$ depend on the singlet-admixture
of the SM-like Higgs boson which, in turn, cannot be too large to not spoil
the measured signal strengths at the LHC.
Nevertheless, we will give here a rough idea of how large
the remnant effect of non-universal $\lambda_i$ on the SM-like
Higgs-boson mass can be while $\pmb{\lambda}^2$ is kept constant.
We performed a parameter scan over all possible
values of the $\lambda_i$ in a scenario in which
$\widetilde{\nu}_{\mu R}^{\mathcal{R}}$
has a mass between $92$ and $115\gev$ and mixes substantially
with the SM-like Higgs boson.
The free parameters were set to the values shown in \refta{tab:lamsqscan}.

\begin{table}
\centering
\renewcommand{\arraystretch}{1.3}
\begin{tabular}{c c c c c c c c}
 $\tan\beta$ & $\pmb{\lambda}^2$ & $\kappa$ & $v_{1,3R}$ & $v_{2R}$ & 
   $A^\lambda$ & $A^\kappa$ & $A^\nu$ \\
 \hline
 $5$ & $3\cdot 0.168^2$ & $0.5$ & $1000$ & $765$ &
   $1000$ & $-1000$ & $-1000$  \\
 \hline
 \hline
 $A^u_3$ & $A^u_{1,2}$ & $A^{(d,e)}$ &
   $m_{\widetilde{Q},\widetilde{u},\widetilde{d}}$ &
   $m_{\widetilde{e}}$ & $M_3$ & & \\
 \hline
 $-2000$ & $-1500$ & $-1500$ & $1500$ & $200$ & $2700$ \\
 \hline
 \hline
 $v_{1L}/10^{-4}$ & $v_{2L}/10^{-4}$ & $v_{3L}/10^{-4}$ & $Y^\nu_{11}/10^{-7}$ &
   $Y^\nu_{22}/10^{-7}$ & $Y^\nu_{33}/10^{-7}$ &
   $M_1$ & $M_2$ \\
 \hline
 $1.497$ & $6.179$ & $4.946$ & $4.388$ & $1.759$ & $6.258$ &
   $1228$ & $2814$
\end{tabular}
\caption{The same as in \refta{tab:scanlam} for the scan
over $\lambda_i$ while $\pmb{\lambda}^2=\text{const}$.}
\label{tab:lamsqscan}
\renewcommand{\arraystretch}{1.0}
\end{table}

\begin{figure}
  \centering
  \includegraphics[height=8cm,keepaspectratio]{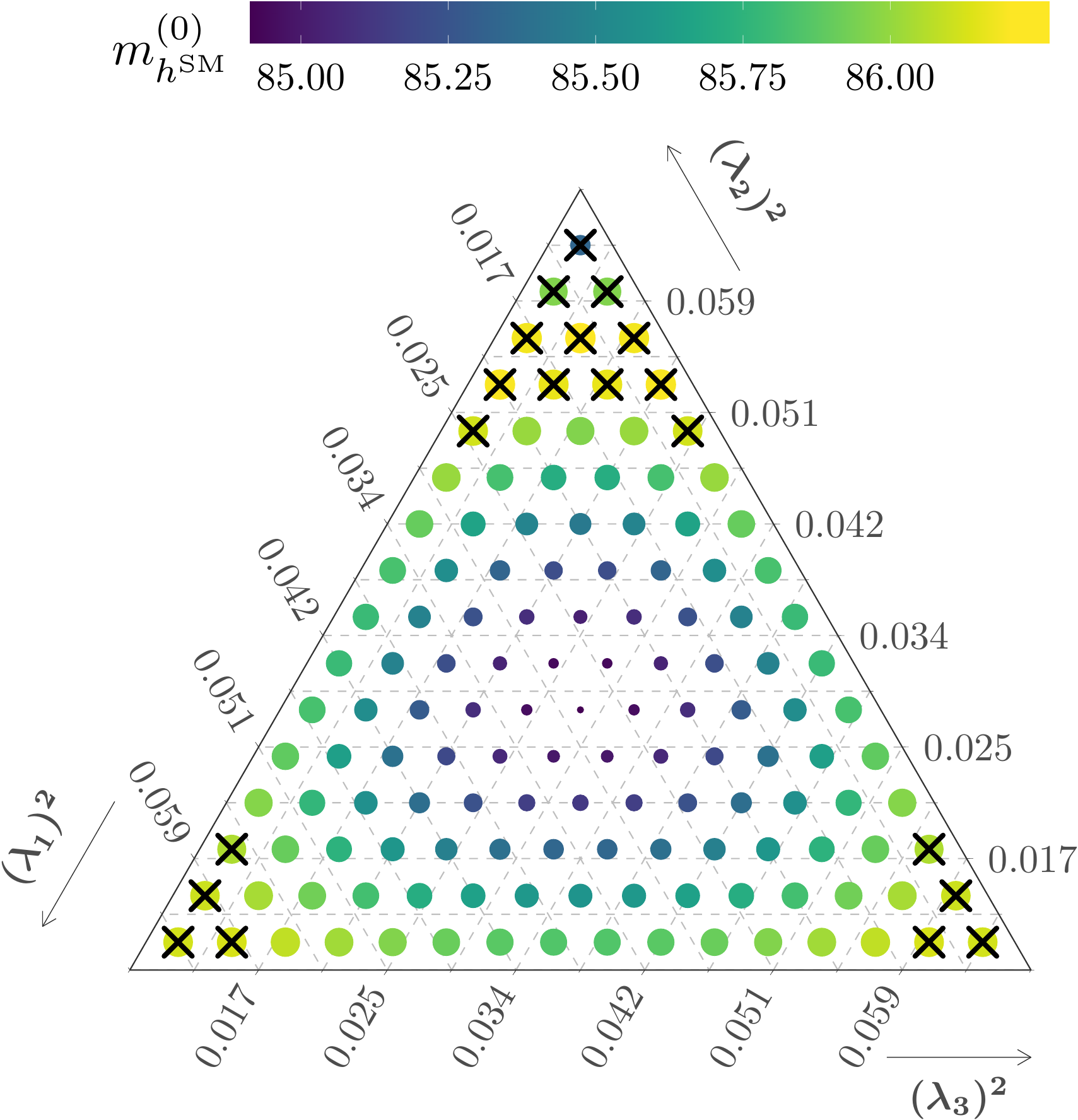}~
  \includegraphics[height=8cm,keepaspectratio]{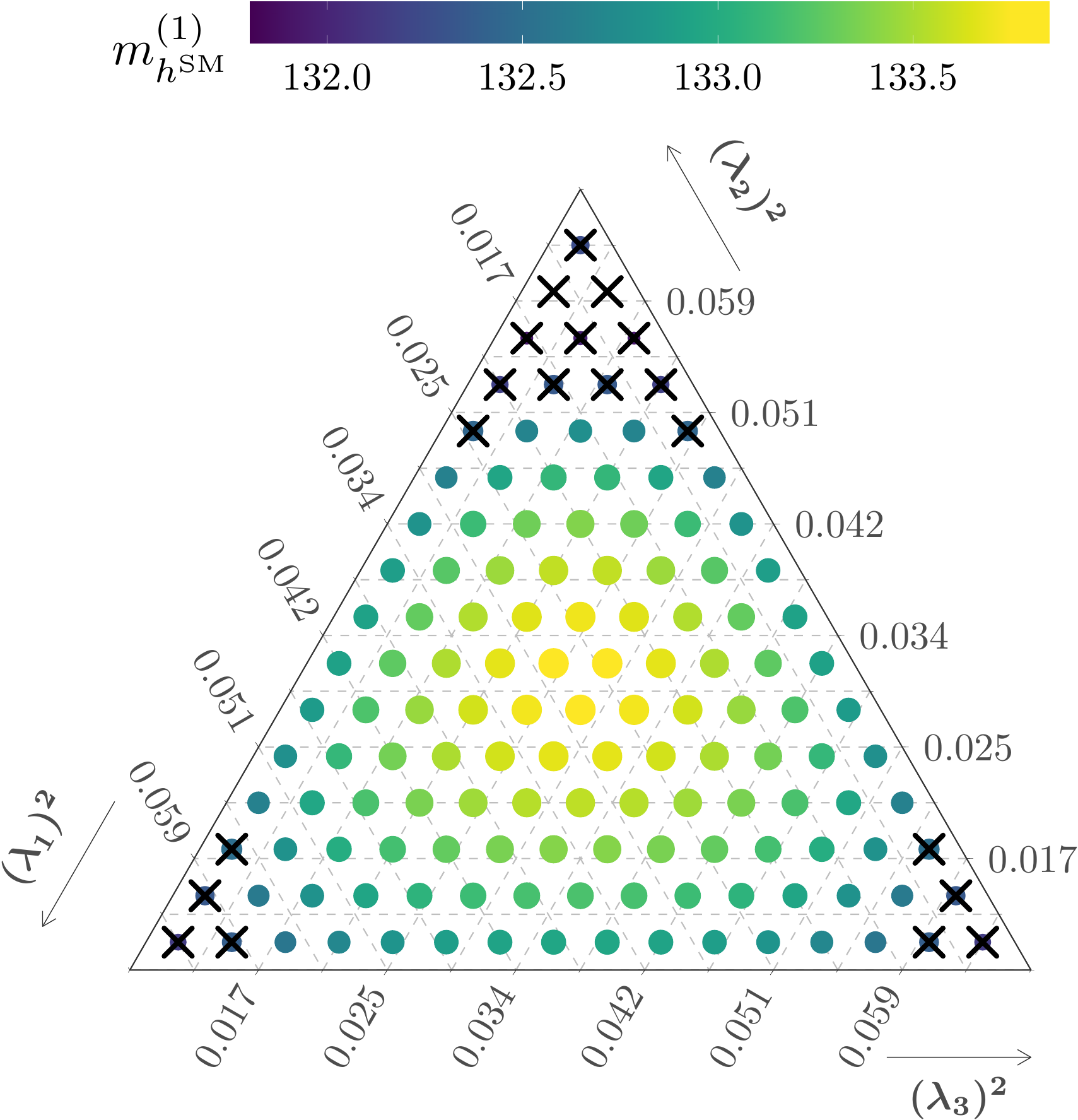} \\[2em]
  \includegraphics[height=8cm,keepaspectratio]{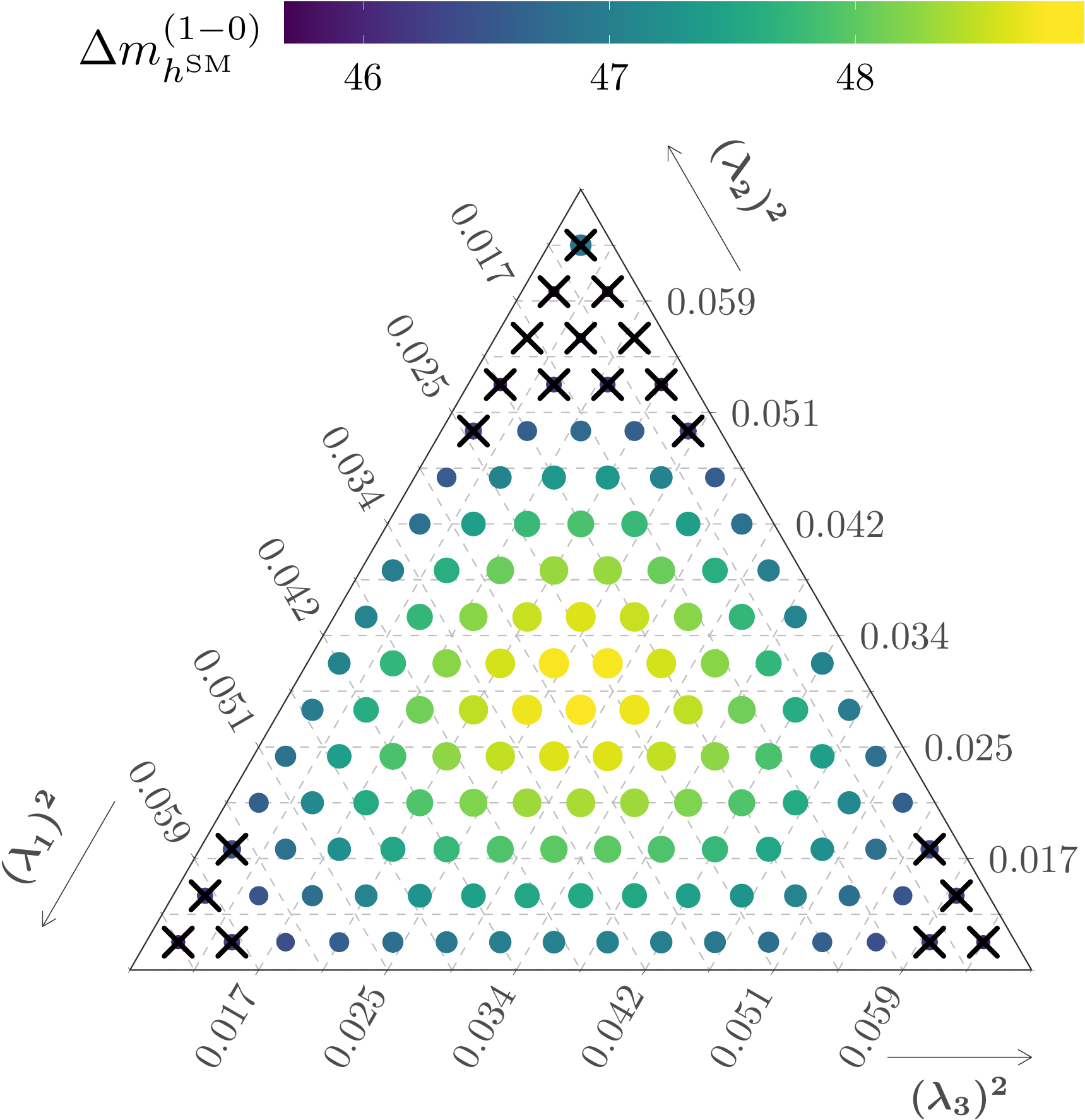}~
  \includegraphics[height=8cm,keepaspectratio]{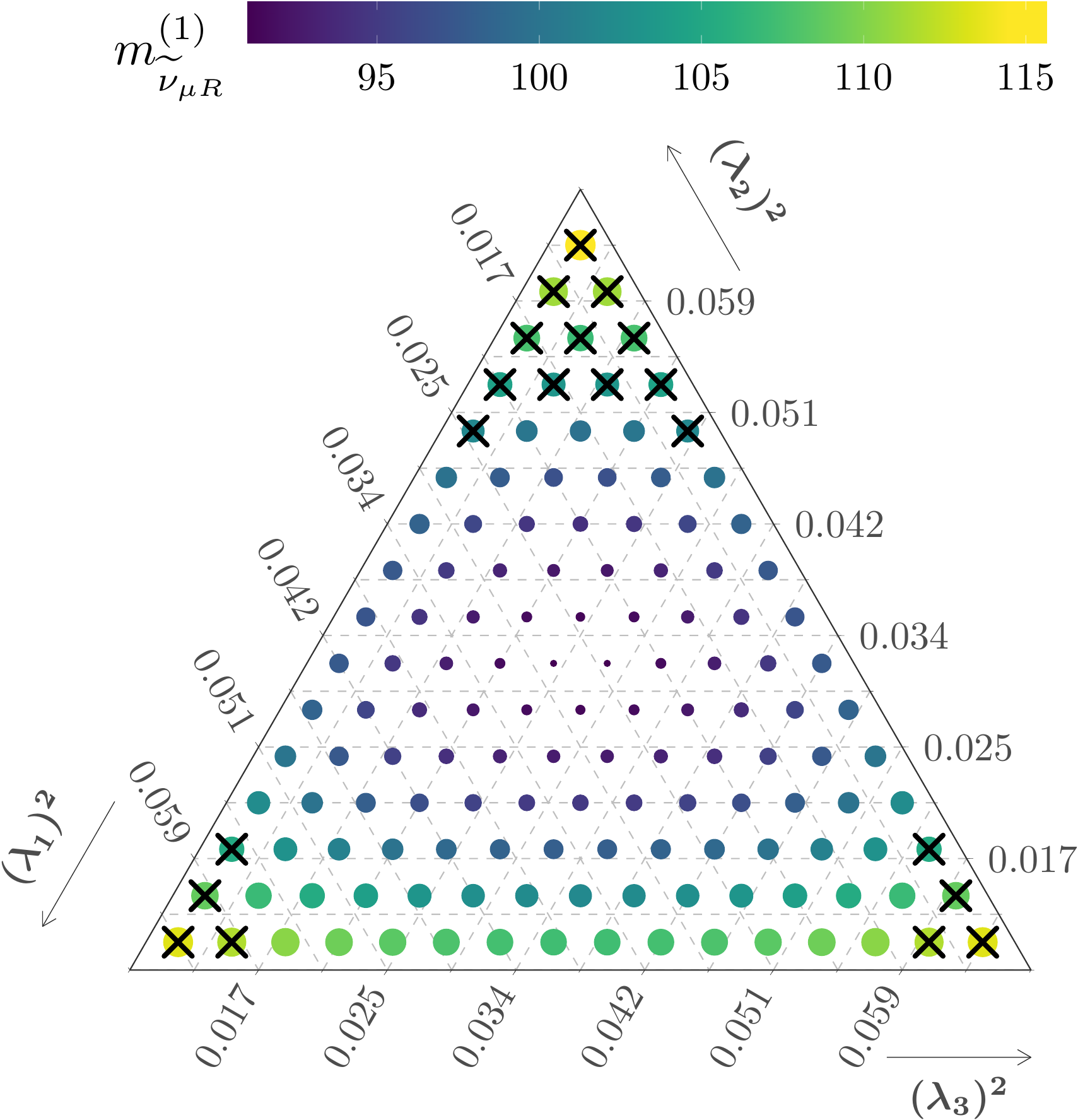}
    \caption{SM-like Higgs-boson mass
    at tree level
    $m_{h^{\mathrm{SM}}}^{(0)}=m_{h_1^{(0)}}$
    (\textit{top left}),
    and at one-loop level
    $m_{h^{\mathrm{SM}}}^{(1)}=m_{h_2^{(1)}}$
    (\textit{top right}), and their difference
    $\Delta m_{h^{\mathrm{SM}}}^{(1-0)} = m_{h^{\mathrm{SM}}}^{(1)}
    - m_{h^{\mathrm{SM}}}^{(0)}$ (\textit{bottom left}),
    and the one-loop mass of $\widetilde{\nu}_{\mu R}$
    (\textit{bottom right})
    for fixed $\pmb{\lambda}^2$ but varying
    $\lambda_i$, indicated by the colors (and sizes
    for better visibility). Crossed points
    are excluded by either \texttt{HiggsBounds},
    \texttt{HiggsSignals} with $\chi^2_{\rm red}\geq 150/101=1.485$ for the
    $13\tev$ data set, or because the
    SM-like Higgs boson mass including the two-loop corrections
    is smaller than 123 GeV or larger than 127 GeV.}
  \label{fig:lamsqscan}
\end{figure}

On can see that the scenario is very similar to the one in \refse{sec:scanlam}.
When the $\lambda_i$ are chosen uniformly we recover the BP
at $\lambda = 0.168$ in \reffi{fig:scanlam} which lies in the middle of the
experimentally allowed parameter region.
In \reffi{fig:lamsqscan} we illustrate the dependence of
the SM-like Higgs-boson mass on the individual values of $\lambda_i$.
We show triangle plots~\cite{ggtern}
with the values of $\lambda_i^2$ on the axes
and with their sum $\pmb{\lambda}^2=\text{const}$.
The colors of the points indicate the mass of the
SM-like Higgs boson at tree level (top left), at one-loop level (top right),
the difference of the SM-like Higgs-boson mass
at tree level and one-loop level (bottom left),
and the one-loop mass of the right-handed $\mu$-sneutrino (bottom right)
which is the lightest $\cp$-even scalar in this scenario.
We do not show the two-loop
mass of the SM-like Higgs boson, because the two-loop corrections
supplemented from \FH\ are purely MSSM-like corrections independent of
$\lambda_i$, thus not playing a role in the following discussion.
However, the parameters are chosen such that
the corrections beyond the one-loop level
shift the SM-like Higgs-boson mass into the
vicinity of $\sim 125\gev$ (see below).
In the upper right plot one sees that the one-loop mass of the SM-like
Higgs boson is the largest in the central point in which all $\lambda_i$
are equal. The tree-level mass, on the other hand, shows the opposite
behavior and is the largest in the corners of the upper left plot in which
one of the $\lambda_i$ is practically zero. The one-loop mass varies in
the experimentally allowed region by more than $1\gev$.
This demonstrates that for an accurate
prediction of the SM-like Higgs-boson mass it is
crucial to include the independent contributions of all three
$\lambda_i$ to the radiative corrections, when mixing effects between
the right-handed sneutrinos and the SM-like Higgs boson are sizable.

Note that the variation of the one-loop mass would be even larger if
we neglect the experimental constrains. In this scenario, the main exclusion
limit is the requirement to have the SM-like Higgs-boson
mass above $123\gev$ at two-loop.
In the corners of the plots, the mass of $\widetilde{\nu}_{\mu R}^{\mathcal{R}}$,
shown in the lower right plot, increases to
values very close to the SM-like Higgs-boson mass. This increases the mixing
between both scalars which, in turn, reduces the radiative one-loop
corrections to the SM-like Higgs boson. Practically speaking, parts of the
loop-corrections ``are lost'' to $\widetilde{\nu}_{\mu R}^{\mathcal{R}}$. This is why the
difference between the tree-level and the one-loop mass of the SM-like Higgs boson,
shown in the lower left plot of \reffi{fig:lamsqscan},
is the smallest when the mass of $\widetilde{\nu}_{\mu R}^{\mathcal{R}}$ is the largest.

We also emphasize that the variation of the difference between tree-level
and one-loop mass, as a measure for the size of genuine one-loop corrections,
is more than twice as
large as the variation of the one-loop mass of
the SM-like Higgs boson. One can see a compensation
of lower tree-level mass, but larger loop corrections in the center of the plots,
leading to a more stable one-loop mass of the SM-like Higgs boson.
This is due to the fact that the loop-corrected mass eigenstate
of the SM-like Higgs boson is the physical state, whose properties
are constrained by the experimental measurements.
Therefore, the singlet-admixture cannot be very large at loop-level, and the parameter
dependence of the SM-like Higgs-boson mass induced by this admixture
necessarily cannot be too large.
The tree-level state, on the other hand, is not physical and can have larger
mixing with $\widetilde{\nu}_{\mu R}^{\mathcal{R}}$, leading to a stronger dependence
on parameters related to the sneutrinos, like $\lambda_i$ in this case.
Thus, even though the overall dependence of the SM-like Higgs-boson
mass on parameters beyond the MSSM can be very large at tree level,
the dependence will be diminished
at loop-level in the parameter space that
fulfills experimental constraints on
the SM-like Higgs-boson properties.

\subsection{CMS and LEP excess at $\sim96\gev$}
\label{sec:lep}

Searches for the SM Higgs boson at LEP can nowadays be used to
constrain the parameter space of models with additional scalar
particles with masses below the SM-like Higgs-boson mass.
Interestingly, an excess over the SM background was observed
at a mass of $95$-$98\gev$
in the Higgsstrahlung production channel with an associated
decay of the Higgs boson to a pair of $b$-quarks~\cite{Barate:2003sz}.
Assuming that this excess can be explained by an additional
Higgs boson, the signal strength,
i.e, the cross section times branching ratio to $b$-quarks
normalized to the SM prediction for a SM-like Higgs boson at the same mass,
was extracted in \citere{Cao:2016uwt} to be
\begin{equation}
 \mu_{\mathrm{LEP}} = 0.117 \pm 0.057 \; .
\end{equation}
At about the same mass, CMS observed an excess over the SM background
in the $pp$ production with associated decay
to diphotons in the $8\tev$ and the $13\tev$
data~\cite{CMS:2015ocq,Sirunyan:2018aui}.
For the CMS excess, the signal strength was reported in
\citeres{Shotkin:2017,Sirunyan:2018aui} to be
\begin{equation}
 \mu_{\mathrm{CMS}} = 0.6 \pm 0.2 \; .
\end{equation}

In a previous publication, we
showed that it is possible to accommodate both excesses
simultaneously at the $1\sigma$ level in the $\mnSSM$ with just
one right-handed neutrino~\cite{Biekotter:2017xmf}.
We described how a right-handed sneutrino
at $\sim 96\gev$ can acquire substantial couplings to SM particles
via its mixing with the SM-like Higgs boson.\footnote{Similar solutions
were published in
supersymmetric~\cite{Cao:2016uwt,Domingo:2018uim,Choi:2019yrv} and
non-supersymmetric~\cite{Haisch:2017gql,Liu:2018xsw,Biekotter:2019kde,Biekotter:2019mib}
models with extended Higgs sectors.
See \citeres{Heinemeyer:2018jcd,Heinemeyer:2018wzl} for a review.}
However, we did not include an accurate prediction of the properties
of the neutrinos in our analysis in the \mnSSM\ with one generation
of right-handed neutrinos, because in that case at least one
neutrino mass has to be generated via quantum corrections.
On the contrary, in the \mnSSM\ with three generations of right-handed neutrinos,
we can describe the mass differences and mixings of
the neutrinos at tree level, which is of course much more feasible.

\begin{table}
\centering
\renewcommand{\arraystretch}{1.3}
\begin{tabular}{c c c c c c c c c c c}
 $\tan\beta$ & $\lambda_{1,2}$ & $\lambda_3$ & $\kappa_{111,222}$ &
   $\kappa_{333}$ & $v_{1,2R}$ & $v_{3R}$ & 
     $A^\lambda_{1,2}$ & $A^\lambda_{3}$ & \\ 
 \hline
 $1.945$ & $0.01$ & $[0.538,0.542]$ & $0.3$ & $0.05$ & $1200$ &
   $[884,888]$ & $1000$ & $[806,814]$ \\
 \hline
 \hline
 $A^\nu_{ii}$ & $A^\kappa_{111,222}$ & $A^\kappa_{333}$ & $A^u_3$ &
   $A^{u}_{1,2}$ & $A^{d,e}$ & 
     $m_{\widetilde{Q},\widetilde{u},\widetilde{d},\widetilde{e}}$ &
       $M_1$ & $M_2$ & $M_3$\\
 \hline
 $-1000$ & $-300$ & $[-124,-100]$ & $-650$ & $0$ & $0$ & $1000$ &
   $400$ & $800$ & $2700$ \\
\end{tabular}
\caption{Parameters of the scan to fit the LEP and the CMS excesses.
Dimensionful parameters are given in GeV. If the family index is
omitted the parameter has a universal value independent of the index.}
%The parameters in the last row are fitted to neutrino oscillation data.}
\label{tab:lep}
\renewcommand{\arraystretch}{1.0}
\end{table}

The values of the free parameters to fit the excesses are shown
in \refta{tab:lep}. The scalar used to fit the excesses is the
right-handed $\tau$-sneutrino $\widetilde{\nu}_{\tau R}^{\mathcal{R}}$.
This is assured by setting $\kappa_{333} \ll \kappa_{111,222}$, so
that $\widetilde{\nu}_{\tau R}^{\mathcal{R}}$ has a smaller mass than the other
two right-handed sneutrinos. Then, $v_{3R}$ and $A^\kappa_{333}$
are used to tune the mass of $\widetilde{\nu}_{\tau R}^{\mathcal{R}}$
to be at $\sim 95$--$98\gev$. A sufficiently
large mixing of $\widetilde{\nu}_{\tau R}^{\mathcal{R}}$
with the SM-like Higgs boson is achieved with a large value
of $\lambda_3 \sim 0.54$, while $\lambda_{1,2}=0.01$ are very small
to avoid that the effective $\mu$-term becomes very large.
Alternatively, one could have used smaller values for $v_{1,2 R}$, but then
the other two $\cp$-even sneutrinos
$\widetilde{\nu}_{e,\mu \; R}^{\mathcal{R}}$ would have been very light as well,
potentially carrying away some of the mixing between the SM-like
Higgs boson and the right-handed sneutrinos.\footnote{A scenario in which
several right-handed sneutrinos give rise to the observed excesses
is beyond the scope of our paper.}
Since $\lambda_3$ is large, the SM-like Higgs boson receives additional
contribution to the tree-level mass. This is why $\tan\beta$ is set to
a small value, and, besides $A^u_3 = -650\gev$, the soft trilinear
parameters $A^{u,d,e}$ can be set to zero.
Apart from that, $\tan\beta$ shall not be much larger than one to not
suppress the coupling of $\widetilde{\nu}_{\tau R}^{\mathcal{R}}$ to $t$-quarks,
which scales with the inverse of $\sin\beta$ (see \refeq{eq:effcpl}).
Finally, we choose a range for $A^\lambda_3$ in which the mixing
between $\widetilde{\nu}_{\tau R}^{\mathcal{R}}$ and the SM-like Higgs boson
is sufficiently large.
We will show results for small ranges of the parameters
$\lambda_3$, $v_{3R}$, $A^\lambda_3$ and $A^\kappa_{333}$.
While $v_{3R}$ and $A^\kappa_{333}$ are mainly correlated to
the mass of $\widetilde{\nu}_{\tau R}^{\mathcal{R}}$, $\lambda_3$ and
$A^\lambda_3$ affect the doublet composition of $\widetilde{\nu}_{3 R}$.
This certainly is not an exhaustive parameter scan covering
the complete parameter space, but the scan gives an idea of how
the excesses can be accommodated within the \mnSSM, and
it resembles the solution we found in the \mnSSM\ with just
one right-handed neutrino~\cite{Biekotter:2017xmf}.

\begin{figure}
  \centering
  \includegraphics[width=0.8\textwidth]{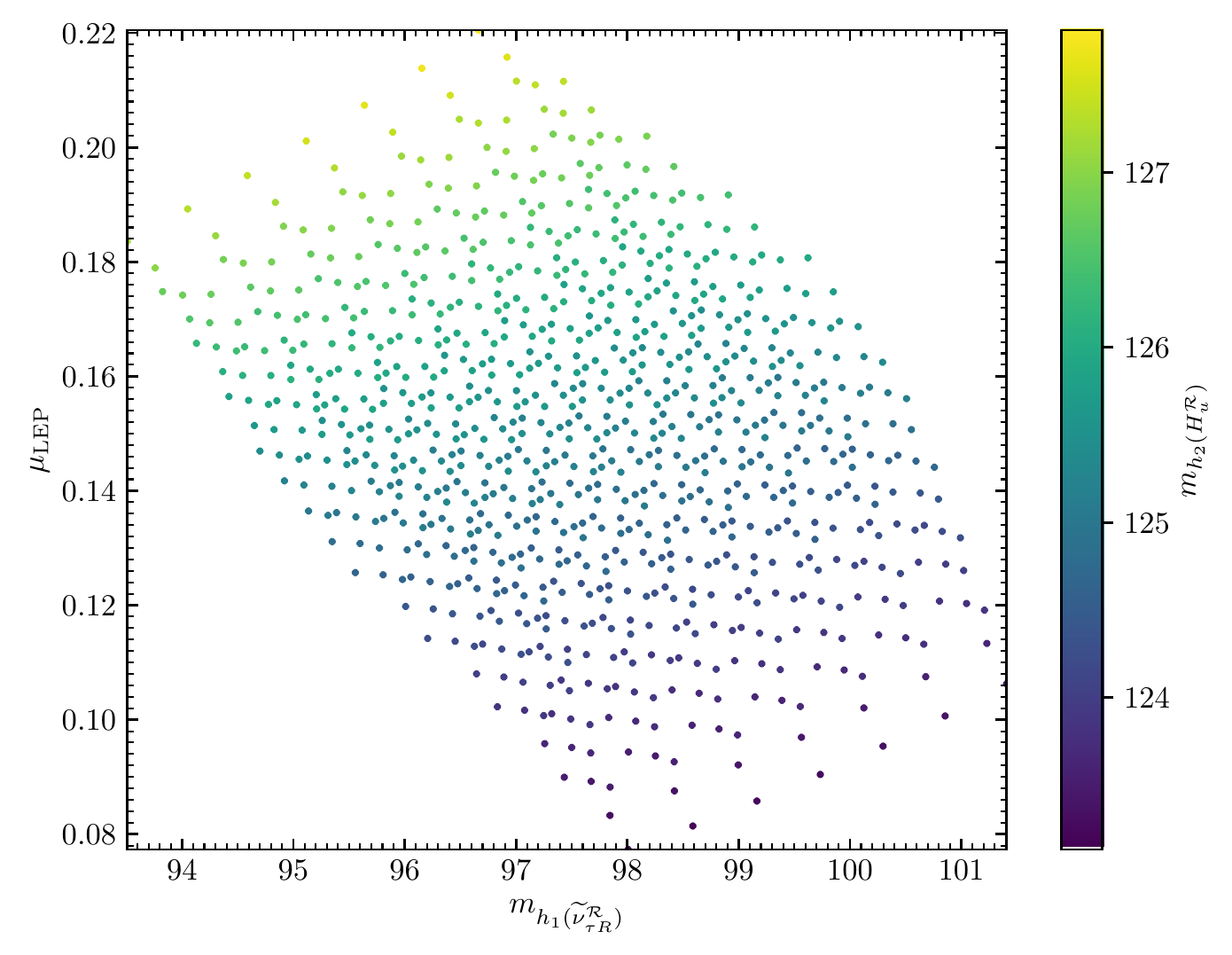}
  \\
  \includegraphics[width=0.8\textwidth]{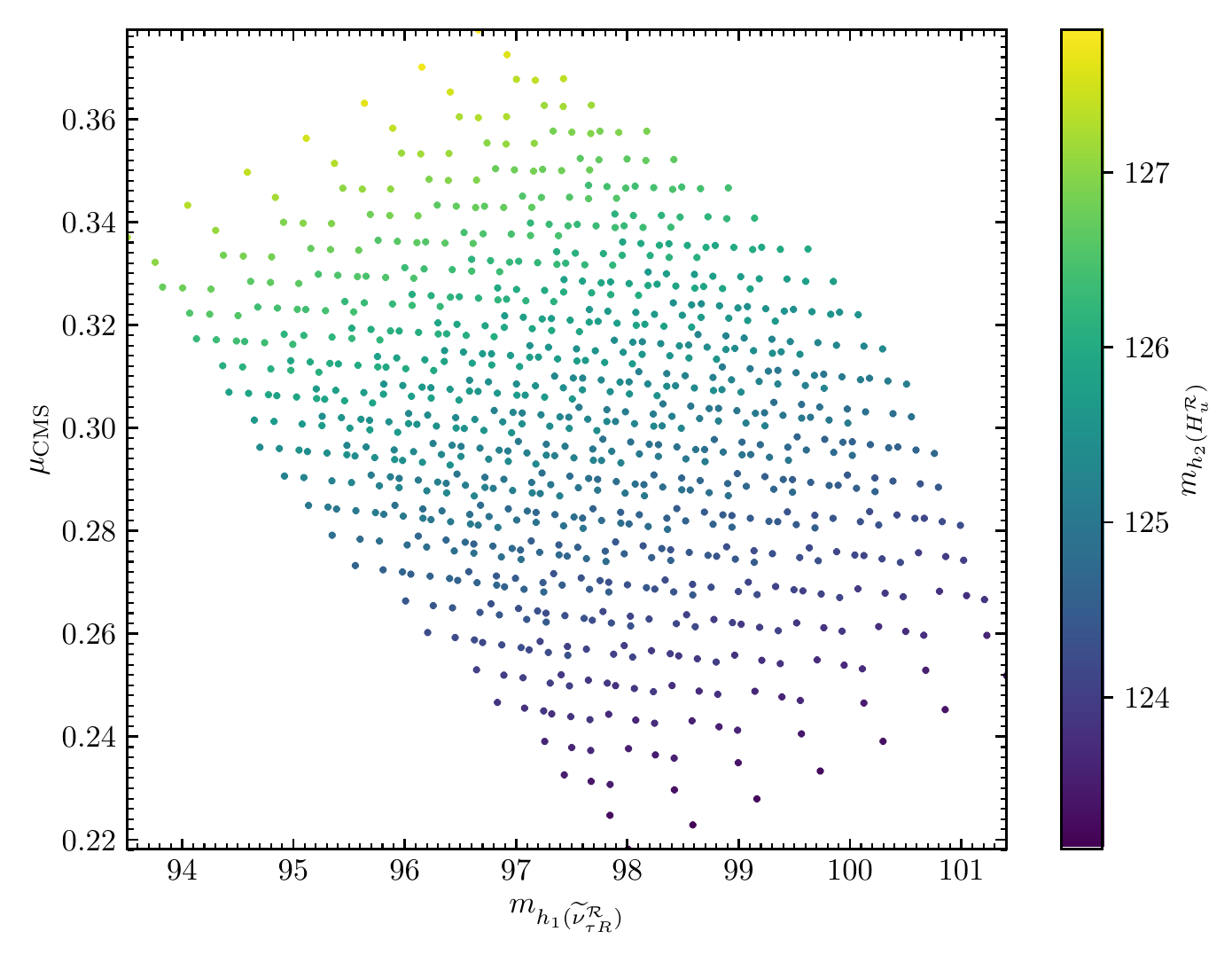}
  \caption{Values for $\mu_{\mathrm{LEP}}$ \textit{(top)}
  and for $\mu_ {\mathrm{CMS}}$ \textit{(bottom)} 
  for each parameter point versus the mass of $\widetilde{\nu}_{\tau R}^{\mathcal{R}}$.
  The colors indicate the mass of the SM-like Higgs boson.}
  \label{fig:lepcms}
\end{figure}

In \reffi{fig:lepcms} we show the results for the
signal strength of the LEP excess $\mu_{\mathrm{LEP}}$
(top) and of the CMS excess $\mu_{\mathrm{CMS}}$ (bottom).
In both plots the colors of the points indicate the
SM-like Higgs-boson mass, while the mass of
$\widetilde{\nu}_{\tau R}^{\mathcal{R}}$
is shown on the horizontal axis.
The signal strengths were calculated in the narrow-width
approximation, and the branching ratio and cross section
ratios w.r.t.\ the SM where calculated using the
effective coupling approximation as explained
in \citere{Biekotter:2017xmf}.
One can immediately see that it is rather easy to
achieve the experimental value of $\mu_{\mathrm{LEP}}$, whereas
the largest values for $\mu_{\mathrm{CMS}}$ reached in our scan
are just below the lower limit of $\mu_{\mathrm{CMS}}=0.4$,
i.e., $1\sigma$ below the central value.
This is due to the
fact that the main decay channel of $\widetilde{\nu}_{\tau R}^{\mathcal{R}}$
is the decay to a pair of bottom quarks, and it is
harder to achieve a substantial branching ratio to
diphotons required for the CMS excess.
Nevertheless, both excesses are fitted at the $1\sigma$ level
considering the experimental uncertainties, while
fitting the neutrino data and
being in agreement with the experimental constraints
on the SM-like Higgs boson, which we again checked
with \texttt{HiggsSignals} assuming a theoretical uncertainty
of the SM-like Higgs-boson mass of $3\gev$.

\begin{figure}
  \centering
  \includegraphics[width=0.8\textwidth]{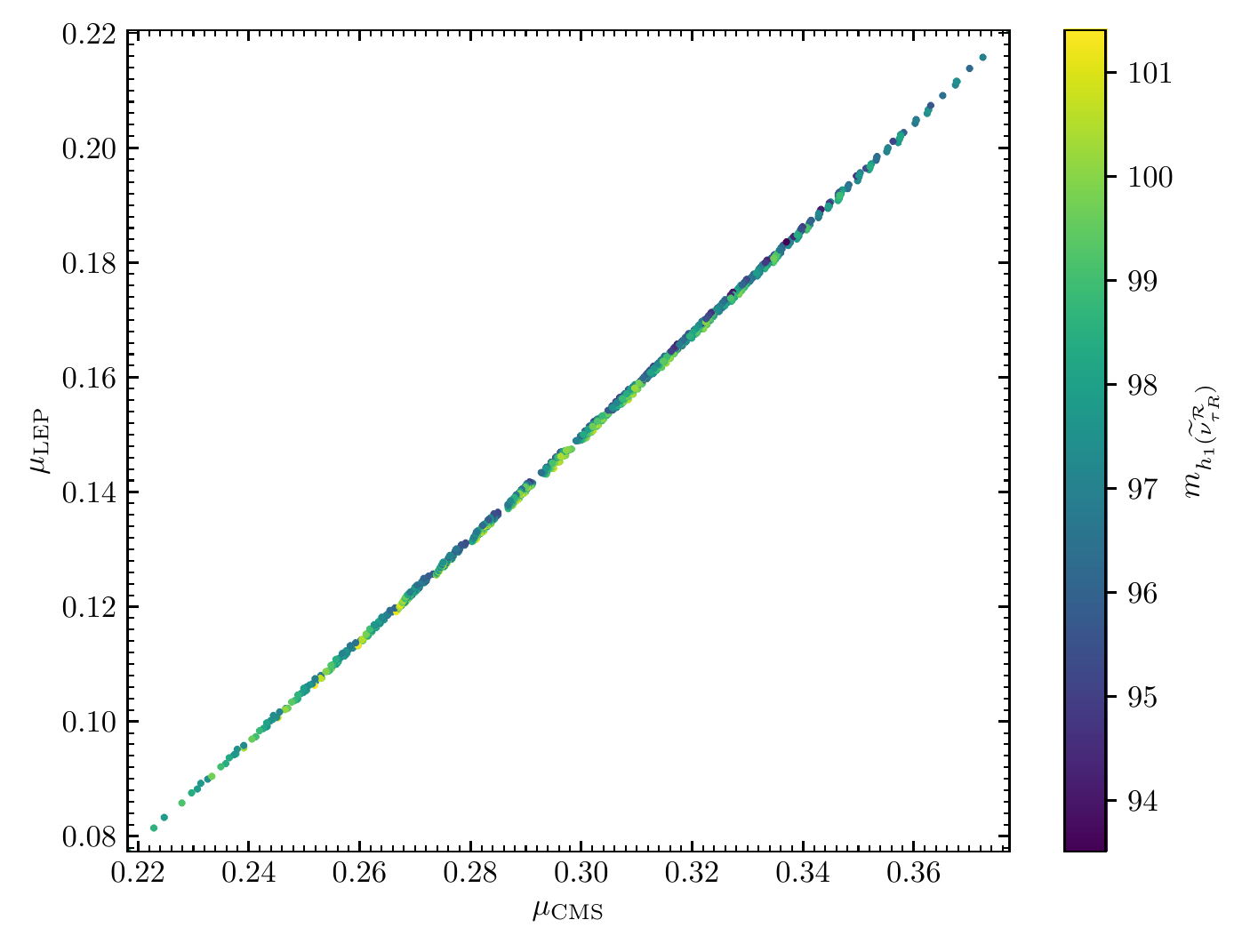}
  %~
  %\includegraphics[width=0.48\textwidth]{mumumh2.pdf}
  \caption{Correlation of both signal strengths,
  with the colors encoding the mass of $\widetilde{\nu}_{\tau R}^{\mathcal{R}}$.}
  %\textit{(left)} and the SM-like Higgs boson mass \textit{(right)}.}
  \label{fig:lepmh1mh2}
\end{figure}

In \reffi{fig:lepmh1mh2} we show the correlation of
both signal strengths, with the colors encoding the
mass of $\widetilde{\nu}_{\tau R}^{\mathcal{R}}$.
%on the left and
%the SM-like Higgs boson mass on the right.
The strong correlation one can see has its origin in the
fact that both signal strengths increase with the amount
of doublet-component of $\widetilde{\nu}_{\tau R}^{\mathcal{R}}$.
In principle, one could achieve a further enhancement of $\mu_{\mathrm{CMS}}$
and a suppression of $\mu_{\mathrm{LEP}}$
by suppressing the down-type doublet component of $\widetilde{\nu}_{\tau R}^{\mathcal{R}}$.
Then, the branching ratio to bottom quarks becomes smaller,
and the diphoton branching ratio increases because of the
smaller total decay width due to the reduction of the decay
width to bottom quarks.
However, finding such points is difficult,
because the dominant terms mixing the right-handed sneutrinos
$\widetilde{\nu}_{i R}^{\mathcal{R}}$
with the doublet fields $H_d^{\mathcal{R}}$ and
$H_u^{\mathcal{R}}$ scale equally with
$\lambda_i$, $A^\lambda_i$, $\kappa_{ijk}$ and $v_{iR}$
at tree level, as can be seen in \refeqs{eq:mixdr} and (\ref{eq:mixur}).
The only difference are the factors $v_d$ and $v_u$
in each equation, respectively,
which cannot be exploited too much, because, as mentioned before,
$\tan\beta$ should not be too far from one.
This is why an extensive scan of the
vast parameter space of the \mnSSM\ would be necessary
to find parameter points in which $\mu_{\mathrm{CMS}}$ is further
enhanced without increasing $\mu_{\mathrm{LEP}}$ too much,
which however lies beyond the scope of this paper.

Instead, we will focus on the rest of the spectrum, which heavily
depends on the values of the neutrino Yukawa couplings $Y^\nu_{ij}$
and the vevs of the left-handed sneutrinos $v_{iL}$, once
the remaining parameters are fixed to the values listed in \refta{tab:lep}.
As an example, we show in \refta{tab:lepneu} two possible sets of
parameters that accommodate accurate neutrino masses and mixings in
the parameter scan of this section.
In contrast to the other scenarios we presented before,
here we will make use of non-zero values of the non-diagonal elements
of $Y^\nu_{ij}$ in one of the BPs.
Naturally, this simplifies the accommodation of neutrino
properties in agreement with experimental data, because there
are six more free parameters that can be adjusted.
The price to pay is that there is usually more than one
set of parameters of $Y^\nu_{ij}$ and $v_{iL}$ that give accurate
predictions for the neutrino sector.
The reason for showing two distinct scenarios is that
in the \mnSSM\ the scalar
sector is deeply related to the neutrino sector. Thus, different sets
of parameters predict fundamentally different scalar spectra,
and since there is no theoretical argument that the neutrino Yukawa
couplings have to be diagonal, we used the additional freedom to
present a point in which, on top of the explanation of the LEP and the
CMS excesses, there are a several other light particles possibly in
reach of future colliders.

\begin{table}
\centering
\renewcommand{\arraystretch}{1.3}
\begin{tabular}{c|c c c c c c}
& $Y^\nu_{11}$ & $Y^\nu_{12}$ & $Y^\nu_{13}$ & $Y^\nu_{21}$ & $Y^\nu_{22}$ &
   $Y^\nu_{23}$ \\
\hline
\textbf{BP1} & $8.109\cdot 10^{-8}$ & $0$ & $0$ &
   $0$ & $1.154\cdot 10^{-7}$ & $0$ \\
\textbf{BP2} & $7.088\cdot 10^{-8}$ & $1.181\cdot 10^{-8}$ & $-3.404\cdot 10^{-9}$ &
  $1.902\cdot 10^{-8}$ & $1.238\cdot 10^{-7}$ & $1.783\cdot 10^{-8}$ \\
\hline
\hline
& $Y^\nu_{31}$ & $Y^\nu_{32}$ & $Y^\nu_{33}$ & $v_{1L}$ & $v_{2L}$ & $v_{3L}$ \\
\hline
\textbf{BP1} & $0$ & $0$ & $8.855\cdot 10^{-7}$ &
   $1.890\cdot 10^{-5}$ & $2.601\cdot 10^{-4}$ & $1.871\cdot 10^{-4}$ \\
\textbf{BP2} & $-2.103\cdot 10^{-9}$ & $6.923\cdot 10^{-9}$ & $1.383\cdot 10^{-8}$ &
  $1.792\cdot 10^{-6}$ & $2.072\cdot 10^{-4}$ & $3.673\cdot 10^{-4}$ \\
\hline
\hline
\multirow{2}{*}{\textbf{BP1}} & $m_{\lambda_4} (\nu_{\tau R})$ &
  $m_{A_1} (\widetilde{\nu}_{\tau R}^{\mathcal{I}})$ &
   $m_{H^+_1} (\widetilde{\mu}_{L})$ &
     $m_{A_2} (\widetilde{\nu}_{\mu L}^{\mathcal{I}})$ &
       $m_{h_3} (\widetilde{\nu}_{\mu L}^{\mathcal{R}})$ &
         $m_{\lambda_5} (\widetilde{H}_{d,u})$ \\
\cline{2-7}
& 78 & 97-109 & 283 & 285 & 285 & $323-326$ \\
\hline
\hline
\multirow{2}{*}{\textbf{BP2}} & $m_{\lambda_4} (\nu_{\tau R})$ &
  $m_{h_1} (\widetilde{\nu}_{\tau L}^{\mathcal{R}})$ &
    $m_{A_1} (\widetilde{\nu}_{\tau L}^{\mathcal{I}})$ &
      $m_{H^+_1} (\widetilde{\tau}_L)$ &
        $m_{A_2} (\widetilde{\nu}_{\tau R}^{\mathcal{I}})$ &
          $m_{\lambda_5} (\widetilde{H}_{d,u})$ \\
\cline{2-7}
& 78 & 79 & 79 & 98 & 97-109 & $323-326$
\end{tabular}
\caption{Parameter sets BP1 and BP2
used to fit the neutrino oscillation data accurately
in the scan reproducing the LEP and the CMS excesses.
In the last four rows we list the masses of the
six lightest non-SM particles (in addition to
$\widetilde{\nu}_{\tau R}^{\mathcal{R}}$ at $\sim 96\gev$)
for each BP.
Dimensionful parameters are given in GeV.}
\label{tab:lepneu}
\renewcommand{\arraystretch}{1.0}
\end{table}

In both BPs the lightest BSM particle
is the right-handed
$\tau$-neutrino. This is because $\kappa_{333}$ has to be small to
decrease the mass of the corresponding sneutrino $\widetilde{\nu}_{\tau R}^{\mathcal{R}}$.
Consequently, also the Majorana mass term for the neutrino will be small and
$m_{\nu_{\tau R}}$ is small.
However, its mass is still above half the SM-like Higgs-boson mass, so
the decay of the SM-like Higgs boson into $\nu_{\tau R}$ is forbidden.
The striking difference between both BPs is the mass
scale of the left-handed $\tau$-sneutrinos and sleptons.
In BP2 the Yukawa coupling $Y^\nu_{33}$ is the smallest diagonal
element of $Y^\nu$, and the vev corresponding to the third
family of left-handed sneutrinos $v_{3L}$ is the largest of
the three.
This reduces the masses of the left-handed $\cp$-even and $\cp$-odd
$\tau$-sneutrino (see \refeq{eq:nuLapprx})
and the $\tau$-slepton to values below the
SM-like Higgs-boson mass.
In BP1, on the other hand, $v_{2L}$ is the largest of the
three left-handed vevs, and therefore the left-handed $\mu$-sneutrinos
and $\mu$-slepton are the lightest left-handed sfermions,
although still more than twice
as heavy as the SM-like Higgs boson.
In this way, the phenomenology of both BPs
is distinct, even though the properties of the
right-handed $\tau$-sneutrino, which is the particle used to fit the
excesses, are not affected by the left-handed sector. This is because
its branching ratios are dominantly given by the mixing-effects with
the SM-like Higgs boson, which is not suppressed by the small
neutrino Yukawa couplings $Y^\nu_{ij}$.

Nonetheless, BP2 can give rise to additional interesting signal at
colliders. A dedicated analysis of the collider phenomenology of
light left-handed $\tau$-sneutrinos/sleptons
at the LHC can be found in
\citeres{Ghosh:2017yeh,Lara:2018rwv},
where it was shown
that there are no direct bounds from
LEP/LHC searches that can be used to
set lower limits on the masses of
these particles
in the framework
of the \mnSSM.
This analysis made use of the fact that the charged
$\widetilde{\tau}_L$
can be produced, and thus
provide a source for the left-handed sneutrinos,
since it decays into lighter
$\widetilde{\nu}_{\tau L}^{\mathcal{R}}$ or
$\widetilde{\nu}_{\tau L}^{\mathcal{I}}$.
An important feature
is that the subsequent decays of
$\widetilde{\nu}_{\tau L}^{\mathcal{R,I}}$
can be prompt or displaced.
However, in \citeres{Ghosh:2017yeh,Lara:2018rwv}
it is assumed that the left-handed sneutrino is the LSP.
This is not the case here, since the right-handed $\tau$-neutrino
is even lighter. Signals at colliders from the lightest BSM particle
$\nu_{\tau R}$ are not expected, because it is a gauge singlet,
thus cannot be produced directly.
In principle, it can be produced indirectly
via the decay of the sfermions.
However, the spectrum is very compressed, such that a
pair production of $\nu_{\tau R}$ from the decays of
$\widetilde{\nu}_{\tau R}$ at $\sim 96\gev$ or the SM-like
Higgs boson at $\sim 125\gev$ is kinematically forbidden,
and the production of a pair of a right- and a left-handed
$\tau$-neutrino is suppressed by the size of $Y^\nu_{33}$.

%%%%%%%%%%%%%%%%%%%%%%%%%%%%%%%%%%%%%%%%%%%%%%%%%%%%%%%%%%%%%%%%%%%%%%%%%%%%%%%
%%%%%%%%%%%%%%%%%%%%%%%%%%%%%%%%%%%%%%%%%%%%%%%%%%%%%%%%%%%%%%%%%%%%%%%%%%%%%%%

\section{Conclusion and Outlook}
\label{sec:concl}
% The \mnSSM\ is a simple SUSY extension
% of the SM that is capable of describing neutrino physics in agreement with 
% experimental data. As in other SUSY models,
% higher-order corrections are crucial to reach a
% theoretical uncertainty at the same level of the (anticipated) experimental
% accuracy. So far, higher-order corrections in the \mnSSM\ had been restricted
% to \DRbar\ calculations, which suffer from the disadvantage that they cannot
% be directly connected to (possibly future observed) new BSM particles.
The \mnSSM\ is a simple SUSY extension of the SM that is capable of describing
neutrino physics in agreement with experimental data. As in other SUSY models,
higher-order corrections are crucial to reach a theoretical uncertainty at the
same level of the (anticipated) experimental accuracy. So far, higher-order
corrections in the \mnSSM\ had been restricted to \DRbar\ calculations, which
suffer from the disadvantage that they cannot be directly connected to
(possibly future observed) new BSM particles. More recently we had evaluated
the corrections to the neutral scalar masses and mixings, but restricting
ourselves to one generation of heavy neutrinos~\cite{Biekotter:2017xmf}. 

% In this paper we have performed the complete one-loop renormalization of the
% neutral scalar sector of the \mnSSM\ with three generation of right-handed
% neutrinos in a mixed on-shell/\DRbar\ scheme. The renormalization procedure
% was discussed in detail for each of the free parameters appearing in the
% \mnSSM\ Higgs sector.
In this paper we have presented the complete one-loop renormalization of the
neutral scalar sector of the \mnSSM\ with three generation of right-handed
neutrinos in a mixed on-shell/\DRbar\ scheme. In this way, for the first time,
it is possible to evaluate the masses and mixings in the neutral scalar sector
with high precision, while simultaneously describe correctly the experimental
neutrino data, such as mass differences and mixing angles.
An on-shell (OS) renormalization has been chosen for parameters that can be
directly identified with (potentially) observable states, whereas a
\DRbar\ renormalization has been chosen for the remaining parameters. 
We provide details on the full set renormalization parameters, which were
implemented into a \FA\ model file (that can be provided by the authors upon
request). 

We have performed the calculation of the masses of the neutral scalars in the
\mnSSM\ with three generations of heavy neutrinos, taking into account the
full set of one-loop corrections as described above. These corrections have
been supplemented by contributions to the neutral Higgs-boson sector of the
MSSM at and beyond the two-loop level as provided by \FH. These corrections are
crucial to obtain a reliable prediction of the mass of an SM-like Higgs boson
around $\sim 125 \gev$.

The masses of the neutral scalar bosons have been evaluated in a set of
scenarios that exemplify the relevant dependences on the underlying
\mnSSM\ parameters. The scenarios are in agreement with all available searches
for additional Higgs bosons (via the code \texttt{HiggsBounds}),
as well with the properties of
the SM-like Higgs boson as measured at the LHC (via the code
\texttt{HiggsSignals}), while at
the same time reproducing correctly the measured values of neutrino mass
differences and mixing angles. In a first scenario we varied the assumed to be
universal parameter $\la$. We find that large one-loop
corrections for the right-handed sneutrinos arise, that are
crucial to accurately account for possible mixing effects between
them and the SM-like Higgs boson.
In a second scenario we have varied the assumed to be universal vev of the
right-handed sneutrinos. Here we find that all three $\cp$-even
right-handed sneutrinos can have masses below $\sim 125\gev$,
without being excluded by cross-section limits from direct searches
for additional Higgs bosons.
In our third scenario we deviate from the intergenerational universality
assumptions. We take $\pmb{\la}^2 := \la_i \la_i$
to be constant, but vary instead the
individual $\la_i$. We find that the non-universality of the $\la_i$ has an
important impact on the predictions of the neutral scalar masses. It has been
shown that in the experimentally allowed parameter space the
non-universality of the $\la_i$ can account for deviations
of the SM-like Higgs-boson mass of $\sim 1 \gev$, emphasizing
the importance to consider the full set of \mnSSM\ parameters
at the one-loop level.

As a final example we have discussed how the \mnSSM\ can describe two excesses
in the searches for light Higgs bosons in the vicinity of $\sim 96 \gev$.
These are a $\sim 3\, \sigma$ excess in the diphoton final state as reported by
CMS and a $\sim 2\, \sigma $ excess in the $b\bar b$ final state as published by
LEP. We demonstrated that the \mnSSM\ can account at the $\sim 1\,\sigma$ level
for both excesses, while being in agreement with all available Higgs-boson
searches and measurements, as well as the available neutrino data. We are
eagerly awaiting updated experimental analyses from ATLAS and CMS to confirm
or refute these excesses.

Further explorations of the scalar sector of the \mnSSM\ are necessary to
cover the wealth of (possible) phenomenology that this model offers. This
includes further studies dropping the (artificial) intergenerational
universality assumptions, where in this paper we have taken only the first
step. It furthermore includes the evaluation of the charged scalar sector at
the one-loop level and beyond, which is expected to be important for the
phenomenology at the LHC. Also studies going beyond the LHC searches may
become relevant, such as analyses of the possibility of a first order phase
transition, leading to gravitational waves created in the early universe. We
leave these studies for future work.

\subsection*{Acknowledgements}
We thank Daniel E. Lopez-Fogliani for helpful discussions
concerning the neutrino physics.
This work was supported in part by the Spanish Agencia Estatal de
Investigaci\'on through the grants
FPA2016-78022-P MINECO/FEDER-UE (TB and SH),
FPA2015-65929-P MINECO/FEDER-UE
and PGC2018-095161-B-I00 (CM),
by the ``Spanish Red Consolider MultiDark'' FPA2017-90566-REDC,
and IFT Centro de
Excelencia Severo Ochoa SEV-2016-0597. The work of TB was
funded by Fundaci\'on La Caixa under `La Caixa-Severo Ochoa' international
predoctoral grant.

%%%%%%%%%%%%%%%%%%%%%%%%%%%%%%%%%%%%%%%%%%%%%%%%%%%%%%%%%%%%%%%%%%%%%%%%%%%%%%%
%%%%%%%%%%%%%%%%%%%%%%%%%%%%%%%%%%%%%%%%%%%%%%%%%%%%%%%%%%%%%%%%%%%%%%%%%%%%%%%

\appendix

\section{Mass matrices}

Here we give the entries of the
following scalar mass matrices.

\subsection{\protect\boldmath $\cp$-even scalars}
\label{app:cpeven}

In the interaction basis $\varphi^T=(H_d^{\mathcal{R}},H_u^{\mathcal{R}},
\widetilde{\nu}_{iR}^{\mathcal{R}},\widetilde{\nu}_{jL}^{\mathcal{R}})$
the mass matrix for the $\cp$-even scalars $m_{\varphi}^2$ is defined by:
\begin{align}
m_{H_{d}^{\mathcal{R}}H_{d}^{\mathcal{R}}}^{2}&=m_{H_d}^2+
\frac{1}{8}\left(g_1^2+g_2^2\right)\left( 3v_d^2+v_{iL}v_{iL}-v_u^2\right)+
\frac{1}{2}\left( v_u^2\lambda_i\lambda_i +
\left( v_{iR}\lambda_i \right)^2\right) \; , \\[4pt]
m_{H_{u}^{\mathcal{R}}H_{u}^{\mathcal{R}}}^{2}&=m_{H_u}^2+
\frac{1}{8}\left(g_1^2+g_2^2\right)\left(3v_u^2-v_d^2-v_{iL}v_{iL}\right)+
\frac{1}{2}\lambda_i\lambda_i v_d^2 - v_d Y^\nu_{ji}\lambda_i v_{jL} \notag \\
&+\frac{1}{2} \left( v_{iR}\lambda_i \right)^2
+\frac{1}{2}Y^\nu_{ji} Y^\nu_{ki} v_{jL} v_{kL}
+\frac{1}{2}Y^\nu_{ij} Y^\nu_{ik} v_{jR} v_{kR}
 \; , \\[4pt]
m_{H_{u}^{\mathcal{R}}H_{d}^{\mathcal{R}}}^{2}&=
-\frac{1}{4}\left(g_1^2+g_2^2\right)v_dv_u
+ v_d v_u \lambda_i \lambda_i
- \frac{1}{\sqrt{2}} T^\lambda_i v_{iR}
- v_u Y^\nu_{ji} \lambda_i v_{jL}
- \frac{1}{2} \kappa_{ijk} \lambda_i v_{jR} v_{kR} \; , \\[4pt]
\label{eq:mixdr}
m_{\widetilde{\nu}^{\mathcal{R}}_{iR} H_{d}^{\mathcal{R}}}^{2}&=
- v_u \kappa_{ijk} \lambda_j v_{kR}
- \frac{1}{2} v_{jR} \lambda_j v_{kL} Y^\nu_{ki}
+ v_d v_{jR} \lambda_j \lambda_i
- \frac{1}{2} v_{jL} Y^\nu_{jk} v_{kR} \lambda_i
- \frac{1}{\sqrt{2}} v_u T^\lambda_i \; , \\[4pt]
\label{eq:mixur}
m_{\widetilde{\nu}^{\mathcal{R}}_{iR} H_{u}^{\mathcal{R}}}^{2}&=
\frac{1}{\sqrt{2}} v_{jL} T^\nu_{ji}
+ v_u Y^\nu_{ji} Y^\nu_{jk} v_{kR}
- v_d \kappa_{ijk} \lambda_j v_{kR}
+ Y^\nu_{lj} \kappa_{ijk} v_{kR} v_{lL}
+ v_u v_{jR} \lambda_j \lambda_i
- \frac{1}{\sqrt{2}} v_d T^\lambda_i \; , \\[4pt]
m_{\widetilde{\nu}^{\mathcal{R}}_{iR} \widetilde{\nu}^{\mathcal{R}}_{jR}}^{2}&=
\left( m_{\widetilde{v}_R}^2 \right)_{ij}
+ \frac{1}{2} v_u^2 Y^\nu_{ki} Y^\nu_{kj}
- v_d v_u \kappa_{ijk} \lambda_k
+ \sqrt{2} v_{kR} T^\kappa_{ijk}
+ v_u Y^\nu_{lk} \kappa_{ijk} v_{lL}
+ \frac{1}{2} v_{kL} Y^\nu_{ki} v_{lL} Y^\nu_{lj} \notag \\
& + 2 \kappa_{ikl} \kappa_{jkm} v_{lR} v_{mR}
+ \kappa_{ijk} \kappa_{klm} v_{lR} v_{mR}
- \frac{1}{2} v_d v_{kL} Y^\nu_{kj} \lambda_i
- \frac{1}{2} v_d v_{kL} Y^\nu_{ki} \lambda_j \notag \\
& + \frac{1}{2} \left( v_d^2 + v_u^2 \right) \lambda_i \lambda_j \; , \\[4pt]
m_{\widetilde{\nu}_{iL}^{\mathcal{R}} H_{d}^{\mathcal{R}}}^{2}&=
\left(m_{H_d\widetilde{L}}^2\right)_i+
\frac{1}{4}\left(g_1^2+g_2^2\right)v_dv_{iL}
- \frac{1}{2} v_u^2 Y^\nu_{ij} \lambda_j
- \frac{1}{2} v_{jR} \lambda_j v_{kR} Y^\nu_{ik} \; , \\[4pt]
m_{\widetilde{\nu}_{iL}^{\mathcal{R}} H_{u}^{\mathcal{R}}}^{2}&=
-\frac{1}{4}\left(g_1^2+g_2^2\right)v_uv_{iL}
- v_d v_u Y^\nu_{ij} \lambda_j
+ \frac{1}{\sqrt{2}} v_{jR} T^\nu_{ij}
+ v_u Y^\nu_{ij} Y^\nu_{kj} v_{kL}
+ \frac{1}{2} Y^\nu_{ij} \kappa_{jkl} v_{kR} v_{lR} \; , \\[4pt]
m_{\widetilde{\nu}_{iR}^{\mathcal{R}} \widetilde{\nu}^{\mathcal{R}}_{jL}}^{2}&=
v_u Y^\nu_{jk} \kappa_{ikl} v_{lR}
+ \frac{1}{2} v_{kL} Y^\nu_{ki} v_{lR} Y^\nu_{jl}
- \frac{1}{2} v_d v_{kR} \lambda_k Y^\nu_{ji}
+ \frac{1}{2} v_{kL} Y^\nu_{kl} v_{lR} Y^\nu_{ji} \; , \notag \\
& - \frac{1}{2} v_d v_{kR} Y^\nu_{jk} \lambda_i
+ \frac{1}{\sqrt{2}} v_u T^\nu_{ji} \; , \\[4pt]
m_{\widetilde{\nu}_{iL}^{\mathcal{R}} \widetilde{\nu}_{jL}^{\mathcal{R}}}^{2}&=
\left(m_{\widetilde{L}}^2\right)_{ij}+
\frac{1}{8}\delta_{ij}\left(g_1^2+g_2^2\right)\left(v_d^2-v_u^2+v_{kL}v_{kL}\right)
+ \frac{1}{4}\left(g_1^2+g_2^2\right)v_{iL}v_{jL} \notag \\
& + \frac{1}{2} v_u^2 Y^\nu_{ik} Y^\nu_{jk}
+ \frac{1}{2} v_{kR} Y^\nu_{jk} v_{lR} Y^\nu_{il} \; .
\end{align}

%%%%%%%%%%%%%%%%%%%%%%%%%%%%%%%%%%%%%%%%%%%%%%%%%%%%%%%%%%%%%%%%%%%%%%%%%%%%%%%

\subsection{\protect\boldmath $\cp$-odd scalars}
\label{app:cpodd}

In the interaction basis $\sigma^T=(H_d^{\mathcal{I}},H_u^{\mathcal{I}},\widetilde{\nu}_{iR}^{\mathcal{I}},\widetilde{\nu}_{jL}^{\mathcal{I}})$ the mass matrix for the $\cp$-odd scalars $m_{\sigma}^2$ is defined by:
\begin{align}
m_{H_{d}^{\mathcal{I}}H_{d}^{\mathcal{I}}}^{2} &=
m_{H_d}^2+
\frac{1}{8}\left(g_1^2+g_2^2\right)\left(v_d^2+v_{iL}v_{iL}-v_u^2\right)
+ \frac{1}{2} \left( v_u^2 \lambda_i \lambda_i
+ \left( v_{iR} \lambda_i \right)^2 \right) \; , \\[4pt]
m_{H_{u}^{\mathcal{I}}H_{u}^{\mathcal{I}}}^{2}&=
m_{H_u}^2+
\frac{1}{8}\left(g_1^2+g_2^2\right)\left(v_u^2-v_d^2-v_{iL}v_{iL}\right)
+ \frac{1}{2} v_d^2 \lambda_i \lambda_i
- v_d Y^\nu_{ji} \lambda_i v_{jL}
+ \frac{1}{2} \left( v_{iR} \lambda_i \right)^2 \notag \\
& + \frac{1}{2} Y^\nu_{ji} Y^\nu_{ki} v_{jL} v_{kL}
+ \frac{1}{2} Y^\nu_{ij} Y^\nu_{ik} v_{jR} v_{kR} \; , \\[4pt]
m_{H_{u}^{\mathcal{I}}H_{d}^{\mathcal{I}}}^{2}&=
\frac{1}{2} \kappa_{ijk} \lambda_i v_{jR} v_{kR} +
\frac{1}{\sqrt{2}} v_{iR} T^\lambda_i \; , \\[4pt]
m_{\widetilde{{\nu}}^{\mathcal{I}}_{iR} H_{d}^{\mathcal{I}}}^{2}&=
v_u \kappa_{ijk} \lambda_j v_{kR}
+ \frac{1}{2} v_{jR} \lambda_j v_{kL} Y^\nu_{ki}
- \frac{1}{2} v_{jL} Y^\nu_{jk} v_{kR} \lambda_i
- \frac{1}{\sqrt{2}} v_u T^\lambda_i \; , \\[4pt]
m_{\widetilde{{\nu}}^{\mathcal{I}}_{iR} H_{u}^{\mathcal{I}}}^{2}&=
\frac{1}{\sqrt{2}} v_{jL} T^\nu_{ji}
+ v_d \kappa_{ijk} \lambda_j v_{kR}
- Y^\nu_{lj} \kappa_{ijk} v_{kR} v_{lL}
- \frac{1}{\sqrt{2}} v_d T^\lambda_i \; , \\[4pt]
m_{\widetilde{{\nu}}^{\mathcal{I}}_{iR} \widetilde{{\nu}}^{\mathcal{I}}_{jR}}^{2}&=
\left( m_{\widetilde{\nu}}^2 \right)_{ij}
+ \frac{1}{2} v_u^2 Y^\nu_{ki} Y^\nu_{kj}
+ v_d v_u \kappa_{ijk} \lambda_k
- \sqrt{2} v_{kR} T^\kappa_{ijk}
- v_u Y^\nu_{lk} \kappa_{ijk} v_{lL}
+ \frac{1}{2} v_{kL} Y^\nu_{ki} v_{lL} Y^\nu_{lj} \notag \\
& + \kappa_{ikm} \kappa_{jkl} v_{lR} v_{mR}
- \kappa_{ijk} \kappa_{klm} v_{lR} v_{mR}
- \frac{1}{2} v_d v_{kL} Y^\nu_{kj} \lambda_i
- \frac{1}{2} v_d v_{kL} Y^\nu_{ki} \lambda_j \notag \\
& + \frac{1}{2} \left( v_d^2 + v_u^2 \right) \lambda_i \lambda_j \; , \\[4pt]
m_{\widetilde{\nu}_{iL}^{\mathcal{I}} H_{d}^{\mathcal{I}}}^{2}&=
\left(m_{H_d\widetilde{L}}^2\right)_i
- \frac{1}{2} v_u^2 Y^\nu_{ij} \lambda_j
- \frac{1}{2} v_{jR} \lambda_j v_{kR} Y^\nu_{ik} \; , \\[4pt]
m_{\widetilde{\nu}_{iL}^{\mathcal{I}} H_{u}^{\mathcal{I}}}^{2}&=
- \frac{1}{\sqrt{2}} v_{jR} T^\nu_{ij}
- \frac{1}{2} Y^\nu_{ij} \kappa_{jkl} v_{kR} v_{lR} \; , \\[4pt]
m_{\widetilde{\nu}_{iR}^{\mathcal{I}} \widetilde{\nu}^{\mathcal{I}}_{jL}}^{2}&=
- v_u Y^\nu_{jk} \kappa_{ikl} v_{lR}
- \frac{1}{2} v_{kL} Y^\nu_{ki} v_{lR} Y^\nu_{jl}
- \frac{1}{2} v_d v_{kR}  \lambda_k Y^\nu_{ji}
+ \frac{1}{2} v_{kL} Y^\nu_{kl} v_{lR} Y^\nu_{ji} \notag \\
& + \frac{1}{2} v_d v_{kR} Y^\nu_{jk} \lambda_i
+ \frac{1}{\sqrt{2}} v_u T^\nu_{ji} \; , \\[4pt]
m_{\widetilde{\nu}_{iL}^{\mathcal{I}} \widetilde{\nu}_{jL}^{\mathcal{I}}}^{2}&=
\left(m_{\widetilde{L}}^2\right)_{ij}+
\frac{1}{8}\delta_{ij}\left(g_1^2+g_2^2\right)\left(v_d^2+v_{kL}v_{kL}-v_u^2\right)
+ \frac{1}{2} v_u^2 Y^\nu_{ik} Y^\nu_{jk}
+ \frac{1}{2} v_{kR} Y^\nu_{ik} v_{lR} Y^\nu_{jl} \; . \label{eq:mlinmps}
\end{align}

\subsection{Charged scalars}\label{app:charged}
In the gauge eigenstate basis $C^T=
({H^-_d}^*,{H^+_u},\widetilde{e}_{iL}^*,\widetilde{e}_{jR}^*)$ the entries
of $m_{H^+}^2$ are given by:
\begin{align}
m_{H_{d}^-{H^{-}_d}^{*}}^{2}&=m_{H_d}^2+
\frac{1}{8}g_1^2\left( v_d^2+v_{iL}v_{iL}-v_u^2\right)+
\frac{1}{8}g_2^2\left( v_d^2-v_{iL}v_{iL}+v_u^2\right)
+ \frac{1}{2} \left( v_{iR} \lambda_i \right)^2 \notag \\
& + \frac{1}{2} Y^e_{ij} Y^e_{ik} v_{jL} v_{kL} \; , \\[4pt]
m_{{H_{u}^+}^* {H_u^+}}^{2}&=m_{H_u}^2+
\frac{1}{8}g_1^2\left( v_u^2-v_d^2-v_{iL}v_{iL}\right)+
\frac{1}{8}g_2^2\left( v_u^2+v_d^2+v_{iL}v_{iL}\right)
+ \frac{1}{2} \left( v_{iR} \lambda_i \right)^2 \notag \\
& + \frac{1}{2} Y^\nu_{ij} Y^\nu_{ik} v_{jR} v_{kL} \; , \\[4pt]
m_{H_{d}^-{H^{+}_u}}^{2}&=
\frac{1}{4}g_2^2v_dv_u
- \frac{1}{2} v_d v_u \lambda_i \lambda_i
+ \frac{1}{\sqrt{2}} v_{iR} T^\lambda_i
+ \frac{1}{2} v_u Y^\nu_{ji} \lambda_i v_{jL}
+ \frac{1}{2} \kappa_{ijk} \lambda_i v_{jR} v_{kR} \; , \\[4pt]
m_{{{H^{-}_d}} {\widetilde{e}_{iL}}^*}^2&=
\left(m_{H_d\widetilde{L}}^2\right)_i+
\frac{1}{4}g_2^2v_dv_{iL}-
\frac{1}{2}v_d Y^e_{ji}Y^e_{jk}v_{kL}
- \frac{1}{2} v_{jR} Y^\nu_{ij} v_{kR} \lambda_k \; , \\[4pt]
m_{{H_u^+}^* {\widetilde{e}_{iL}}^*}^2&=
\frac{1}{4}g_2^2v_u v_{iL}
+ \frac{1}{2} v_d v_u Y^\nu_{ij} \lambda_j
- \frac{1}{\sqrt{2}} v_{jR} T^\nu_{ij}
- \frac{1}{2} v_u Y^\nu_{ij} Y^\nu_{kj} v_{kL}
- \frac{1}{2} Y^\nu_{ij} \kappa_{jkl} v_{kR} v_{lR} \; , \\[4pt]
m_{{{H^{-}_d}} {\widetilde{e}_{iR}}^*}^2&=
-\frac{1}{\sqrt{2}}v_{jL}T^e_{ij}
- \frac{1}{2} v_u Y^e_{ij} Y^\nu_{jk} v_{jR} \; , \\[4pt]
m_{{H_u^+}^* {\widetilde{e}_{iR}}^*}^2&=
- \frac{1}{2} v_d Y^e_{ij} Y^\nu_{jk} v_{kR}
- \frac{1}{2} v_{jR} \lambda_j v_{kL} Y^e_{ik} \; , \\[4pt]
m_{\widetilde{e}_{iL} \widetilde{e}_{jL}^{*}}^{2}&=
\left(m_{\widetilde{L}}^2\right)_{ij}+
\frac{1}{8}\delta_{ij}\left(g_1^2-g_2^2\right)
\left( v_d^2-v_u^2+v_{kL}v_{kL}\right)+
\frac{1}{4}g_2^2v_{iL}v_{jL}+
\frac{1}{2}v_d^2 Y^e_{ki}Y^e_{kj} \notag \\
&+\frac{1}{2} v_{kR} Y^\nu_{jk} v_{lR} Y^\nu_{il} \; , \label{eq:mlinmch} \\[4pt]
m_{\widetilde{e}_{iR} \widetilde{e}_{jR}^{*}}^{2}&=
\left(m_{\widetilde{e}}^2\right)_{ij}+
\frac{1}{4}\delta_{ij}g_1^2\left( v_u^2-v_d^2-v_{kL}v_{kL}\right)+
\frac{1}{2}v_d^2Y^e_{ik}Y^e_{jk}+
\frac{1}{2}v_{kL}Y^e_{ik}v_{lL}Y^e_{jl} \; , \\[4pt]
m_{\widetilde{e}_{iL} \widetilde{e}_{jR}^{*}}^{2}&=
\frac{1}{\sqrt{2}}v_d T^e_{ij}-
\frac{1}{2} v_u v_{kR} \lambda_k Y^e_{ij} \; .
\end{align}

%%%%%%%%%%%%%%%%%%%%%%%%%%%%%%%%%%%%%%%%%%%%%%%%%%%%%%%%%%%%%%%%%%%%%%%%%%%%%%%

%%%%%%%%%%%%%%%%%%%%%%%%%%%%%%%%%%%%%%%%%%%%%%%%%%%%%%%%%%%%%%%%%%%%%%%%%%%%%%%
%%%%%%%%%%%%%%%%%%%%%%%%%%%%%%%%%%%%%%%%%%%%%%%%%%%%%%%%%%%%%%%%%%%%%%%%%%%%%%%

\section{Explicit expressions for counterterms}\label{app:counters}

In this section we will state the one-loop counterterms that were calculated diagrammatically in the $\DRbar$ scheme and checked against master formulas for the one-loop beta functions and anomalous dimensions of soft SUSY
breaking parameters~\cite{Martin:1993zk,Yamada:1994id,Luo:2002ti}, superpotential 
parameters~\cite{Machacek:1983fi,Luo:2002ti}, vacuum expectation
values~\cite{Sperling:2013eva} and wave-functions with kinetic
mixing~\cite{Machacek:1983tz,Fonseca:2013bua}. The master formulas were evaluated using the mathematica package \texttt{SARAH}~\cite{Staub:2010jh}.

%%%%%%%%%%%%%%%%%%%%%%%%%%%%%%%%%%%%%%%%%%%%%%%%%%%%%%%%%%%%%%%%%%%%%%%%%%%%%%%

\subsection{Field renormalization counterterms}
\label{app:fieldcounters}

We list the field renormalization counterterms defined in \refeq{eq:drfieldcounterdef} in the $\DRbar$ scheme in the interaction
basis $(H_d,H_u,\widetilde{\nu}_{1R},\widetilde{\nu}_{2R},
\widetilde{\nu}_{3R},\widetilde{\nu}_{1L},
\widetilde{\nu}_{2L},\widetilde{\nu}_{3L})$:
\begin{align}
\delta Z_{11}&=-\frac{\Delta}{16 \pi^2}
	\left( \lambda_i\lambda_i + Y^e_{ij}Y^e_{ij}
	+3\left(Y^d_iY^d_i\right) \right) \; , \\
\delta Z_{1,5+i}&=\frac{\Delta}{16 \pi^2}
	\lambda_j Y^\nu_{ij} \; , \\
\delta Z_{22}&=-\frac{\Delta}{16 \pi^2}
	\left( \lambda_i\lambda_i
	+ Y^\nu_{ij}Y^\nu_{ij}
	+ 3\left( Y^u_{i}Y^u_{i}\right) \right) \; , \\
\delta Z_{2+i,2+j}&=-\frac{\Delta}{8 \pi^2}
	\left( \lambda_i\lambda_j
	+ \kappa_{ikl} \kappa_{jkl}
	+ Y^\nu_{ki} Y^\nu_{kj} \right) \; , \\
\delta Z_{5+i,5+j}&=-\frac{\Delta}{16 \pi^2}
	\left( Y^e_{ki}Y^e_{kj} + Y^\nu_{ik} Y^\nu_{jk} \right) \; ,
\end{align}
where the indices run from 1 to 3.
We checked that the coefficients of the divergent part of the field renormalization counterterms are equal to the one-loop anomalous dimensions of the corresponding superfields $\gamma_{ij}^{(1)}$, neglecting the terms proportional to the gauge couplings $g_1$ and $g_2$, and divided by the loop factor $16 \pi^2$, i.e.,
\begin{equation}
\delta Z_{ij}=\left.\frac{\gamma_{ij}^{(1)}\Delta}{16
  \pi^2}\right|^{g1,g2\to 0}
 \; ,
\end{equation} 
which is the same relation that holds in the (N)MSSM.

%%%%%%%%%%%%%%%%%%%%%%%%%%%%%%%%%%%%%%%%%%%%%%%%%%%%%%%%%%%%%%%%%%%%%%%%%%%%%%%

\subsection{Parameter counterterms}
\label{app:paracounters}

\subsubsection{Renormalization conditions}
In the following we list the renormalization conditions for
the counterterms of the parameter
used to renormalize the neutral scalar potential:
\begin{align}
\delta g_2 &= \frac{g_2}{2} \Biggl(
    \frac{\delta M_W^2}{M_W^2} -
    \frac{\delta v^2}{v^2} \Biggr) \; , \\[4pt]
\delta g_1 &= \frac{2}{g_1 v^2}\left( \delta M_Z^2-\delta M_W^2 \right)
  -\frac{g_1}{2}\frac{\delta v^2}{v^2} \; , \\[4pt]
\delta v^2 &= \frac{4 \SW^2 \MW^2}{e^2}
\left(
\frac{\delta \SW^2}{\SW^2}+\frac{\delta \MW^2}{\MW^2}-
2 \left. \delta Z_e \right|^{\rm div} \right) \; , \\[4pt]
\delta v_d^2 &= \frac{4 v_d}{g_2}
    \Biggl(
        \left. \Sigma_{\widetilde{W}^0\widetilde{H}_d^0}^S
            \right|^{\rm div}
        - \frac{g_2}{4}
        \biggl(
            v_d
            \bigl(
                \delta Z^\chi_{55} +
                \delta Z^\chi_{66}
            \bigr) +
            v_{iL} \delta Z^\chi_{i6}
        \biggr) -
        \frac{v_d}{2} \left. \delta g_2 \right|^{\rm div}
    \Biggr) \; , \\[4pt]
\delta v_u^2 &=
\frac{4 v_u}{g_2}
\Biggl(
    - \left. \Sigma_{\widetilde{W}^0\widetilde{H}_u^0}^S
        \right|^{\rm div}
    - \frac{g_2 v_u}{4}
    \biggl(
        \delta Z^\chi_{55} + \delta Z^\chi_{77}
    \biggr)
    - \frac{v_u}{2} \left. \delta g_2 \right|^{\rm div}
\Biggr) \; , \\[4pt]
\delta Y^\nu_{ij} &=
\frac{\sqrt{2}}{v_u} \left. \Sigma_{\nu_{iL} \nu_{jR}^*}^S \right|^{\rm div}
- \frac{1}{2} \left(
              \delta Z_{ik}^\chi Y^\nu_{kj} -
              \delta Z_{i6}^\chi \lambda_j +
              \delta Z_{7+k,7+j}^\chi Y^\nu_{ik}
        \right)
- \frac{Y^\nu_{ij}}{2} \frac{\delta v_u^2}{v_u^2} \; ,
\\[4pt]
\delta \lambda_i &=
- \frac{\sqrt{2}}{v_u} \left.
  \Sigma_{\widetilde{H}_d^0 \nu_{iR}^*}^S \right|^{\rm div}
+ \frac{1}{2} \left(
                  \delta Z_{j6}^\chi Y^\nu_{ji} -
                  \delta Z_{66}^\chi \lambda_i -
                  \delta Z_{7+i,7+j}^\chi \lambda_j
              \right)
- \frac{\lambda_i}{2} \frac{\delta v_u^2}{v_u^2} \; , \\[4pt]
\delta v_{iL} &=
- \frac{2}{g_1} \left.
  \Sigma_{\nu_{iL} \widetilde{B}}^S \right|^{\rm div}
- \frac{1}{2} \left(
                \delta Z_{ij}^\chi v_{jL} +
                \delta Z_{i6}^\chi v_d +
                \delta Z_{44}^\chi v_{iL}
              \right)
- v_{iL} \frac{\delta g_1}{g_1} \; , \\[4pt]
Y^\nu_{ij} \delta v_{jR} &=
\sqrt{2} \left. \Sigma_{\nu_{iL} \widetilde{H}^0_u}^S \right|^{\rm div} -
\frac{1}{2} \left(
            \delta Z_{ij}^\chi v_{kR} Y^\nu_{jk} -
            \delta Z_{i6}^\chi v_{jR} \lambda_j +
            \delta Z_{77}^\chi v_{jR} Y^\nu_{ij}
            \right) -
            v_{jR} \delta Y^\nu_{ij} \; , \\[4pt]
- v_u Y^\nu_{jk} \delta \kappa_{iij} &=
\left. \Gamma_{\widetilde{\nu}_{iR}^{\mathcal{R}} \widetilde{\nu}_{iR}^{\mathcal{R}}
  \widetilde{\nu}_{jL}^{\mathcal{R}}}^{(1)} \right|^{\rm div} -
\frac{1}{2} \biggl( 2 \delta Z_{2+k,2+i}
                     \Gamma_{\widetilde{\nu}_{kR}^{\mathcal{R}}
                        \widetilde{\nu}_{iR}^{\mathcal{R}}
                           \widetilde{\nu}_{jL}^{\mathcal{R}}}^{(0)} +
                     \delta Z_{1,5+j}
                     \Gamma_{\widetilde{\nu}_{iR}^{\mathcal{R}}
                         \widetilde{\nu}_{iR}^{\mathcal{R}}
                           H_d^{\mathcal{R}}}^{(0)} \notag \\
                 & + \delta Z_{5+k,5+j}
                   \Gamma_{\widetilde{\nu}_{iR}^{\mathcal{R}}
                       \widetilde{\nu}_{iR}^{\mathcal{R}}
                           \widetilde{\nu}_{kL}^{\mathcal{R}}}^{(0)}
          \biggr) -
\biggl(
\lambda_i Y^\nu_{ji} \delta v_d + v_d Y^\nu_{ji} \delta \lambda_i +
  v_d \lambda_i \delta Y^\nu_{32} - v_u \kappa_{iik} \delta Y^\nu_{jk} \notag \\
& - \kappa_{iik} Y^\nu_{jk} \delta v_u - v_u Y^\nu_{jk} \delta \kappa_{iik}
   - v_{kL} Y^\nu_{ki} \delta Y^\nu_{ji}
  - v_{kL} Y^\nu_{ji} \delta Y^\nu_{ki}- Y^\nu_{ji} Y^\nu_{ki} \delta v_{kL}
\biggr) \; , \\[4pt]
\delta \kappa_{123} &=
\frac{\sqrt{2}}{v_{1R}} \left. \Sigma_{\nu_{2R} \nu_{3R}}^S \right|^{\rm div} -
\frac{1}{2 v_{1R}} \biggl(
           \delta Z_{9,7+i}^\chi v_{jR} \kappa_{i3j} +
           v_{jR} \kappa_{i2j} \delta Z_{7+i,10}^\chi
           \biggr) \notag \\ & -
\frac{1}{v_{1R}} \biggl(
                 v_{2R} \delta \kappa_{223} +
                 v_{3R} \delta \kappa_{233} +
                 \kappa_{i23} \delta v_{iR}
                 \biggr)
\; , \\[4pt]
\delta T^\lambda_i & =
\sqrt{2} \left. 
  \Gamma_{H_d^{\mathcal{R}} H_u^{\mathcal{R}}
    \widetilde{\nu}_{iR}^{\mathcal{R}}}^{(1)}
    \right|^{\rm div} -
\frac{1}{\sqrt{2}}
  \biggl(
  \delta Z_{11} \Gamma_{H_d^{\mathcal{R}} H_u^{\mathcal{R}}
    \widetilde{\nu}_{iR}^{\mathcal{R}}} +
  \delta Z_{5+j,1} \Gamma_{ \widetilde{\nu}_{jL}^{\mathcal{R}} H_u^{\mathcal{R}}
    \widetilde{\nu}_{iR}^{\mathcal{R}}} +
  \delta Z_{22} \Gamma_{H_d^{\mathcal{R}} H_u^{\mathcal{R}}
    \widetilde{\nu}_{iR}^{\mathcal{R}}} \notag \\ & +
  \delta Z_{2+j,2+i} \Gamma_{H_d^{\mathcal{R}} H_u^{\mathcal{R}}
    \widetilde{\nu}_{jR}^{\mathcal{R}}}
  \biggr) -
\sqrt{2}
  \biggl(
  \kappa_{ijk} \lambda_j \delta v_{kR} +
  \kappa_{ijk} v_{kR} \delta \lambda_j +
  \lambda_j v_{kR} \delta \kappa_{ijk}
  \biggr)
\; , \\[4pt]
\delta T^\nu_{ij} &=
- \sqrt{2} \left. \Gamma_{H_u^{\mathcal{R}} \widetilde{\nu}_{iL}^{\mathcal{R}}
  \widetilde{\nu}_{jR}^{\mathcal{R}}}^{(1)} \right|^{\rm div}
+ \frac{1}{\sqrt{2}}
  \biggl(
  \delta Z_{22} \Gamma_{H_u^{\mathcal{R}} \widetilde{\nu}_{iL}^{\mathcal{R}}
  \widetilde{\nu}_{jR}^{\mathcal{R}}}^{(0)} +
  \delta Z_{2+k,2+j} \Gamma_{H_u^{\mathcal{R}} \widetilde{\nu}_{iL}^{\mathcal{R}}
  \widetilde{\nu}_{kR}^{\mathcal{R}}}^{(0)} \notag \\ & +
  \delta Z_{1,5+i} \Gamma_{H_u^{\mathcal{R}} H_d^{\mathcal{R}}
  \widetilde{\nu}_{jR}^{\mathcal{R}}}^{(0)} +
  \delta Z_{5+k,5+i} \Gamma_{H_u^{\mathcal{R}} \widetilde{\nu}_{kL}^{\mathcal{R}}
  \widetilde{\nu}_{jR}^{\mathcal{R}}}^{(0)}
  \biggr) -
\sqrt{2}
  \biggl(
  Y^\nu_{ik} \kappa_{jkl} \delta v_{lR} +
  \kappa_{jkl} v_{lR} \delta Y^\nu_{ik} \notag \\ & +
  Y^\nu_{ik} v_{lR} \delta \kappa_{jkl}
  \biggr)
\; , \\[4pt]
\delta T^\kappa_{ijk} &=
- \frac{1}{\sqrt{2}} \left. \Gamma_{\widetilde{\nu}_{iR}^{\mathcal{R}}
  \widetilde{\nu}_{jR}^{\mathcal{R}}
  \widetilde{\nu}_{kR}^{\mathcal{R}}}^{(1)} \right|^{\rm div}
+ \frac{1}{2\sqrt{2}}
  \biggl(
  \delta Z_{li} \Gamma_{\widetilde{\nu}_{lR}^{\mathcal{R}}
    \widetilde{\nu}_{jR}^{\mathcal{R}}
    \widetilde{\nu}_{kR}^{\mathcal{R}}}^{(0)} +
  \delta Z_{lj} \Gamma_{\widetilde{\nu}_{iR}^{\mathcal{R}}
    \widetilde{\nu}_{lR}^{\mathcal{R}}
    \widetilde{\nu}_{kR}^{\mathcal{R}}}^{(0)} +
  \delta Z_{lk} \Gamma_{\widetilde{\nu}_{iR}^{\mathcal{R}}
    \widetilde{\nu}_{jR}^{\mathcal{R}}
    \widetilde{\nu}_{lR}^{\mathcal{R}}}^{(0)}
  \biggr)  \notag \\ & -
2 \bigl(
  \kappa_{klm} v_{mR} \delta \kappa_{ijl} +
  \kappa_{ijl} v_{mR} \delta \kappa_{klm} +
  \kappa_{ijl} \kappa_{klm} \delta v_{mR} +
  \kappa_{jkl} v_{mR} \delta \kappa_{ilm} \notag \\ & +
  \kappa_{ilm} v_{mR} \delta \kappa_{jkl} +
  \kappa_{ilm} \kappa_{jkl} \delta v_{mR} +
  \kappa_{jlm} v_{mR} \delta \kappa_{ikl} +
  \kappa_{ikl} v_{mR} \delta \kappa_{jlm} \notag \\ & +
  \kappa_{ikl} \kappa_{jlm} \delta v_{mR}
  \bigr)
\; , \\[4pt]
\delta \left( m_{H_d \widetilde{L}}^2 \right)_i &=
\left. \delta \left( m_\sigma^2 \right)_{1,5+i}\right|^{\rm div} +
\frac{1}{2} \biggl(
Y^\nu_{ij} \lambda_j \delta v_u^2 +
v_u^2 \lambda_j \delta Y^\nu_{ij} +
v_u^2 Y^\nu_{ij} \delta \lambda_j +
\lambda_j v_{kR} Y^\nu_{ik} \delta_{jR} \notag \\ & +
v_{jR} v_{kR} Y^\nu_{ik} \delta \lambda_j +
v_{jR} \lambda_j Y^\nu_{ik} \delta v_{kR} +
v_{jR} \lambda_j v_{kR} \delta Y^\nu_{ik}
\biggr)
\; , \\[4pt]
\delta \left( m_{\widetilde{\nu}}^2 \right)_{ij} &=
\left. \delta \left( m_\sigma^2 \right)_{2+i,2+j}\right|^{\rm div} -
\frac{1}{2} \biggl(
Y^\nu_{ki} Y^\nu_{kj} \delta v_u^2 +
v_u^2 Y^\nu_{kj} \delta Y^\nu_{ki} +
v_u^2 Y^\nu_{ki} \delta Y^\nu_{kj}
\biggr) -
v_u \kappa_{ijk} \lambda_k \delta v_d \notag \\ & -
v_d \kappa_{ijk} \lambda_k \delta v_u -
v_d v_u \lambda_k \delta \kappa_{ijk} -
v_d v_d \kappa_{ijk} \delta \lambda_k +
\sqrt{2} \biggl(
T^\kappa_{ijk} \delta v_{kR} +
v_{kR} \delta T^\kappa_{ijk}
\biggr) \notag \\ & +
Y^\nu_{lk} \kappa_{ijk} v_{lL} \delta v_u +
v_u \kappa_{ijk} v_{lL} \delta Y^\nu_{lk} +
v_u Y^\nu_{lk} v_{lL} \delta \kappa_{ijk} +
v_u Y^\nu_{lk} \kappa_{ijk} \delta v_{lL} \notag \\ & -
\frac{1}{2} \biggl(
Y^\nu_{ki} v_{lL} Y^\nu_{lj} \delta v_{kL} +
v_{kL} v_{lL} Y^\nu_{lj} \delta Y^\nu_{ki} +
v_{kL} Y^\nu_{ki} Y^\nu_{lj} \delta v_{lL} +
v_{kL} Y^\nu_{ki} v_{lL} \delta Y^\nu_{lj}
\biggr) \notag \\ & -
\kappa_{jkl} v_{lR} v_{mR} \delta \kappa_{ikm} -
\kappa_{ikm} v_{lR} v_{mR} \delta \kappa_{jkl} -
\kappa_{ikm} \kappa_{jkl} v_{mR} \delta v_{lR} -
\kappa_{ikm} \kappa_{jkl} v_{lR} \delta v_{mR} \notag \\ & +
\kappa_{klm} v_{lR} v_{mR} \delta \kappa_{ijk} +
\kappa_{ijk} v_{lR} v_{mR} \delta \kappa_{klm} +
\kappa_{ijk} \kappa_{klm} v_{mR} \delta v_{lR} +
\kappa_{ijk} \kappa_{klm} v_{lR} \delta v_{mR} \notag \\ & +
\frac{1}{2} \biggl(
v_{kL} Y^\nu_{kj} \lambda_i \delta v_d +
v_d Y^\nu_{kj} \lambda_i \delta v_{kL} +
v_d v_{kL} \lambda_i \delta Y^\nu_{kj} +
v_d v_{kL} Y^\nu_{kj} \delta \lambda_i
\biggr) \notag \\ & +
\frac{1}{2} \biggl(
v_{kL} Y^\nu_{ki} \lambda_j \delta v_d +
v_d Y^\nu_{ki} \lambda_j \delta v_{kL} +
v_d v_{kL} \lambda_j \delta Y^\nu_{ki} +
v_d v_{kL} Y^\nu_{ki} \delta \lambda_j
\biggr) \notag \\ & -
\frac{1}{2} \biggl(
\lambda_i \lambda_j \delta v_d^2 +
\lambda_i \lambda_j \delta v_u^2 +
v_d^2 \lambda_j \delta \lambda_i +
v_d^2 \lambda_i \delta \lambda_j +
v_u^2 \lambda_j \delta \lambda_i \notag \\ & +
v_u^2 \lambda_i \delta \lambda_j
\biggr)
\quad \text{for}
    \quad i\neq j \; , \\[4pt]
\delta \left( m_{\widetilde{L}}^2 \right)_{ij} &=
\left. \delta \left( m_\sigma^2 \right)_{5+i,5+j} \right|^{\rm div} -
\frac{1}{2} \biggl(
Y^\nu_{ik} Y^\nu_{jk} \delta v_u^2 +
v_u^2 Y^\nu_{jk} \delta Y^\nu_{ik} +
v_u^2 Y^\nu_{ik} \delta Y^\nu_{jk} +
Y^\nu_{ik} v_{lR} Y^\nu_{jl} \delta v_{kR} \notag \\ & +
v_{kR} v_{lR} Y^\nu_{jl} \delta Y^\nu_{ik} +
v_{kR} Y^\nu_{ik} Y^\nu_{jl} \delta v_{lR} +
v_{kR} Y^\nu_{ik} v_{lR} \delta Y^\nu_{jl}
\biggr)
\quad \text{for} \quad i\neq j \; ,
\end{align}
where
\begin{equation}
\delta Z_e=
\left[
\frac{1}{2} \left( \frac{\partial \Sigma_{\gamma\gamma}^T}{\partial p^2} \left( 0 \right) \right) +
\frac{\SW}{\CW \MZ^2} \Sigma_{\gamma Z}^T \left( 0 \right)
\right] \; ,
\end{equation}
with $\Sigma_{\gamma\gamma}^T(0)$ the transverse part of the photon
self-energy and $\Sigma_{\gamma Z}^T$ the transverse part of the
mixed photon-Z boson self-energy, and
\begin{align}
\Gamma_{\widetilde{\nu}_{iR}^{\mathcal{R}} \widetilde{\nu}_{kR}^{\mathcal{R}}
\widetilde{\nu}_{jL}^{\mathcal{R}}}^{(0)} &=
\frac{1}{2} v_d \lambda_k Y^\nu_{ji} +
\frac{1}{2} v_d \lambda_i Y^\nu_{jk} -
\frac{1}{2} v_{lL} Y^\nu_{li} Y^\nu_{jk} -
\frac{1}{2} v_{lL} Y^\nu_{lk} Y^\nu_{ji} -
v_u \kappa_{ikl} Y^\nu_{jl} \; , \\[4pt]
\Gamma_{\widetilde{\nu}_{iR}^{\mathcal{R}} \widetilde{\nu}_{jR}^{\mathcal{R}}
 H_d^{\mathcal{R}}}^{(0)} &=
-v_d \lambda_i \lambda_j +
v_u \kappa_{ijk} \lambda_k +
\frac{1}{2} \left(
      Y^\nu_{ki} \lambda_j v_{kL} +
      Y^\nu_{kj} \lambda_i v_{kL}
      \right)
\; , \\[4pt]
\Gamma_{H_d^{\mathcal{R}} H_u^{\mathcal{R}}
  \widetilde{\nu}_{iR}^{\mathcal{R}}}^{(0)} &=
\frac{T^\lambda_i}{\sqrt{2}} +
\kappa_{ijk} \lambda_j v_{kR}
\; , \\[4pt]
\Gamma_{H_u^{\mathcal{R}} \widetilde{\nu}_{iL}^{\mathcal{R}}
  \widetilde{\nu}_{jR}^{\mathcal{R}}}^{(0)} &=
- \frac{T^\nu_{ij}}{\sqrt{2}}
- Y^\nu_{ik} \kappa_{jkl} v_{lR}
\; , \\[4pt]
\Gamma_{\widetilde{\nu}_{iR}^{\mathcal{R}}
\widetilde{\nu}_{jR}^{\mathcal{R}} \widetilde{\nu}_{kR}^{\mathcal{R}}}^{(0)} &=
- \sqrt{2} T^\kappa_{ijk}
- 2 \bigl(
  \kappa_{ijl} \kappa_{klm} v_{mR} +
  \kappa_{ilm} \kappa_{jkl} v_{mR} +
  \kappa_{ikl} \kappa_{jlm} v_{mR}
  \bigr) \; .
\end{align}

\subsubsection{Explicit form}
\label{app:explicitCT}
We list the explicit form of the counterterms of the free parameters renormalized in the $\DRbar$ scheme:
\begin{align}
\delta\kappa_{ijk}&=
\frac{\Delta}{16\pi^2}\Bigl(
Y^\nu_{lk} Y^\nu_{lm} \kappa_{ijm}
+ Y^\nu_{lj} Y^\nu_{lm} \kappa_{ikm}
+ Y^\nu_{li} Y^\nu_{lm} \kappa_{jkm}
+ \kappa_{ikl} \kappa_{lmn} \kappa_{jmn}
+ \kappa_{ijl} \kappa_{lmn} \kappa_{kmn} \notag \\
& + \kappa_{ilm} \kappa_{lmn} \kappa_{jkn}
+ \kappa_{jkl} \lambda_l \lambda_i
+ \kappa_{ikl} \lambda_l \lambda_j
+ \kappa_{ijl} \lambda_l \lambda_k
\Bigr) \; , \label{eq:dkap} \\[4pt]
\delta\lambda_i&=
\frac{\Delta}{32\pi^2}\Biggl(
\biggl(
- \frac{4 \pi \alpha (\SW^2 + 3 \CW^2)}{\CW^2 \SW^2}
+ 4 \lambda_j \lambda_j
+ 3 (Y^d_j Y^d_j + Y^u_j Y^u_j)
+ Y^e_{jk} Y^e_{jk}
+ Y^\nu_{jk} Y^\nu_{jk}
\biggr) \lambda_i \notag \\
& + 3 Y^\nu_{ji} Y^\nu_{jk} \lambda_k
+ 2 \kappa_{ijk} \kappa_{jkl} \lambda_l
\Biggr) \; , \\[4pt]
\delta Y^\nu_{ij}&=
\frac{\Delta}{32\pi^2}\Biggl(
\biggl(
-\frac{4\pi\alpha Y^\nu_i\left( \SW^2+3\CW^2 \right)}{\CW^2\SW^2}
+ \lambda_k \lambda_k
+ 3 Y^u_k Y^u_k
+ Y^\nu_{kl} Y^\nu_{kl}
\biggr) Y^\nu_{ij}
+ Y^e_{ki} Y^e_{kl} Y^\nu_{lj} \notag \\
& + 3 Y^\nu_{ik} Y^\nu_{lk} Y^\nu_{lj}
+ 2 Y^\nu_{ik} \kappa_{klm} \kappa_{jlm}
+ 3 Y^\nu_{ik} \lambda_k \lambda_j
\Biggr) \; , \\[4pt]
\delta T^\kappa_{ijk}&=
\frac{\Delta}{16\pi^2}\biggl(
2 Y^\nu_{ml} \kappa_{jkl} T^\nu_{mi}
+ 2 Y^\nu_{ml} \kappa_{ikl} T^\nu_{mj}
+ 2 Y^\nu_{ml} \kappa_{ijl} T^\nu_{mk}
+ Y^\nu_{lk} Y^\nu_{lm} T^\kappa_{ijm}
+ Y^\nu_{lj} Y^\nu_{lm} T^\kappa_{ikm} \notag \\
& + Y^\nu_{li} Y^\nu_{lm} T^\kappa_{jkm}
+ 2 \kappa_{jkl} \kappa_{lmn} T^\kappa_{imn}
+ 2 \kappa_{ikl} \kappa_{lmn} T^\kappa_{jmn}
+ 2 \kappa_{ijl} \kappa_{lmn} T^\kappa_{kmn}
+ \kappa_{klm} \kappa_{lmn} T^\kappa_{ijn} \notag \\
& + \kappa_{jlm} \kappa_{lmn} T^\kappa_{ikn}
+ \kappa_{ilm} \kappa_{lmn} T^\kappa_{jkn}
+ \lambda_l T^\kappa_{jkl} \lambda_i
+ \lambda_l T^\kappa_{ikl} \lambda_j
+ \lambda_l T^\kappa_{ijl} \lambda_k
+ 2 \kappa_{jkl} \lambda_l T^\lambda_i \notag \\
& + 2 \kappa_{ikl} \lambda_l T^\lambda_j
+ 2 \kappa_{ijl} \lambda_l T^\lambda_k
\biggr) \; , \\[4pt]
\delta T^\lambda_i &=
\frac{\Delta}{32\pi^2}\Biggl(
\biggl(
- \frac{4 \pi \alpha (\SW^2 + 3 \CW^2)}{\CW^2 \SW^2}
+ 6 \lambda_j \lambda_j
+ 3 (Y^d_j Y^d_j + Y^u_j Y^u_j)
+ Y^e_{jk} Y^e_{jk}
+ Y^\nu_{jk} Y^\nu_{jk}
\biggr) T^\lambda_i \notag \\
& + \biggl( \frac{8 \pi \alpha(M_1 \SW^2 + 3 M_2 \CW^2)}{\CW^2 \SW^2}
+ 6 \lambda_j T^\lambda_j
+ 6 Y^d_{j} T^d_{j}
+ 2 Y^e_{jk} T^e_{jk}
+ 6 Y^u_{j} T^u_{j}
+ 2 Y^\nu_{jk} T^\nu_{jk} \biggr) \lambda_i \notag \\
& + 5 Y^\nu_{kj} \lambda_j T^\nu_{ki}
+ 4 Y^\nu_{ji} Y^\nu_{jk} T^\lambda_k
+ 4 \kappa_{jkl} \lambda_j T^\kappa_{ikl}
+ 2 \kappa_{ijk} \kappa_{jkl} T^\lambda_l
\Biggr) \; , \\[4pt]
\delta T^\nu_{ij}&=
\frac{\Delta}{32\pi^2}\Biggl(
\biggl(
- \frac{4 \pi \alpha (\SW^2 + 3 \CW^2)}{\CW^2 \SW^2}
+ \lambda_k \lambda_k
+ 3 Y^u_k Y^u_k
+ Y^\nu_{kl}
\biggr) T^\nu_{ij}
+ \biggl(
\frac{8 \pi \alpha(M_1 \SW^2 + 3 M_2 \CW^2)}{\CW^2 \SW^2} \notag \\
& + 2 \lambda_k T^\lambda_k
+ 6 Y^u_k T^u_k
+ 2 Y^\nu_{kl} T^\nu_{kl}
\biggr) Y^\nu_{ij}
+ 2 Y^e_{lk} Y^\nu_{kj} T^e_{li}
+ 4 Y^\nu_{kj} Y^\nu_{kl} T^\nu_{il}
+ Y^e_{ki} Y^e_{kl} T^\nu_{lj} \notag \\
& + 5 Y^\nu_{ik} Y^\nu_{lk} T^\nu_{lj}
+ 4 Y^\nu_{ik} \kappa_{klm} T^\kappa_{jlm}
+ 2 \kappa_{jkl} \kappa_{klm} T^\nu_{im}
+ 4 \lambda_k T^\nu_{ik} \lambda_j
+ 5 Y^\nu_{ik} \lambda_k T^\lambda_j
\Biggr) \; , \\[4pt]
\delta v_{iR}^2 &=
- \frac{\Delta}{8\pi^2} v_{iR} \delta_{ij}
\biggl(
Y^\nu_{kj} Y^\nu_{kl} v_{lR}
+ \kappa_{jkm} \kappa_{klm} v_{lR}
+ v_{kR} \lambda_k \lambda_j
\biggr) \; , \\[4pt]
\delta v_{iL}^2 &=
\frac{\Delta}{16\pi^2} v_{iL}  \delta_{ij} \left(
\frac{2 \pi \alpha \left( \SW^2 + 3 \CW^2 \right) v_{jL}}{\SW^2 \CW^2}
+ v_d Y^\nu_{jk} \lambda_k
- \left( Y^e_{kj} Y^e_{kl}
- Y^\nu_{jk} Y^\nu_{lk} \right) v_{lL}
\right) \; , \\[4pt]
\delta v^2 &=
- \frac{\Delta}{16 \pi^2} \Biggl(
- \frac{2 \pi \alpha \left( \SW^2+3\CW^2\right) v^2}{\SW^2\CW^2}
+ Y^e_{ij} Y^e_{ik} v_{jL} v_{kL}
+ v_d^2 \biggl( 
\lambda_i \lambda_i
+ 3 Y^d_i Y^d_i
+ Y^e_{ij} Y^e_{ij}
\biggr) \notag \\
& + Y^\nu_{ji} Y^\nu_{ki} v_{jL} v_{kL}
+ v_u^2 \biggl(
\lambda_i \lambda_i
+ 3 Y^u_i Y^u_i
+ Y^\nu_{ij} Y^\nu_{ij}
\biggr)
- 2 v_d \lambda_i Y^\nu_{ji} v_{jL}
\Biggr) \; , \\[4pt]
\delta\tb &=
\frac{\Delta}{32\pi^2} \tb \left(
3 \left( Y^d_i Y^d_i 
- Y^u_iY^u_i \right)
+ Y^e_{ij} Y^e_{ij}
- Y^\nu_{ij} Y^\nu_{ij}
- \frac{1}{v_d} \lambda_i Y^\nu_{ji} v_{jL}
\right) \label{eq:dtbdrbar} \; , \\[4pt]
\delta\left(m_{H_d\widetilde{L}}^2\right)_i &=
-\frac{\Delta}{32\pi^2} \left(
\left( m_{H_d}^2 + 2 m_{H_u}^2 \right) Y^\nu_{ij} \lambda_j
+ 2 T^\nu_{ij} T^\lambda_j
+ \left(
2 \left( m_{\widetilde{\nu}}^2 \right)_{jk} Y^\nu_{ij}
+ \left( m_{\widetilde{L}}^2 \right)_{ji} Y^\nu_{jk}
\right) \lambda_k
\right) \label{eq:dmhl2} \; , \\[4pt]
\delta \left(m_{\widetilde{L}}^2\right)_{ij} &=
\frac{\Delta}{32\pi^2} \Biggl(
2 m_{H_d}^2 Y^e_{ki} Y^e_{kj}
+ 2 m_{H_u}^2 Y^\nu_{ik} Y^\nu_{jk}
- \left(m_{H_d\widetilde{L}}^2\right)_j Y^\nu_{ik} \lambda_k
+ 2 T^e_{ki} T^e_{kj}
+ 2 T^\nu_{ik} T^\nu_{jk} \notag \\
& + \left( m_{\widetilde{L}}^2 \right)_{jk} Y^e_{lk} Y^e_{li}
+ 2 \left( m_{\widetilde{e}}^2 \right)_{kl} Y^e_{ki} Y^e_{lj}
+ \left( m_{\widetilde{L}}^2 \right)_{ki} Y^e_{lk} Y^e_{lj}
+ \left( m_{\widetilde{L}}^2 \right)_{jk} Y^\nu_{kl} Y^\nu_{il} \notag \\
& + 2 \left( m_{\widetilde{\nu}}^2 \right)_{kl} Y^\nu_{ik} Y^\nu_{jl}
+ \left( m_{\widetilde{L}}^2 \right)_{ki} Y^\nu_{kl} Y^\nu_{jl}
\Biggr) \quad \text{for} \quad i\neq j \; , \label{eq:dmv2} \\[4pt]
\delta \left(m_{\widetilde{\nu}}^2\right)_{ij} &=
\frac{\Delta}{16\pi^2} \Biggr(
2 m_{H_u}^2 Y^\nu_{ki} Y^\nu_{kj}
+ 2 T^\nu_{ki} T^\nu_{kj}
+ 2 T^\kappa_{ikl} T^\kappa_{jkl}
+ 2 \left( m_{\widetilde{L}}^2 \right)_{kl} Y^\nu_{kj} Y^\nu_{li}
+ \left( m_{\widetilde{\nu}}^2 \right)_{kj} Y^\nu_{lk} Y^\nu_{li} \notag \\
& + \left( m_{\widetilde{\nu}}^2 \right)_{ik} Y^\nu_{lk} Y^\nu_{lj}
+ 4 \left( m_{\widetilde{\nu}}^2 \right)_{kl} \kappa_{jkm} \kappa_{ilm}
+ \left( m_{\widetilde{\nu}}^2 \right)_{kj} \kappa_{klm} \kappa_{ilm}
+ \left( m_{\widetilde{\nu}}^2 \right)_{ik} \kappa_{klm} \kappa_{jlm} \notag \\
& - 2 \left(m_{H_d\widetilde{L}}^2\right)_k Y^\nu_{kj} \lambda_i
+ \left( m_{\widetilde{\nu}}^2 \right)_{kj} \lambda_k \lambda_i
+ \left( m_{\widetilde{\nu}}^2 \right)_{ik} \lambda_k \lambda_j \notag \\
& + 2 \biggl(
\bigl( m_{H_d}^2 + m_{H_u}^2 \bigr) \lambda_i \lambda_j
+ T^\lambda_i T^\lambda_j
\biggr)
\Biggl) \quad \text{for} \quad i\neq j \; . \label{eq:dmv2}
\end{align}
The counterterms in \refeqs{eq:dkap}-(\ref{eq:dtbdrbar}) were
calculated diagrammatically using our \FA\ model file and afterwards checked
to fulfill the one-loop relation
\begin{equation}
\delta X = \frac{\beta_X^{(1)}\Delta}{32\pi^2} \; ,
\end{equation}
where $\delta X$ stands for one of the counterterms just mentioned, and $\beta_X^{(1)}$ is the one-loop coefficient of the beta function of the parameter $X$, which could be obtained by the help of the
mathematica package \texttt{SARAH}~\cite{Staub:2010jh}.

On the contrary, the counterterms of the soft masses stated in 
\refeqs{eq:dmhl2}-(\ref{eq:dmv2}) are the ones derived from
the one-loop beta function we obtained with \texttt{SARAH}, which were then
numerically checked to be equal to the counterterms
for $(m_{H_d\widetilde{L}}^2)_i$ and $(m_{\widetilde{L}}^2)_{ij}$ we
calculated diagrammatically in the $\cp$-odd scalar sector.

%%%%%%%%%%%%%%%%%%%%%%%%%%%%%%%%%%%%%%%%%%%%%%%%%%%%%%%%%%%%%%%%%%%%%%%%%%%%%%%
%%%%%%%%%%%%%%%%%%%%%%%%%%%%%%%%%%%%%%%%%%%%%%%%%%%%%%%%%%%%%%%%%%%%%%%%%%%%%%%

\section{Standard model values}\label{app:smvalues}

\refta{tab:smsum} summarizes the values for the SM-like parameters we chose in our calculation.
%%%%%%%%%%%%%%%%%%%%%%%%%% T A B L E %%%%%%%%%%%%%%%%%%%%%%%%%%%%%%%%%%%%%%%%%%
\begin{table}[H]
\renewcommand{\arraystretch}{1.7}
\centering
\begin{tabular}{c c c c}
 $m_t^{\text{OS}}$ & $m_t^{\overline{\text{MS}}}(m_t)$ &
   $ m_b^{\overline{\text{MS}}}(m_b)$ &
   $m_\tau$ \\
 \hline
 $172.5$ & $167.48$ & $4.16$ & $\num{1.7792}$ \\
 \hline
 \hline
 $\MW$ & $\MZ$ & $v$ \\
 \hline
 $\num{80.385}$ &
 	$\num{91.1875}$ & $\num{246.2196}$ 
\end{tabular}
\caption{Values for parameters of the standard model in GeV.}
\label{tab:smsum}
\renewcommand{\arraystretch}{1.0}
\end{table}
%%%%%%%%%%%%%%%%%%%%%%%%%% T A B L E %%%%%%%%%%%%%%%%%%%%%%%%%%%%%%%%%%%%%%%%%%
\noindent
The value for $v$ corresponds to a value for the Fermi constant of
$G_F=\SI{1.16638e-5}{GeV^{-2}}$. The values for the
gauge boson masses
define the cosine of the weak mixing angle to be
$\CW=\num{0.881535}$. Note that since the SM leptons mix
with the Higgsinos and gauginos in the \mnSSM, the lepton masses are not
the real phyiscal input parameters. However, the mixing is tiny, so there
will always be three mass eigenstates in the charged fermion sector
corresponding to the three standard model leptons, having
approximately the masses $m_e$, $m_\mu$ and $m_\tau$. This is why we use
the measured values for these masses, such as $m_\tau$ in
\refta{tab:smsum}, and then calculate the
real input parameters, which are the Yukawa couplings
\begin{equation}
Y^e_1=\frac{\sqrt{2}m_e}{v_d} \; , \quad Y^e_2=\frac{\sqrt{2}m_\mu}{v_d} \; , 
\quad Y^e_3=\frac{\sqrt{2}m_\tau}{v_d} \; .
\end{equation}

\newpage

\bibliography{munuSSM}

\end{document}